\documentclass[12pt]{article}
\usepackage{geometry}
\usepackage[onehalfspacing]{setspace}
\usepackage{amssymb}
\usepackage{amsfonts}
\usepackage{graphicx}
\usepackage{amsmath}
\usepackage{epstopdf}
\usepackage{natbib}
\usepackage{bibunits}
\usepackage{morefloats}
\usepackage{float}
\usepackage{comment}
\usepackage{pdflscape}
\usepackage{adjustbox}
\usepackage{subcaption}
\usepackage{longtable}
\usepackage{threeparttablex}
\usepackage[dvipsnames]{xcolor}
\usepackage{txfonts}

\usepackage{booktabs}

\usepackage[final]{pdfpages}
\usepackage[nice]{nicefrac}
\usepackage{amsmath}
\usepackage{xfrac}

\usepackage{mathpazo} 
\usepackage[scaled]{helvet} 
\usepackage{courier} 
\normalfont
\usepackage[T1]{fontenc}

\usepackage[pdftex,bookmarks,colorlinks]{hyperref}
\hypersetup{
  colorlinks,
  citecolor=blue,
  linkcolor=red,
  urlcolor=Green}

\usepackage[compact]{titlesec}
\titlespacing{\section}{0pt}{0pt}{8pt}
\titlespacing{\subsection}{0pt}{2pt}{5pt}
\titlespacing{\subsubsection}{0pt}{3pt}{5pt}

\usepackage{enumitem}
\setlist[enumerate]{nosep,topsep={2pt},partopsep={2pt}}

\usepackage[bottom]{footmisc}

  {\list{}{\leftmargin=#1\rightmargin=#1}\item[]}%
  {\endlist}

\begin{document}

\title{\Huge Macroeconomic Data Transformations Matter\thanks{%
We thank the Editor Esther Ruiz, two anonymous referees, and Hugo Couture who
provided excellent research assistance. We acknowledge
financial support from the Chaire en macro\'{e}conomie et pr\'{e}visions ESG
UQAM.}}
\author{Philippe Goulet Coulombe$^1$\thanks{%
Corresponding Author: \href{mailto:gouletc@sas.upenn.edu}{{gouletc@sas.upenn.edu}}. Department of Economics, UPenn.}  \and Maxime Leroux$^2$ \and Dalibor Stevanovic$^2$\thanks{%
Corresponding Author: \href{mailto:dstevanovic.econ@gmail.com}{{dstevanovic.econ@gmail.com}}. D\'{e}partement des sciences \'{e}conomiques, UQAM.} \and St\'{e}phane Surprenant$^2$ }
\date{%
    $^1$University of Pennsylvania\\%
    $^{2}$Universit\'{e} du Qu\'{e}bec \`{a} Montr\'{e}al\\[2ex]%
This version: \today \\
\vspace{0.4cm}}
\maketitle

\begin{abstract}
\noindent In a low-dimensional linear regression setup, considering linear transformations/combinations of predictors does not alter predictions. However, when the forecasting technology either uses shrinkage or is nonlinear, it does. This is precisely the fabric of the machine learning (ML) macroeconomic forecasting environment. Pre-processing of the data translates to an alteration of the regularization -- explicit or implicit -- embedded in ML algorithms. We review old transformations and propose new ones, then empirically evaluate their merits in a substantial pseudo-out-sample exercise. It is found that traditional factors should almost always be included as predictors and moving average rotations of the data can provide important gains for various forecasting targets.  Also, we note that while predicting directly the average growth rate is equivalent to averaging separate horizon forecasts when using OLS-based techniques, the latter can substantially improve on the former when regularization and/or nonparametric nonlinearities are involved.

\end{abstract}

\vspace{2cm}


\clearpage



\section{Introduction}
 
Following the recent enthusiasm for Machine Learning (ML) methods and widespread availability of big data, macroeconomic forecasting research gradually evolved further and further away from the traditional tightly specified OLS regression. Rather, nonparametric non-linearity and regularization of many forms are slowly taking the center stage, largely because they can provide sizable forecasting gains when compared with traditional methods (see, among others, \cite{kim2018mining,medeiros2019forecasting,gclss2020,goulet2020macroeconomy}), even during the Covid-19 episode \citep{gcms2021}. In such environments, different linear transformations of the informational set $X$ \textit{can} change the prediction and taking first differences may \textit{not} be the optimal transformation for many predictors, despite the fact that it guarantees viable frequentist inference. For instance, in penalized regression problems -- like Lasso or Ridge --, different rotations of $X$ imply different priors on $\beta$ in the original regressor space. Moreover, in tree-based models algorithms, since the problem of inverting a near singular matrix $X'X$ simply does not happen, making the use of more persistent (and potentially highly cross-correlated regressors) much less harmful. In sum, in the ML macro forecasting environment, traditional data transformations -- such as those designed to enforce stationarity \citep{mccracken2016fred} -- may leave some forecasting gains on the table. To provide guidance for the growing number of researchers and practitioners in the field, we conduct an extensive pseudo-out-of-sample forecasting exercise to evaluate the virtues of standard and newly proposed data transformations. 

From the ML perspective, it is often suggested that a "feature engineering" step may improve algorithms' performance \citep{kuhn2019feature}. This is especially true of Random Forests (RF) and Boosted Trees (BT), two regression tree ensembles widely regarded as the most performing off-the-shelf algorithms within the modern ML canon \citep{hastie2009elements}. Among other things, both successfully handle a high-dimensional $X$ by recruiting relevant predictors in a sea of useless ones. This implies the data scientist leveraging some domain knowledge can create plausibly more salient features out of the original data matrix, and let the algorithm decide whether to use them or not. Of course, an extremely flexible model, like a neural network with many layers, could very well create those relevant transformations internally in a data-driven way. Yet, this idyllic scenario is a dead end when data points are few, regressors are numerous, and a noisy $y$ serves as a prediction target. This sort of environment, of which macroeconomic forecasting is a notable example, will often  benefit from any prior knowledge one can incorporate in the model. Since transforming the data transforms the prior, doing so properly by including well-motivated rotations of $X$ has the power to increase ML performance on such challenging data sets.

Macroeconomic modelers have been thinking about designing successful priors for a long time. There is a wide literature on Bayesian Vector Autoregressions (VAR) starting with \cite{litterman1984}. Even earlier on, the penalized/restricted estimation of lag polynomials was extensively studied \citep{almon1965distributed,shiller1973distributed}. The motivation for both strands of work is the large ratio of parameters to observations. Forty years later, many more data points are available, but models have grown in complexity. Consequently, large VARs \citep{banbura2010large} and MIDAS regression \citep{ghysels2004midas} still use those tools to regularize over-parametrized models. ML algorithms, usually allowing for sophisticated functional forms, also critically rely on shrinkage. However, when it comes to nonlinear nonparametric methods -- especially Boosting and Random Forests -- there are no explicit parameters to penalize.  Nevertheless, in the case of RF, the ensuing ensemble averaging prediction benefits from ridge-like shrinkage as randomization allows each feature to contribute to the prediction, albeit in a moderate way  \citep{hastie2009elements,mentch2019randomization}. Just like rotating regressors changes the prior in a Ridge regression (see discussion in \cite{GC2019}), rotating regressors in such algorithms will alter the implicit shrinkage scheme -- i.e., move the prior mean away from the traditional zero. This motivates us to propose two rotations of $X$ that implicitly implement a more time-series-friendly prior in ML models: moving average factors (MAF) and moving average rotation of $X$ (MARX). Other than those motivated above, standard transformations are also being studied. This includes factors extracted by principal components of $X$ and the inclusion of variables in levels to retrieve low frequency information. 

We are interested in predicting stationary targets through a \textit{direct} (in opposition to iterated) forecasting approach. There are at least two ways one can construct direct forecasts of the \textit{average growth} rate of a variable over the next $h>1$ months -- an important quantity for the conduct of monetary policy and fiscal planning. 
A popular approach is to forecast the final object of interest by projecting it directly on the informational set $X$ (e.g., \citealt{stock2002forecasting}).  An alternative is the path average approach where every step until the final horizon is predicted separately. A potential benefit of fitting the whole path first and then constructing the final target is to allow for the selected predictors, the harshness of regularization, and the type of nonlinearities to fully adapt  when different relationships arise among the variables during the path.\footnote{An obvious drawback is that implies estimating and tuning $h$ models rather than one.} Since those three modeling elements are wildly nonlinear operations in the original input, averaging the path before or after ML is performed can produce very different results. 

To evaluate the contribution of data transformations for macroeconomic prediction, we conduct an extensive pseudo-out-of-sample forecasting experiment (38 years, 10 key monthly macroeconomic indicators, 6 horizons) with three linear and two nonlinear ML methods (Elastic Net, Adaptive Lasso, Linear Boosting, Random Forests, and Boosted Trees), and two standard econometric reference models (autoregressive and factor-augmented autoregression).

Main results can be summarized as follows. \textbf{First}, combining non-standard data transformations, \textit{MARX}, \textit{MAF} and \textit{Level}, minimizes the RMSE for 8 and 9 variables out of 10 when respectively predicting at short horizons 1 and 3-month ahead.  They remain resilient at longer horizons as they are part of best RMSE specifications around 80\% of time.  \textbf{Second}, their contribution is magnified when combined with nonlinear ML models -- 38 out of 47 cases\footnote{There are 47 cases where at least one of these transformations is used.} -- with an advantage for Random Forests over Boosted Trees. Both algorithms allow for nonlinearities via tree base learners and make heavy use of shrinkage via ensemble averaging. This is precisely the algorithmic environment we conjectured could benefit most from non-standard transformations of $X$. \textbf{Third}, traditional factors can help tremendously. The overwhelming majority of best information sets for each target included factors. On that regard, this amounts to a clear takeaway message: while ML methods can handle the high-dimensional $X$ (both computationally and statistically), extracting common factors remains straightforward feature engineering that pays off. \textbf{Fourth}, the path average approach is preferred to the direct counterpart for almost all real activity variables and at most horizons. Combined with high-dimensional methods that use some form of regularization improves predictability by as much as 30\%. 

The rest of the paper is organized as follows. In section \ref{sec:ML}, we present the ML predictive framework and detail the data transformations and forecasting models. In section \ref{sec:fcst}, we detail the forecasting experiment and in section \ref{sec:results} we present main results. Section \ref{sec:conclusion} concludes. 

\section{Machine Learning Forecasting Framework}\label{sec:ML}

Machine learning algorithms offer ways to approximate unknown and potentially complicated functional forms with the objective of minimizing the expected loss of a forecast over $h$ periods. The focus of the current paper is to construct a feature matrix susceptible to improve the macroeconomic forecasting performance of off-the-shelf ML algorithms. Let $H_t = \left[ H_{1t}, ..., H_{Kt} \right]$ for $t=1,...,T$ be the vector of variables found in a large macroeconomic dataset 
and let $y_{t+h}$ be our target variable that is supposed stationary. The corresponding prediction problem is given by
\begin{equation}
y_{t+h} = g( f_Z(H_t) ) + e_{t+h}. 
\end{equation}
\noindent 
To illustrate the data pre-processing point, define $Z_t \equiv f_Z(H_t)$ as the $N_Z$-dimensional feature vector, formed by combining several transformations of the variables in $H_t$.\footnote{Obviously, in the context of a pseudo-out-of-sample experiment, feature matrices must be built recursively to avoid data snooping.} The function $f_Z$ represents the data pre-processing and/or featuring engineering whose effects on forecasting performance we seek to investigate. The training problem for $f_Z = I()$ is 
\begin{equation}
	\underset{g \in \mathcal{G}}{\text{min}} \left\{ \sum_{t=1}^T \left( y_{t+h}- g\left(H_t\right) \right)^2 + \text{pen}(g;\tau) \right\}. 
\end{equation} 
\noindent The function $g$, chosen as a point in the functional space $\mathcal{G}$, maps transformed inputs into the transformed targets. $\text{pen()}$ is the regularization function whose strength depends on some vector/scalar hyperparameter(s) $\tau$. Let $\circ$ denote the function product and $\tilde{g} := g \circ f_Z$. Clearly, introducing a general $f_Z$ leads to
\begin{align*}
	\underset{g \in \mathcal{G}}{\text{min}} \left\{ \sum_{t=1}^T \left( y_{t+h}- g\left(f_Z(H_t)\right) \right)^2 + \text{pen}(g;\tau) \right\} \enskip \leftrightarrow \enskip \underset{\tilde{g} \in \mathcal{G}}{\text{min}} \left\{ \sum_{t=1}^T \left( y_{t+h}- \tilde{g}\left(H_t\right) \right)^2 + \text{pen}(f_Z^{-1} \circ \tilde{g};\tau) \right\} 
\end{align*} 
which is, simply, a change of regularization. Now, let $g^*(f_Z^*(H_t))$ be the "oracle" combination of best transformation $f_Z$ and true function $g$. Let $g(f_Z(H_t))$ be a functional form and data pre-processing selected by the practitioner. In addition, denote $\hat{g}(Z_t)$ and $\hat y_{t+h}$ the fitted model and its forecast. The forecast error can be decomposed as 
\begin{equation}\label{eq2}
		y_{t+h} - \hat y_{t+h} = \underbrace{ g^*(f_Z^*(H_t)) - g(f_Z(H_t))}_{\text{approximation error}} + \underbrace{ g(Z_t) - \hat{g}(Z_t)}_{\text{estimation error}} + e_{t+h}.
\end{equation}
While the intrinsic error $e_{t+h}$ is not shrinkable, the estimation error can be reduced by either adding more relevant data points or restricting the domain $\mathcal{G}$. The benefits of the latter can be offset by a corresponding increase of the approximation error. Thus, an optimal $f_Z$ is one that entails a prior that reduces estimation error at a minimal approximation error cost. Additionally, since most ML algorithms perform variable selection, there is the extra possibility of pooling different $f_Z$'s together and let the algorithm itself choose the relevant restrictions.\footnote{More concretely, a factor $F$ is a linear combination of $X$. If an algorithm pick $F$ rather than creating its own combination of different elements of $X$, it is implicitly imposing a restriction.}


The marginal impact of the increased domain $\mathcal{G}$ has been explicitly studied in  \cite{gclss2020}, with $Z_t$ being factors extracted from the stationarized version of FRED-MD. The primary objective of this paper is to study the relevance of the choice of $f_Z$, combined with popular ML approximators $g$.\footnote{There are many recent contributions considering the macroeconomic forecasting problem with econometric and machine learning methods in a big data environment \citep{kim2018mining, kotchoni2019macroeconomic}. However, they are done using the standard stationary version of FRED-MD database. Recently, \cite{mccracken2020fred} studied the relevance of unit root tests in the choice of stationarity transformation codes for macroeconomic forecasting with factor models.} To evaluate the virtues of standard and newly proposed data transformations, we conduct a pseudo-out-of-sample (POOS) forecasting experiment using various combinations of $f_Z$'s and $g$'s. 

Finally, a question often overlooked in the forecasting literature is how one should construct the forecast for average growth/difference of the level variable $Y_t$, which is the popular target in macroeconomic applications. The usual approach -- and also the least computationally demanding -- is that of fitting the model on $y_{t+h}=\sfrac{\sum_{h'=1}^h \Delta Y_{t+h'}}{h}$ directly and using $\hat{y}_{t+h}^{\text{direct}}$ as prediction, where $\Delta Y_{t+h'} = Y_{t+h'}-Y_{t+h'-1}$ is the simple growth/difference of the variable of interest. Another approach, requiring the estimation of $h$ different functions, is the \emph{path average} approach where each $\Delta Y_{t+h'}$ is fitted separately and the forecast for $y_{t+h}$ is obtained from $\hat{y}_{t+h}^{\text{path-avg}}=\sfrac{\sum_{h'=1}^h \widehat{\Delta Y}_{t+h'}}{h}$. 

The common wisdom -- from OLS -- is that such strategies are interchangeable. But the equivalence does not hold when regularization and nonparametric nonlinearities are involved. For instance, it breaks in the simplest possible departure from OLS, a ridge regression, where
\begin{align}\label{rr1}
\hat{y}_{t+h}^{\text{path-avg}}=\frac{1}{h} \sum_{h'=1}^h Z(Z'Z +\lambda_{h'}I)^{-1}Z' \Delta Y_{t+h'}, 
\end{align}
and only if $\lambda_{h'}=\lambda \enskip \forall h'$ then 
\begin{align}\label{rr2}
\hat{y}_{t+h}^{\text{path-avg}}=Z(Z'Z +\lambda I)^{-1}Z' \frac{\sum_{h'=1}^h  \Delta Y_{t+h'}}{h} = \hat{y}_{t+h}^{\text{direct}}.
\end{align}
This setup naturally includes the known equivalence in the OLS case ($\lambda_{h'}=0 \enskip \forall h')$. We get even further from the equivalence with Lasso, Random Forests, and Boosted Trees which all imply the nonlinear hard-thresholding operation of variable selection -- and basis expansion creation for the last two. With those, we get even further from the equivalence by having a different $Z_{h'}^* \subset Z$ in each prediction function.

Of course, the path average approach can be rather demanding since it implies $h$ estimation (and likely cross-validation) problems --- with the benefit of providing a whole path rather than merely $y_{t+h}$. The second question address then concerns whether those benefits could additionally include forecasting gains. To investigate this and how this choice interacts with the optimal $f_Z$, we conduct the whole forecasting exercise using both schemes.

\subsection{Old News}

Firstly, we consider more traditional candidates for $f_Z$.

\vskip 0.2cm

{\sc \noindent \textbf{Including Factors}.} 
Common practice in the macroeconomic forecasting literature is to rely on some variant of the transformations proposed by \cite{mccracken2016fred} to obtain a stationary $X_t$ out of $H_t$. Letting $X = \left[ X_t \right]_{t=1}^T$ and imposing a linear latent factor structure $X = F\Lambda + \epsilon$, we can estimate $F$ by the principal components of $X$. The feature matrix of the autoregressive diffusion index (FM hereafter) model of \cite{stock2002forecasting, stock2002macroeconomic} can be formed as
\begin{align}
Z_t = \left[ y_t, Ly_t, ..., L^{p_y} y_t, F_t, LF_t, ..., L^{p_f}F_t \right]
\end{align}
where $L$ is the lag operator and $y_t$ is the current value of the target. In \cite{gclss2020}, factors were deemed the most reliable shrinkage method for macroeconomic forecasting, even when considering ML alternatives. Furthermore, the combination of factors (and nothing else) with nonlinear nonparametric methods is (i) easy, (ii) fast, and (iii) often quite successful. Point (iii) is further re-enforced by this paper's results, especially for forecasting inflation, which contrasts with the results found in \cite{medeiros2019forecasting}.

\vskip 0.2cm

{\sc \noindent \textbf{Including Levels}.} In econometrics, debates on the consequences of unit roots for frequentist inference have a long history\footnote{See for example, \cite{phillips1991criticize, phillips1991optimal, sims1988bayesian, sims1990inference, sims1991understanding}.}, just as does the handling of low frequency movements for macroeconomic forecasting \citep{Graham2006}. Exploiting potential cointegration has been found useful to improve forecasting accuracy under some conditions (e.g., \cite{christoffersen1998cointegration, engle1987forecasting, hall1992cointegration}). 
From the perspective of engineering a feature matrix, the error correction term could be obtained from a first step regression \textit{\`{a} la} \cite{engle1987co} 
and is just a specific linear combination of existing variables. When it is unclear which variables should enter the cointegrating vector -- or whether there exist any such vector -- one can alternatively include both variables in levels and differences into the feature matrix. This sort of approach has been pursued most notably by \cite{cook2017macroeconomic} who combine variables in levels, first differences and even second differences in the feature matrix they provide to various neural network architectures in the forecasting of US unemployment data.\footnote{Another approach is to consider factor modeling directly with nonstationary data  \citep{baing2004,PENA20061237,BANERJEE2014589}.} 

From a purely predictive point of view, using  first differences rather than levels is a linear restriction (using the vector $[1, -1]$) on how $H_t$ and $H_{t-1}$ can jointly impact $y_t$. Depending on the prior/regularization being used with a linear regression, this may largely decrease the estimation error or inflate the approximation one.\footnote{A similar comment would apply to all parametric cointegration restrictions. For recent work on the subject, see for example \cite{chan2015nonlinear}.} However, it is often admitted that in a time series context (even if Bayesian inference is left largely unaltered by non-stationarity \citep{sims1988bayesian}), first differences are useful because they trim out low frequencies which may easily be  redundant in large macroeconomic data sets. Using a collection of highly persistent time series in $X$ can easily lead to an unstable $X'X$ inverse (or even a regularized version). Such problems naturally extend to Lasso \citep{lee2018lasso}. In contrast, tree-based approaches like RF and Boosted Trees do not rely on inverting any matrix. Of course, performing tree-like sample splitting on a trending variable like raw GDP (without any subsequent split on lag GDP), is almost equivalent to split the sample according to a time trend and will often be redundant and/or useless. Nevertheless, there are numerous $H_t$'s where opting for first differencing the data is much less trivial. In such cases, there may be forecasting benefits from augmenting the usual $X$ with levels. 


\subsection{New Avenues}

When regressors outnumber observations, regularization, whether explicit or implicit, is necessary.  Hence, the ML algorithms we use all entail a prior which may or may not be well suited for a time series problem. There is a wide Bayesian VAR literature, starting with \cite{litterman1984}, proposing prior structures that are thought for the multiple blocks of lags characteristic of those models. Additionally, there is a whole strand of older literature that seeks to estimate restricted lag polynomials in Autoregressive Distributed Lags (ARDL) models  \citep{almon1965distributed,shiller1973distributed}. 
While the above could be implemented in a parametric ML model with a moderate amount of pain, it is not clear how such priors framed in terms of lag polynomials can be put to use when there is no explicit lag polynomial. A more convenient approach is to \textbf{(i)} observe that most nonparametric ML methods implicitly shrink the individual contribution of each feature to zero in a Ridge-ean fashion \citep{hastie2009elements,elliott2013complete} and \textbf{(ii)} rotating regressors implies a new prior in the original space. Hence, by simply creating regressors that embody the more sophisticated linear restrictions, we obtain shrinkage better suited for time series.\footnote{A cross-section RF-based example is \cite{rodriguez2006rotation} who propose "Rotation Forest" that build an ensemble of trees based on different rotations of $X$.} A first step in that direction is \cite{goulet2020macroeconomy} who proposes Moving Average Factors to specifically enhance RF's prediction and interpretation potential. A second is to find a rotation of the original lag polynomial such that implementing Ridge-ean shrinkage in fact yields \cite{shiller1973distributed} approach to shrinking lag polynomials.

\vskip 0.2cm

{\sc \noindent \textbf{Moving Average Factors}.} Using factors is a standard approach to summarize parsimoniously a panel of heavily cross-correlated variables. Analogously, one can extract a few principal components from each variable-specific panel of lagged values, i.e.
\begin{align}
&\tilde{X}_{t,k} = \left[ X_{t,k}, LX_{t,k}, ..., L^{P_{MAF}} X_{t,k} \right]
&\tilde{X}_{t,k}= M_t \Gamma_k' + \tilde{\epsilon}_{k,t}, \; k=1,...,K \label{MAF:factor model}
\end{align}
to achieve a similar goal on the time axis. Define a moving average factor as the vector $M_{k}$.\footnote{While we work directly with the latent factors, a related decomposition called singular spectrum analysis works with the estimate of the summed common components, i.e. with $M_k \Gamma_k'$. Since this decomposition naturally yields a recursive formula, it has been used to forecast macroeconomic and financial variables \citep{hassani2009forecasting, hassani2013predicting}, usually in an univariate fashion.} Mechanically, we obtain weighted moving averages, where the weights are the principal component estimates of the loadings in $\Gamma_k$. By construction, those extractions form moving averages of the $P_{MAF}$ lags of $X_{t,k}$ so that it summarizes most efficiently its temporal information.\footnote{$P_{MAF}$ is a tuning parameter analogous to the construction of the panel of variables (usually taken as given) in a standard factor model. We pick $P_{MAF}=12$. We keep two MAFs for each series and they are obtained by PCA.}  By doing so, the goal to summarize information in $X_{t,k}^{1:P_{MAF}}$ is achieved without modifying any algorithm: we can use the MAFs which compresses information ex-ante. As it is the case for standard factors, MAF are designed to maximize the explained variance in $X_{t,k}^{1:P_{MAF}}$, not the fit to the final target. It is the learning algorithm's job to select the relevant linear combinations to maximize the fit.

\vskip 0.2cm

{\sc \noindent \textbf{Moving Average Rotation of $X$}.} There are many ways one can penalize a lag polynomial. One, in the Minnesota prior tradition, is to shrink all lags coefficients to \textit{zero} (except for the first self-lag) with increasing harshness in $p$, the order of the lag. Another is to shrink each $\beta_p$ to $\beta_{p-1}$ and $\beta_{p+1}$ rather than to zero. Intuitively, for higher-frequency series (like monthly data would qualify for here) it is more plausible that a simple linear combination of lags impacts $y_t$ rather than a single one of them with all other coefficients set to zero.\footnote{This is basically a dense vs sparse choice. MAFs go all the way with the first view by imposing it via the extraction procedure.} For instance, it seems more likely that the average of March, April, and May employment growth could impact, say, inflation, than only May's. Mechanically, this means we expect March, April, and May 's coefficients to be close to one another, which motivated the prior ${\beta_{p}}\sim N(\beta_{p-1},\sigma_u^2 I_K)$ and more sophisticated versions of it in other works \citep{shiller1973distributed}. Inputting in the ML algorithm a transformed $X$ such that its implicit shrinkage to zero is twisted into this new prior could generate forecasting gains. The only question left is how to make this operational.

The following derivation is a simple translation of \cite{GC2019}'s insights for time-varying parameters model to regularized lag polynomials \`{a} la \cite{shiller1973distributed}.\footnote{Such reparametrization schemes are also discussed for "fused" Lasso in \cite{tibshirani2015statistical} and employed for a Bayesian local-level model in \cite{koop2003bayesian}.} Consider a generic regularized ARDL model with $K$ variables
\begin{align}\label{eqn:model2}
\min_{\beta_{1}\dots\beta_{P}}{\sum_{t=1}^{T}}\left(y_{t}-{\sum_{p=1}^{P}}X_{t-p}{\beta_{p}}\right)^{2}+\lambda{\sum_{p=1}^{P}}\Vert{\beta}_{p}-\beta_{p-1}\Vert^{2}.
\end{align}
where $\beta_p \in {\rm I\!R}^K$, $X_t \in {\rm I\!R}^K$, $u_p \in {\rm I\!R}^{K \times P}$, and both $y_t$ and $\epsilon_t$ are scalars.\footnote{We use $P$ as a generic maximum number of lags for presentation purposes. In Table \ref{sum_table} we define $P_{MARX}$.} While we adopt the $l_2$ norm for this exposition, our main goal is to extend traditional regularized lag polynomial ideas to cases where there is no explicitly specified norm on $\beta_p - \beta_{p-1}$. For instance, \cite{elliott2013complete} prove that their Complete Subset Regression procedure implies Ridge shrinkage in a special case. Moving away from linearity makes formal arguments more difficult. Nevertheless, it has been argued several times that model/ensemble averaging performs shrinkage akin to that of a ridge regression \citep{hastie2009elements}. For instance, random selection of a subset of eligible features at each split encourage each feature to be included in the predictive function, but in a moderate fashion.\footnote{Recently, \citep{MSoRF} argued that ensemble averaging methods \`{a} la RF prunes a latent tree. Following this view, the need for cleverly pre-assembled data combinations is even clearer.}
The resulting "implicit" coefficient is an average of specifications that included the regressor and some that did not. In the latter case, the coefficient is always zero by construction. Hence, the ensemble shrinks contributions towards zero and the so-called \texttt{mtry} hyperparameter guides the level of shrinkage like a bandwidth parameter would \citep{olson2018making}. 

To get implicit regularized lag polynomial shrinkage, we now rewrite problem \eqref{eqn:model2} as a ridge regression. For all derivations to come, it is less tedious to turn to matrix notations. The Fused Ridge problem is now written as
\begin{align*}
\min_{\boldsymbol{\beta}}\left(\boldsymbol{y}-\boldsymbol{X\beta}\right)'\left(\boldsymbol{y}-\boldsymbol{X\beta}\right)+\lambda \boldsymbol{\beta' D'D \beta}
\end{align*}
where $\boldsymbol{D}$ is the first difference operator. The first step is to reparametrize the problem by using the relationship $\beta_k = C \theta_k$ that we have for all $k$ regressors. $C$ is a lower triangular matrix of ones (for the random walk case) and define ${\theta_k} = [{u_k} \quad {\beta_{0,k}}]$. For the simple case of one parameter and $P=4$:
\renewcommand{\arraystretch}{0.8} 
\[
\begin{bmatrix}
    \beta_0  \\
    \beta_1 \\
    \beta_2 \\
   \beta_3
\end{bmatrix}
=
\begin{bmatrix}
    1 & 0 & 0 & 0 \\
    1 & 1 & 0 & 0 \\
    1 & 1 & 1 & 0 \\
   1 & 1 & 1 & 1
\end{bmatrix}
\begin{bmatrix}
    \beta_0  \\
    u_1 \\
    u_2 \\
   u_3
\end{bmatrix} .
\]
For the general case of $K$ parameters, we have 
$$ \boldsymbol{\beta}= \boldsymbol{C} \boldsymbol{\theta}, \quad  \boldsymbol{C} \equiv I_K \otimes C $$
and $\boldsymbol{\theta}$ is just stacking all the $\theta_k$ into one long vector of length $KP$. Using the reparametrization $\boldsymbol{\beta}= \boldsymbol{C} \boldsymbol{\theta}$, the Fused Ridge problem becomes
\begin{align*}
\min_{\boldsymbol{\theta}}\left(\boldsymbol{y}-\boldsymbol{XC \theta}\right)'\left(\boldsymbol{y}-\boldsymbol{XC\theta}\right)+\lambda \boldsymbol{\theta 'C'D'D C\theta}.
\end{align*}
Let $\boldsymbol{Z} \equiv \boldsymbol{XC}$ and use the fact that $\boldsymbol{D} = \boldsymbol{C}^{-1}$ to obtain the Ridge regression problem
\begin{align}\label{eqn:basicridge}
\min_{\boldsymbol{\theta}}\left(\boldsymbol{y}-\boldsymbol{Z\theta}\right)'\left(\boldsymbol{y}-\boldsymbol{Z\theta}\right)+\lambda \boldsymbol{\theta'\theta}.
\end{align}
We arrived at destination. Using $\boldsymbol{Z}$ rather than $\boldsymbol{X}$ in an algorithm that performs shrinkage will implicitly shrink $\beta_p$ to $\beta_{p-1}$ rather than to 0. This is obviously much more convenient than modifying the algorithm itself and is directly applicable to \textit{any algorithm} using time series data as input. One question remains: what is $\boldsymbol{Z}$, exactly? For a single polynomial at time $t$, we have $Z_{t,k} = X_{t,k} C$. $C$ is gradually summing up the columns of $X_{t,k}$ over $p$. Thus, $Z_{t,k,p}=\sum_{p'=1}^{P} X_{t,k,p'}$.  Dividing each $Z_{t,k,p}$ by $p$ (just another linear transformation, $\tilde{Z}_{t,k,p}$ ), it is now clear that $\tilde{\boldsymbol{Z}}$ is a matrix of moving averages. Those are of increasing order (from $p=1$ to $p=P$) and the last observation in the average is always $X_{t-1,k}$. Hence, we refer to this particular form of feature engineering as Moving Average Rotation of $X$ (MARX). 

\vskip 0.2cm

{\sc \noindent \textbf{Recap}.} We summarize our setup in Table \ref{sum_table}. We have five basic sets of transformations to feed the approximation of $f_Z^*$: (1) single-period differences and growth rates following \cite{mccracken2016fred} ($X_t$ and their lags), (2) principal components of $X_t$ ($F_t$ and their lags), (3) variables in levels ($H_t$ and their lags), (4) moving average factors of $X_t$ ($MAF_t$), and (5) sets of simple moving averages of $X_t$ ($MARX_t$). We consider several forecasting models in order to approximate the true functional form: Autoregressive (AR), Factor Model (FM, \`{a} la \cite{stock2002forecasting}), Adaptive Lasso (AL), Elastic Net (EN), Linear Boosting (LB), Random Forest (RF), and Boosted Trees (BT). Lastly, we apply those specifications to forecasting both direct and path-average targets. The details on forecasting models are presented in Appendix \ref{sec:models}. 

Furthermore, most ML methodologies that handle well high-dimensional data perform  some form or another of variable selection. For instance, RF evaluates a certain fraction of predictors at each split and selects the most potent one. Lasso selects relevant predictors and shrinks others perfectly to zero. By rotating $X$, we can get these algorithms (and others) to perform restriction/transformation selection. Thus, one should not refrain from studying different combinations of $f_Z$'s.\footnote{Notwithstanding, some authors have noted that a trade-off emerges between how focused a RF is and its robustness via diversification. \cite{borup2020targeting} sometimes get improvements over plain RF by adding a Lasso pre-processing step to trim $X$.} As a result, all the combinations of $f_Z$ thereof are admissible and 16 of them are included in the exercise. Moreover, there is a long-standing worry that well-accepted transformations may lead to some over-differenced $X_k$'s \citep{mccracken2020fred}. Including MARX or MAF (which are both specific partial sums of lags) with $X$ can be seen as bridging the gap between a first difference and keeping $H_k$ in levels. Hence, interacting many $f_Z$ is not only statistically feasible, but econometrically desirable given the sizable uncertainty surrounding what is a "proper" transformation of the raw data \citep{choi2015almost}.

\begin{table}[!h]
	\caption{Model Specification Summary} \label{sum_table}
	\vspace{-1.5em}
	\begin{center}
	\begin{footnotesize}
	\begin{tabular}{ll}
		\hline \hline 
		\vspace{-.5em}& \\
		Cases& Feature Matrix $Z_t$ \\ \hline
		\vspace{-.5em} & \\
		F   & $Z_t := \left[ \{L^{i-1}F_t\}_1^{p_{f}}\right]$ \\
		F-X & $Z_t := \left[\{L^{i-1}F_t\}_1^{p_{f}},\{L^{i-1}X_t\}_1^{p_{m}}  \right]$ \\
		F-MARX & $Z_t := \left[\{L^{i-1}F_t\}_1^{p_{f}},\{MARX_{yt}^{i}\}_1^{p_{y}},\{MARX_{1t}^{i}\}_1^{p_{m}},\dots,\{MARX_{Kt}^{i}\}_1^{p_{m}}  \right]$ \\
		F-MAF & $Z_t := \left[\{L^{i-1}F_t\}_1^{p_{f}},\{MAF_{yt}^{i}\}_1^{r_K}, \{MAF_{1t}^{i}\}_1^{r_K},\dots,\{MAF_{Kt}^{i}\}_1^{r_K}\right]$ \\
		F-Level & $Z_t := \left[\{L^{i-1}F_t\}_1^{p_{f}},Y_t,H_t  \right]$ \\
		F-X-MARX & $Z_t := \left[\{L^{i-1}F_t\}_1^{p_{f}},\{L^{i-1}X_t\}_1^{p_{m}},\{MARX_{yt}^{i}\}_1^{p_{y}},\{MARX_{1t}^{i}\}_1^{p_{m}},\dots,\{MARX_{Kt}^{i}\}_1^{p_{m}}\right]$ \\
     	F-X-MAF & $Z_t := \left[\{L^{i-1}F_t\}_1^{p_{f}},\{L^{i-1}X_t\}_1^{p_{m}},\{MAF_{yt}^{i}\}_1^{r_K},\{MAF_{1t}^{i}\}_1^{r_K},\dots,\{MAF_{Kt}^{i}\}_1^{r_K}\right]$ \\
		F-X-Level & $Z_t := \left[\{L^{i-1}F_t\}_1^{p_{f}},\{L^{i-1}X_t\}_1^{p_{m}},Y_t,H_t  \right]$ \\
		F-X-MARX-Level & $Z_t := \left[\{L^{i-1}F_t\}_1^{p_{f}},\{L^{i-1}X_t\}_1^{p_{m}},\{MARX_{yt}^{i}\}_1^{p_{y}},\{MARX_{1t}^{i}\}_1^{p_{m}},\dots,\{MARX_{Kt}^{i}\}_1^{p_{m}},Y_t,H_t  \right]$ \\
		X & $Z_t := \left[\{L^{i-1}X_t\}_1^{p_{m}}\right]$ \\
        MARX & $Z_t := \left[\{MARX_{yt}^{i}\}_1^{p_{y}},\{MARX_{1t}^{i}\}_1^{p_{m}},\dots,\{MARX_{Kt}^{i}\}_1^{p_{m}}  \right]$ \\
        MAF & $Z_t := \left[\{MAF_{yt}^{i}\}_1^{r_K}, \{MAF_{1t}^{i}\}_1^{r_K},\dots,\{MAF_{Kt}^{i}\}_1^{r_K}\right]$ \\
        X-MARX & $Z_t := \left[\{L^{i-1}X_t\}_1^{p_{m}},\{MARX_{yt}^{i}\}_1^{p_{y}},\{MARX_{1t}^{i}\}_1^{p_{m}},\dots,\{MARX_{Kt}^{i}\}_1^{p_{m}}\right]$ \\
		X-MAF & $Z_t := \left[\{L^{i-1}X_t\}_1^{p_{m}},\{MAF_{yt}^{i}\}_1^{r_K},\{MAF_{1t}^{i}\}_1^{r_K},\dots,\{MAF_{Kt}^{i}\}_1^{r_K}\right]$ \\
		X-Level & $Z_t := \left[\{L^{i-1}X_t\}_1^{p_{m}},Y_t,H_t  \right]$ \\
		X-MARX-Level & $Z_t := \left[\{L^{i-1}X_t\}_1^{p_{m}},\{MARX_{yt}^{i}\}_1^{p_{y}},\{MARX_{1t}^{i}\}_1^{p_{m}},\dots,\{MARX_{Kt}^{i}\}_1^{p_{m}},Y_t,H_t  \right]$ \\
		\hline \hline
	\end{tabular}
	\end{footnotesize}
\end{center}
{\flushleft
	\begin{scriptsize}
		\vspace{-2.5em}
		{ \singlespacing
			Note: This table show the combinations of data transformation used to assess the individual marginal contribution of each $f_Z$. Lags of month-to-month (log)-change of the series to forecast are always included. \par}   
	\end{scriptsize}
}
\end{table}

\section{Forecasting Setup}\label{sec:fcst}

In this section, we present the results of a pseudo-out-of-sample forecasting experiment for a group of target variables at monthly frequency from the FRED-MD dataset of \cite{mccracken2016fred}. Our target variables are the industrial production index (INDPRO), total nonfarm employment (EMP), unemployment rate (UNRATE), real personal income excluding current transfers (INCOME), real personal consumption expenditures (CONS), retail and food services sales (RETAIL), housing starts (HOUST), M2 money stock (M2),  consumer price index (CPI), and the production price index (PPI). Given that we make predictions at horizons of 1, 3, 6, 9, 12, and 24 months, we are effectively targeting the average growth rate over those periods, except for the unemployment rate for which we target average differences. These series are representative macroeconomic indicators of the US economy, as stated in \cite{kim2018mining}, which is also based on \cite{gclss2020} exercise for many ML models, itself based on \cite{kotchoni2019macroeconomic} and a whole literature of extensive horse races in the spirit of \cite{SW1998comparison}. The POOS period starts in January of 1980 and ends in December of 2017. We use an expanding window for estimation starting from 1960M01. Following standard practice in the literature, we evaluate the quality of point forecasts using the root Mean Square Error (RMSE). For the forecasted value at time $t$ of variable $v$ made $h$ steps before, we compute 
\begin{align}
	RMSE_{v,h,m} = \sqrt{ \frac{1}{\#\text{OOS}}\sum_{t \in \text{OOS}} (y_{t}^v-\hat{y}_{t-h}^{v,h,m})^2}
\end{align}
The standard \cite{dieboldmariano} (DM) test procedure is used to compare the predictive accuracy of each model against the reference factor model (FM). RMSE is the most natural loss function given that all models are trained to minimize the squared loss in-sample. We also implement the Model Confidence Set (MCS)
that selects the subset of best models at a given confidence level \citep{MCS2011}. 

Hyperparameter selection is performed using the BIC for AR and FM and K-fold cross-validation is used for the remaining models. This approach is theoretically justified in time series models under conditions spelled out by \cite{bergmeir2018note}. Moreover, \cite{gclss2020} compared it with a scheme which respects the time structure of the data 
and found K-fold to be performing as well as or better than this alternative scheme.  All models are estimated every month while their hyperparameters are reoptimized every two years.

\section{Results}\label{sec:results}


Table \ref{bestmodels} shows the best RMSE data transformation combinations as well as the associated functional forms for every target and forecasting horizon. It summarizes the main findings and provide important recommendations for practitioners in the field of macroeconomic forecasting. \textbf{First}, including non-standard choices of macroeconomic data transformation, \textit{MARX}, \textit{MAF} and \textit{Level}, minimize the RMSE for 8 and 9 variables out of 10 when respectively predicting 1 and 3-month ahead.  Their overall importance is still resilient at longer horizons as they are part of best specifications most of the variables. \textbf{Second}, their success is often paired with a nonlinear functional form $g$, 38 out of 47 cases, with an advantage for Random Forests over Boosted Trees. The former is used for 26 of those 38 cases. Both algorithms make heavy use of shrinkage and allow for nonlinearities via tree base learners. This is precisely the algorithmic environment that we precedently conjectured to be where data transformations matter.


\begin{footnotesize}
\begin{ThreePartTable}
\begin{longtable}{l|lll|llll|lll}
\caption{Best model specifications - with target type} \\ 
\hline
 & INDPRO & EMP & UNRATE & INCOME & CONS & RETAIL & HOUST & M2 & CPI & PPI \\ 
\hline
\endhead
\hline
\endfoot
H=1 & RF{\color{ForestGreen}$\medbullet$}{\color{Cyan}$\medbullet$}{\color{RubineRed}$\medbullet$}{\color{blue}$\medbullet$} & RF{\color{ForestGreen}$\medbullet$}{\color{Cyan}$\medbullet$}{\color{RubineRed}$\medbullet$}{\color{blue}$\medbullet$} & BT{\color{ForestGreen}$\medbullet$}{\color{RubineRed}$\medbullet$} & RF{\color{RubineRed}$\medbullet$} & FM{\color{ForestGreen}$\medbullet$} & FM{\color{ForestGreen}$\medbullet$} & EN{\color{ForestGreen}$\medbullet$}{\color{blue}$\medbullet$} & RF{\color{Cyan}$\medbullet$}{\color{blue}$\medbullet$} & AL{\color{RubineRed}$\medbullet$} & EN{\color{ForestGreen}$\medbullet$}{\color{RubineRed}$\medbullet$} \\ 
  H=3 & RF{\color{RubineRed}$\medbullet$} & \underline{RF}{\color{ForestGreen}$\medbullet$}{\color{RubineRed}$\medbullet$} & RF{\color{ForestGreen}$\medbullet$}{\color{Cyan}$\medbullet$}{\color{RubineRed}$\medbullet$}{\color{blue}$\medbullet$} & RF{\color{ForestGreen}$\medbullet$}{\color{RubineRed}$\medbullet$} & RF{\color{ForestGreen}$\medbullet$}{\color{blue}$\medbullet$} & \underline{BT}{\color{ForestGreen}$\medbullet$}{\color{Cyan}$\medbullet$}{\color{RubineRed}$\medbullet$} & \underline{EN}{\color{ForestGreen}$\medbullet$}{\color{blue}$\medbullet$} & \underline{AL}{\color{Cyan}$\medbullet$}{\color{blue}$\medbullet$} & \underline{RF}{\color{ForestGreen}$\medbullet$} & \underline{EN}{\color{RubineRed}$\medbullet$} \\ 
  H=6 & \underline{RF}{\color{RubineRed}$\medbullet$} & \underline{BT}{\color{ForestGreen}$\medbullet$}{\color{RubineRed}$\medbullet$} & \underline{RF}{\color{ForestGreen}$\medbullet$}{\color{RubineRed}$\medbullet$} & \underline{RF}{\color{ForestGreen}$\medbullet$}{\color{Cyan}$\medbullet$}{\color{RubineRed}$\medbullet$} & \underline{RF}{\color{ForestGreen}$\medbullet$}{\color{blue}$\medbullet$} & AL{\color{ForestGreen}$\medbullet$}{\color{RubineRed}$\medbullet$} & \underline{RF}{\color{ForestGreen}$\medbullet$}{\color{Cyan}$\medbullet$}{\color{RubineRed}$\medbullet$} & RF{\color{ForestGreen}$\medbullet$}{\color{blue}$\medbullet$} & RF{\color{ForestGreen}$\medbullet$} & RF{\color{ForestGreen}$\medbullet$} \\ 
  H=9 & \underline{RF}{\color{RubineRed}$\medbullet$} & \underline{BT}{\color{ForestGreen}$\medbullet$}{\color{RubineRed}$\medbullet$} & \underline{LB}{\color{ForestGreen}$\medbullet$}{\color{Cyan}$\medbullet$}{\color{RubineRed}$\medbullet$}{\color{blue}$\medbullet$} & \underline{RF}{\color{ForestGreen}$\medbullet$}{\color{RubineRed}$\medbullet$} & \underline{RF}{\color{Orange}$\medbullet$} & BT{\color{ForestGreen}$\medbullet$}{\color{Cyan}$\medbullet$}{\color{RubineRed}$\medbullet$}{\color{blue}$\medbullet$} & BT{\color{ForestGreen}$\medbullet$}{\color{Orange}$\medbullet$} & RF{\color{ForestGreen}$\medbullet$}{\color{blue}$\medbullet$} & RF{\color{ForestGreen}$\medbullet$} & RF{\color{ForestGreen}$\medbullet$} \\ 
  H=12 & \underline{RF}{\color{RubineRed}$\medbullet$} & \underline{BT}{\color{ForestGreen}$\medbullet$}{\color{RubineRed}$\medbullet$} & \underline{LB}{\color{ForestGreen}$\medbullet$}{\color{Cyan}$\medbullet$}{\color{RubineRed}$\medbullet$}{\color{blue}$\medbullet$} & \underline{RF}{\color{ForestGreen}$\medbullet$}{\color{RubineRed}$\medbullet$} & \underline{RF}{\color{ForestGreen}$\medbullet$}{\color{Orange}$\medbullet$} & BT{\color{ForestGreen}$\medbullet$}{\color{Cyan}$\medbullet$}{\color{blue}$\medbullet$} & RF{\color{ForestGreen}$\medbullet$} & BT{\color{ForestGreen}$\medbullet$}{\color{blue}$\medbullet$} & RF{\color{ForestGreen}$\medbullet$} & RF{\color{ForestGreen}$\medbullet$} \\ 
  H=24 & RF{\color{ForestGreen}$\medbullet$}{\color{blue}$\medbullet$} & \underline{BT}{\color{Orange}$\medbullet$} & BT{\color{ForestGreen}$\medbullet$}{\color{Orange}$\medbullet$} & \underline{RF}{\color{ForestGreen}$\medbullet$}{\color{Cyan}$\medbullet$}{\color{RubineRed}$\medbullet$} & \underline{RF}{\color{ForestGreen}$\medbullet$}{\color{Orange}$\medbullet$} & BT{\color{ForestGreen}$\medbullet$}{\color{Cyan}$\medbullet$}{\color{Orange}$\medbullet$} & RF{\color{ForestGreen}$\medbullet$} & \underline{RF}{\color{ForestGreen}$\medbullet$}{\color{blue}$\medbullet$} & \underline{RF}{\color{Cyan}$\medbullet$} & BT{\color{ForestGreen}$\medbullet$}{\color{blue}$\medbullet$}  
\label{bestmodels}
\end{longtable}
\end{ThreePartTable}
\end{footnotesize}
{\flushleft
\begin{scriptsize}
	\vspace{-2.5em}
	{ \singlespacing
		Note: Bullet colors represent data transformations included in the best model specifications: {\color{ForestGreen}\emph{\textbf{F}}}, {\color{RubineRed}\emph{\textbf{MARX}}}, {\color{Cyan}\emph{\textbf{X}}}, {\color{blue}\emph{\textbf{L}}} and {\color{Orange}\emph{\textbf{MAF}}}. Path average specifications are underlined.\par}   
\end{scriptsize}
}

Without a doubt, the most visually obvious feature of Table \ref{bestmodels} is the abundance of green bullets.  As expected, transforming $X$ into factors is probably the most effective form of feature engineering available to the macroeconomic forecaster. Factors are included as part of the optimal specification for the overwhelming majority of targets.  Furthermore, including factors \textit{only} in combination with RF is the best forecasting strategy for both CPI and PPI inflation for the vast majority of horizons. This is in line with findings in \cite{gclss2020} but in contrast with the results found in \cite{medeiros2019forecasting}. The major difference with the latter is that they estimate and evaluate models on the basis of single month inflation rate, which is only the intermediary step in our path average strategy. In addition, we explore the possibility that $F$ alone could be better than $X$, rather than always both together. As it turns out, the winning combination is RF using factors as \textit{sole inputs} to directly target the average growth. 
Finally, the omission of factors from optimal specifications for  industrial production growth 3 to 12 months ahead is naturally surprising. This points out that current wisdom based on linear models may not be directly applicable to nonlinear ones. In fact, alternative rotations will sometimes do better.


There is plentiful of red bullets populating the top rows of Table \ref{bestmodels}. Indeed, our most salient new transformation is \textit{MARX}. In combination with nonlinear tree-based models, it contributes to improve forecasting accuracy for real activity series such as industrial production, employment, unemployment rate, and income, while they are best paired with penalized regressions to predict the CPI and PPI inflation rates. The dominance of MARX is particularly striking for real activity series as the transformation is included in \textit{every} best specification for those variables at \textit{all} horizons ranging from one month to a year. We further investigate how those RMSE gains materialize in terms of forecasts around key periods in section \ref{sec:casestudy}. While \textit{MAF} performance is often positively correlated with  \textit{MARX}, the latter is usually the better of the two, except for longer-run forecasts -- like those 2-years where MAF is featured for four variables.
 
Considering levels is particularly important for the M2 money stock as it is included in the best model for \textit{all} horizons. For other variables, its pertinence is rather sporadic, with at least two horizons featuring it for INDPRO, UNRATE, CONS, and RETAIL. 
 

The preference for $\hat{y}_{t+h}^{\text{direct}}$ vs $\hat{y}_{t+h}^{\text{path-avg}}$ mostly go on a variable by variable basis. However, there is clear consensus $\hat{y}_{t+h}^{\text{path-avg}} \succ \hat{y}_{t+h}^{\text{direct}}$ for all variables which strongly co-move with the business cycle (INDPRO, EMP, UNRATE, INCOME, CONS) with the notable exception of retail sales and housing starts. When it comes to nominal targets (M2, CPI, PPI), $\hat{y}_{t+h}^{\text{path-avg}} \prec \hat{y}_{t+h}^{\text{direct}}$ is unanimous for horizons 6 to 12 months, and so are the affiliated data transformations as well as the $g$ choice (all tree ensembles, with 8 out of 9 being RF). The quantitative importance of both types of gains on both sides is studied in section \ref{sec:tornado}, while section \ref{sec:casestudy} looks at implied forecasts to understand when and why $\hat{y}_{t+h}^{\text{path-avg}} \succ \hat{y}_{t+h}^{\text{direct}}$, or the reverse.

These findings are particularly important given the increasing interest in ML macro forecasting. They suggest that traditional data transformations, meant to achieve stationarity, do leave substantial forecasting gains on the practitioners' table. These losses can be successfully recovered by combining ML methods with well-motivated rotations of predictors such as \textit{MARX} and \textit{MAF} (or sometimes by simply including variables in levels) and by constructing the final forecast by the path average approach.    

The previous results were desirably expeditive. The detailed results on the underlying performance gains and their statistical significance are presented in Appendix \ref{sec:rmse_results}.

\subsection{Marginal Contribution of Data Pre-processing}\label{sec:tornado}

In order to disentangle \textit{marginal} effects of data transformations on forecast accuracy we run the following regression inspired by \cite{CARRIERO20191226} and \cite{gclss2020}:
\begin{equation}\label{r2_eq}
	R^2_{t,h,v,m} = \alpha_{\mathcal{F}} +  \psi_{t,v,h} + v_{t,h,v,m},
\end{equation}
where $R^2_{t,h,v,m} \equiv 1 - \frac{e^2_{t,h,v,m}}{\frac{1}{T} \sum_{t=1}^T (y_{v,t+h} - \bar{y}_{v,h})^2}$ is the pseudo-out-of-sample $R^2$, and $e^2_{t,h,v,m}$ are squared prediction errors of model $m$ for variable $v$ and horizon $h$ at time $t$. $\psi_{t,v,h}$ is a fixed effect term that demeans the dependent variable by ``forecasting target,'' that is a combination of $t$, $v$, and $h$.  $\alpha_{\mathcal{F}}$ is a vector of $\alpha_{\mathit{MARX}}$, $\alpha_{\mathit{MAF}}$, and $\alpha_{\mathit{F}}$ terms associated to each new data transformation considered in this paper, as well as to the factor model. 
$H_0$ is $\alpha_f =0 \quad \forall f \in \mathcal{F} = [\mathit{MARX}, \ \mathit{MAF}, \ \mathit{F}]$. In other words, the null is that there is no predictive accuracy gain with respect to a base model that does not have this particular data pre-processing. While the generality of (\ref{r2_eq}) is appealing, when investigating the heterogeneity of specific partial effects, it will be much more convenient to run specific regressions for the  multiple hypothesis we wish to test. That is, to evaluate a feature $f$, we run
\begin{align}{\label{e_eq2}}
\forall m \in \mathcal{M}_f: \quad  R^2_{t,h,v,m} = \alpha_f +  \psi_{t,v,h} + v_{t,h,v,m}
\end{align}
where $ \mathcal{M}_f$ is defined as the set of models that differs only by the feature under study $f$.

\begin{figure}[h!]
\centering
\caption{Distribution of \textit{MARX} Marginal Effects (Average Targets)}
\label{fig:MARXavgtornado}
\includegraphics[width=\textwidth,height=0.5\textheight]{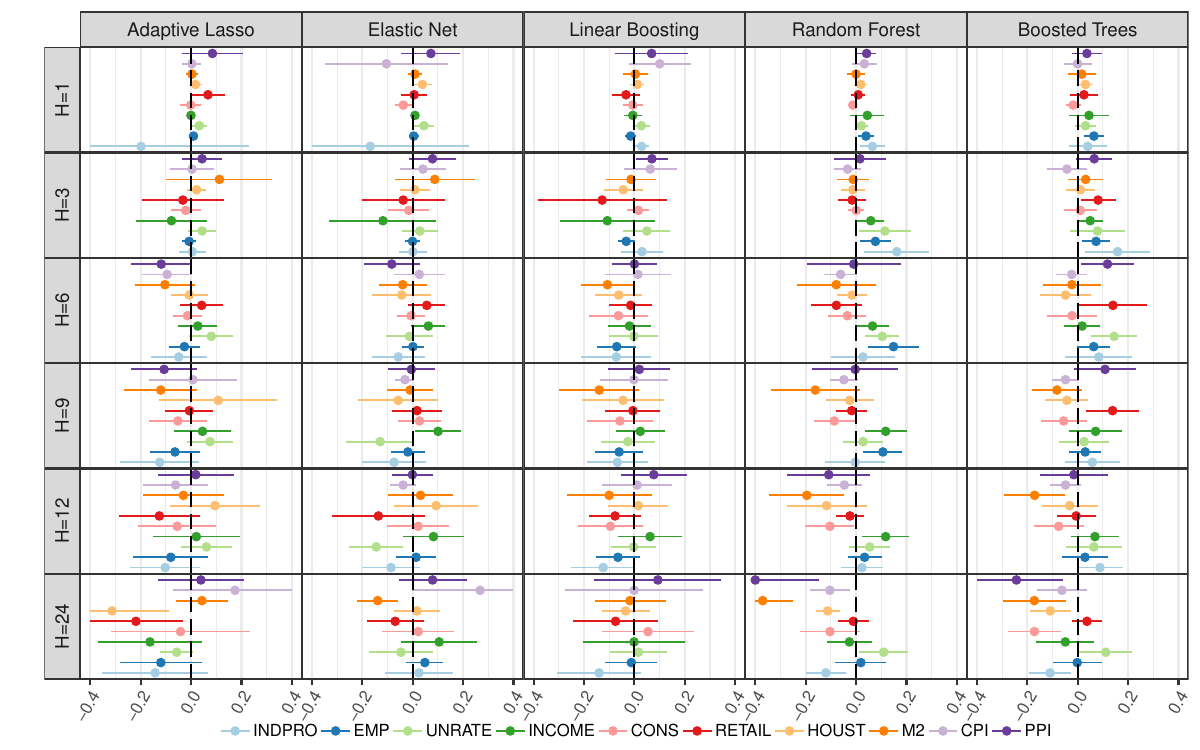}
\flushleft
\begin{scriptsize}
	\vspace{-2em}
	{\singlespacing
	Note: This figure plots the distribution of ${\alpha}_f^{(h,v)}$ from equation (\ref{e_eq2}) done by $(h,v)$ subsets. That is, it shows the average partial effect on the pseudo-$R^2$ from augmenting the model with \textit{MARX} featuring, keeping everything else fixed. SEs are HAC. These are the 95\% confidence bands. \par}  
\end{scriptsize}
\end{figure}

\vspace{0.2em}

\noindent {\sc \textbf{MARX.}} Figure \ref{fig:MARXavgtornado} plots the distribution of $\alpha_{\mathit{MARX}}^{(h,v)}$ from equation (\ref{r2_eq}) done by $(h,v)$ subsets. Hence, we allow for heterogeneous effects of the \textit{MARX} transformation according to 60 different targets. The marginal contribution of \textit{MARX} on the pseudo-$R^2$ depends a lot on models, horizons, and series. However, we remark that at the short-run horizons, when combined with nonlinear methods, it produces positive and significant effects. It particularly improves the forecast accuracy for real activity series like industrial production, labor market series and income, even at larger horizons. For instance, the gains from using \textit{MARX} with RF achieve 16\% when predicting INDPRO at the $h=3$ horizon, and 14\% in the case of employment if $h=6$. When used with linear methods, the estimates are more often on the negative side, except for inflation rates and M2 at short horizons, and a few special cases at the one and two-year ahead horizons.

\begin{figure}[h!]
	\centering
	\caption{Distribution of Marginal Effects of Target Transformation}
	\label{fig:s2a_treatment}
	\includegraphics[width=\textwidth,height=0.5\textheight]{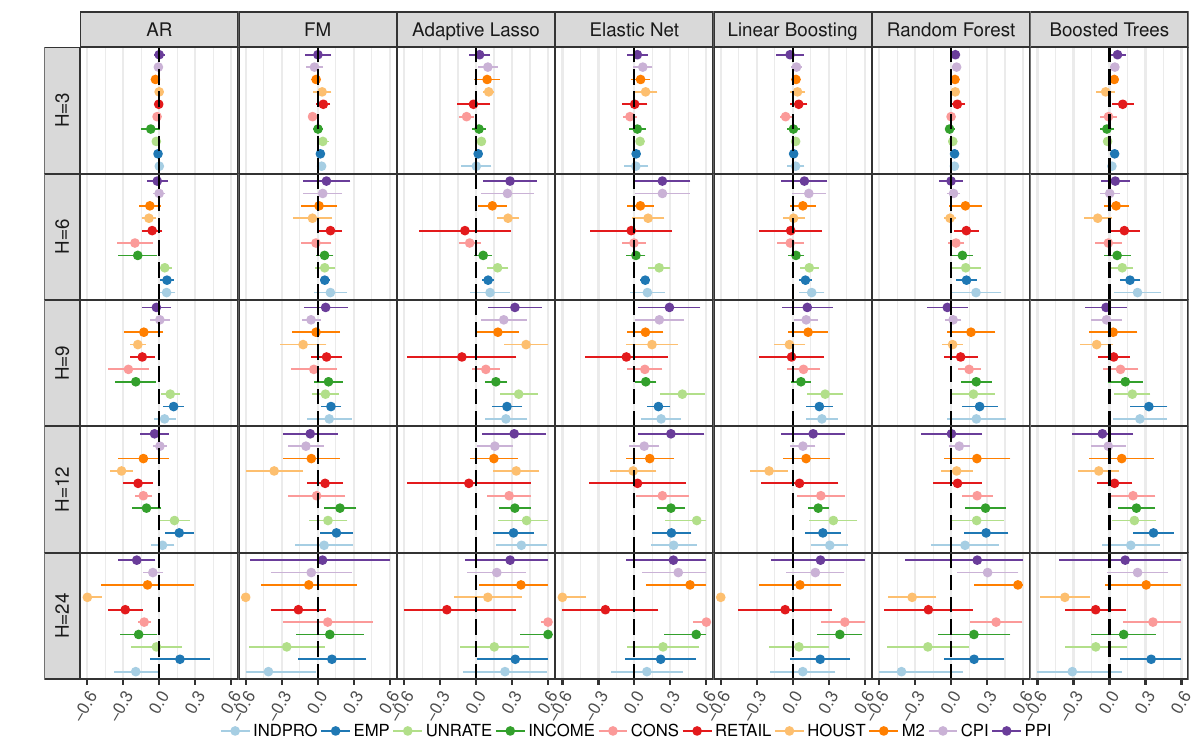}
	\flushleft
	\begin{scriptsize}
		\vspace{-2em}
		{\singlespacing
			Note: This figure plots the distribution of ${\alpha}_f^{(h,v)}$ from equation (\ref{e_eq2}) done by $(h,v)$ subsets. That is, it shows the average partial effect on the pseudo-$R^2$ from accumulating single period predictions ($\hat{y}_{t+h}^{\text{path-avg}}$) instead of targeting the average growth rate directly ($\hat{y}_{t+h}^{\text{direct}}$), keeping everything else fixed. SEs are HAC. These are the 95\% confidence bands. \par}  
	\end{scriptsize}
\end{figure}

\vspace{0.2em}

\noindent {\sc \textbf{Direct vs Path Average.}} Figure \ref{fig:s2a_treatment} reports the most unequivocal result of this paper: $\hat{y}_{t+h}^{\text{direct}}$ can prove largely suboptimal to $\hat{y}_{t+h}^{\text{path-avg}}$. For every method using a high-dimensional $Z_t$ shrunk in some way, i.e., \textit{not} the OLS-based AR and FM,  $\hat{y}_{t+h}^{\text{path-avg}}$ will do significantly better than the direct approach, with  $\alpha_{\mathit{\text{path-avg}}}^{(h,v)}$ sometimes around 30\% and highly statistically significant. As mentioned earlier, those gains are most prevalent for the highly cyclical variables {and longer horizons}. Cases where $\hat{y}_{t+h}^{\text{path-avg}} \prec \hat{y}_{t+h}^{\text{direct}}$ are rare and usually not statistically significant at the 5\% level, except for AR and FM which are both fitted by OLS.

How to explain this phenomenon?  Aggregating separate horizon forecasts allows to leverage the "bet on sparsity" principle of \cite{hastie2015statistical}. Presume the model for $\widehat{\Delta Y}_{t+h'}$ is sparse for each $h'$, yet different. This implies that the \textit{direct} model for $\hat{y}_{t+h}^{\text{direct}}$ is dense, and a much harder problem to learn. RF, BT, and Lasso will all perform better under sparsity, as every model struggle in a truly dense environment (unless it has a factor structure, upon which it becomes sparse in rotated space). An implication of this is that one should, as much as possible, try to make the problem sparse. Yet, whether sparsity will be more prevalent for $\hat{y}_{t+h}^{\text{path-avg}}$ or $\hat{y}_{t+h}^{\text{direct}}$ depends on true DGP. The evidence from Figure \ref{fig:s2a_treatment} suggests that DGPs favoring $\hat{y}_{t+h}^{\text{path-avg}}$ are more prevalent in our experiment. What do those look like?

	
We find it useful to connect this question to recent works on forecasts aggregation, like \cite{Bermingham2014} who forecast the year on year inflation and compare two strategies: forecasting overall inflation directly vs forecasting individual elements of the consumption basket and using a weighted average of forecasts. They find that using more components and aggregating individual forecasts improves performance.\footnote{In a similar vein, \cite{marcellino2003macroeconomic} found that forecasting inflation at
the country level and then aggregating the forecasts increases does better than
forecasting at the aggregate level (Euro).} They provide a simple example to rationalize their result: forecasting an aggregate variable made of two series with differing levels of persistence using only past values of the aggregate will be misspecified. In ML forecasting context, where $Z$ contains "everything" anyway, 
this problem translates from misspecification into making once sparse problems into a dense one, which is harder to learn. Consider a toy multi-horizon problem
\begin{equation}\label{eq:agg}
\begin{aligned}
		\Delta Y_{t+h'} &= \beta_h X_{t,k^*(h')} + \epsilon_{t+h'}, \; h'=1,2 \\
		y_{t+2}    &= \frac{\Delta Y_{t+2} + \Delta Y_{t+1}}{2} \\
		\Rightarrow y_{t+2} &= \frac{\beta_1}{2} X_{t,k^*(1)} + \frac{\beta_2}{2} X_{t,k^*(2)} + \frac{\epsilon_{t+1} + \epsilon_{t+2}}{2}.
\end{aligned}
\end{equation}
where one needs to select a single predictor for each horizon.   In this simple analogy to a high-dimensional problem, unless $k^*(1)=k^*(2)$, that is, the optimally selected regressor is the same for both horizon, the direct approach implies a "denser" problem -- estimating two coefficients rather than one for separate regressions. A scaled-up version of this is that if each horizon along the path implies 25 non-overlapping predictors, then the average growth rate model should have $25 \times h$ predictors, a much harder learning problem. 

Of course, 
the $\hat{y}_{t+h}^{\text{direct}}$ approach might work better, even in a ML environment. For instance, the "aggregated" error term in \eqref{eq:agg} could have a lower variance if $\mathrm{Corr}(\epsilon_{t+1},\epsilon_{t+2}) < 0$. 
Note that this would not imply substantial differences in the OLS paradigm since such errors would rather average out at the aggregation step in $\hat{y}_{t+h}^{\text{path-avg}}$. However, if a regularization level must be picked by cross-validation (like Lasso's $\lambda$), an environment where there is a strong common component across $h'$'s for the conditional mean could favor $\hat{y}_{t+h}^{\text{direct}}$. The reason for this is that choosing a regularization level optimized for a single horizon $h'$ could be different than what may be optimal for the final averaged prediction -- as examplified by our ridge regression case of equations \eqref{rr1} and \eqref{rr2}. This observation is closely related to that of \cite{Granger1987} who shows that the behavior of the aggregate series can easily be dominated by a common component \textbf{even if} it is unimportant for each of the microeconomic unit being aggregated. Translated to our ML-based multi-horizon problem, this means we want to avoid having overly harsh regularization throwing out negligible effects for a given $h'$ whose accumulation over all $h'$'s makes them in fact non-negligible. Thus, if the noise level is much higher for single horizons forecasts, an overly strong $\lambda_{h'}$ for each $h'$ may be chosen whereas $\lambda_{h}$ for $\hat{y}_{t+h}^{\text{direct}}$ could be milder and allow for otherwise neglected signals to come through.

\begin{figure}[t!]
	\begin{center}
	\caption{Variable Importance} \label{VI}
		\begin{subfigure}{.5\textwidth}	\includegraphics[width=\textwidth]{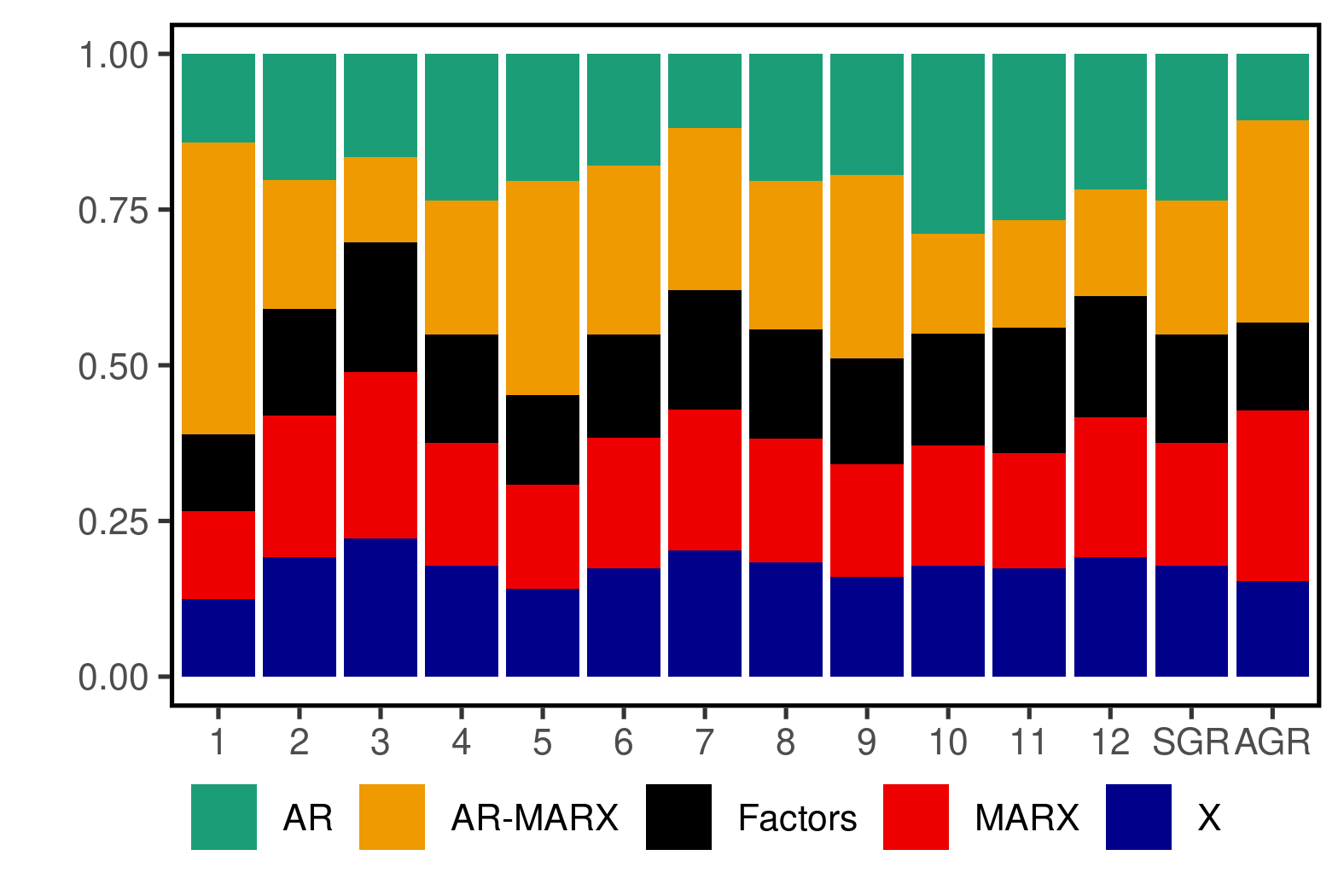}
			\caption*{\hspace{1.5em} Income}
		\end{subfigure}%
		\begin{subfigure}{.5\textwidth}	\includegraphics[width=\textwidth]{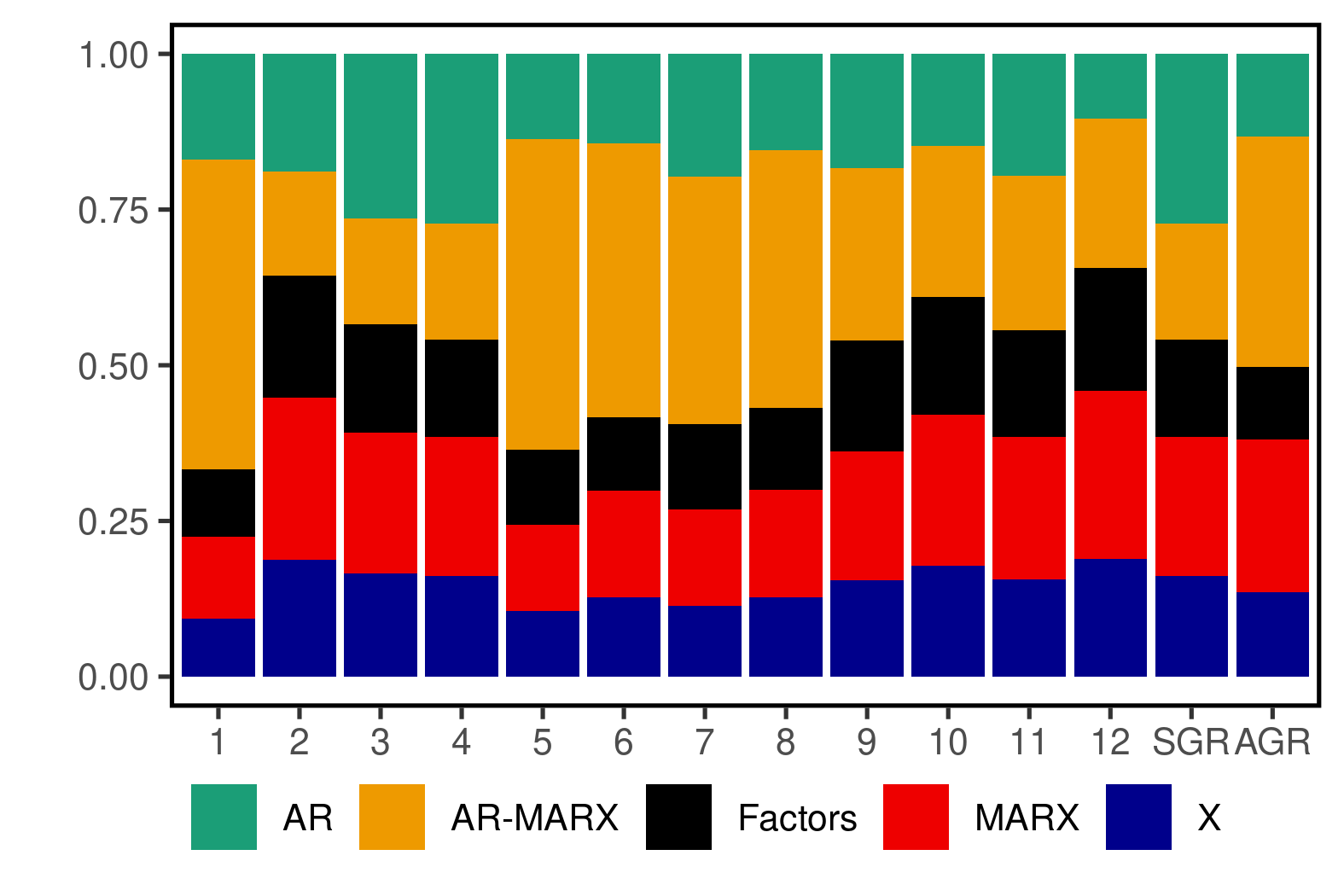}
			\caption*{\hspace{2em} Employment}
		\end{subfigure}
		\begin{subfigure}{.5\textwidth}
			\includegraphics[width=\textwidth]{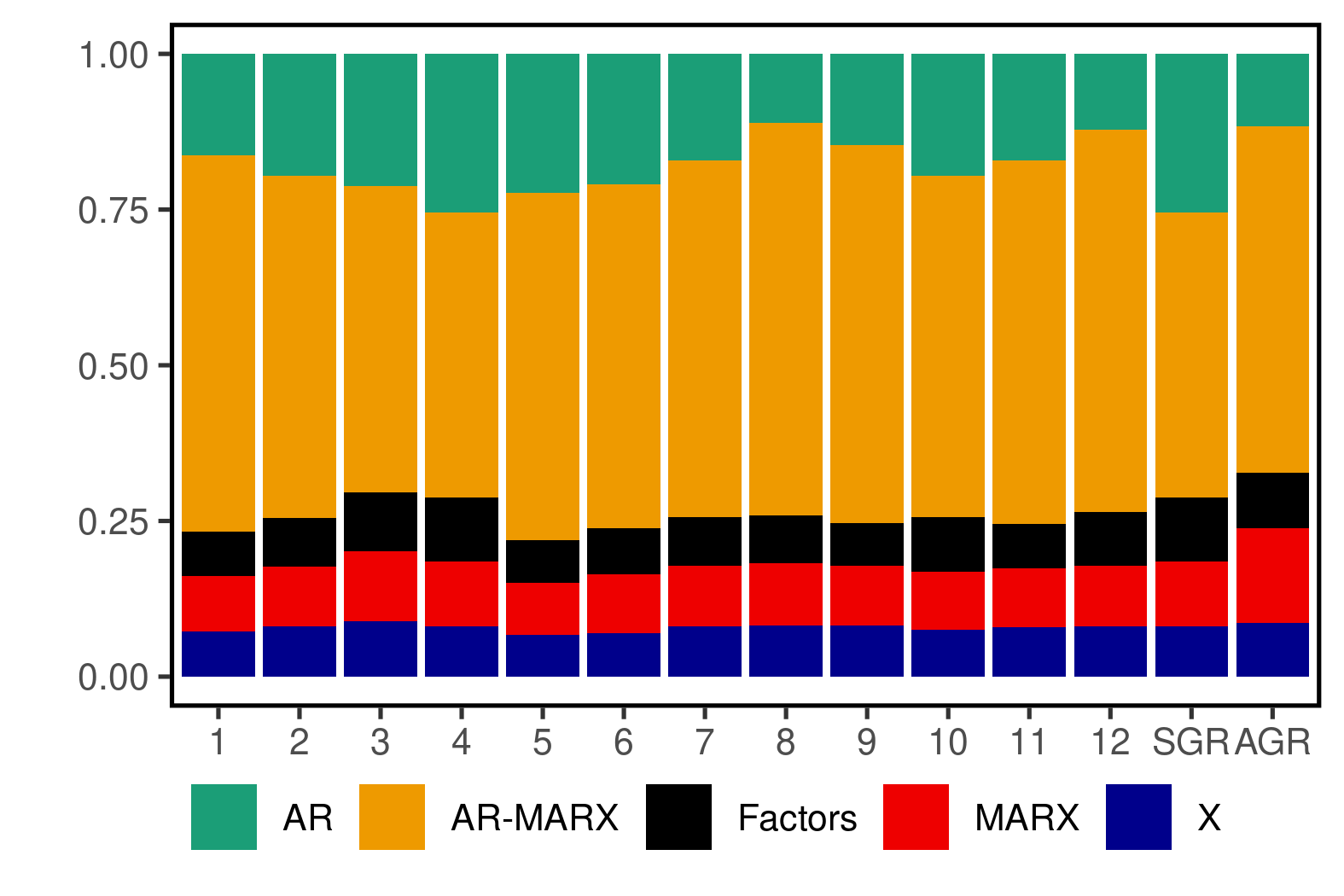}
			\caption*{\hspace{1em} Inflation}
		\end{subfigure}%
		\begin{subfigure}{.5\textwidth}
			\includegraphics[width=\textwidth]{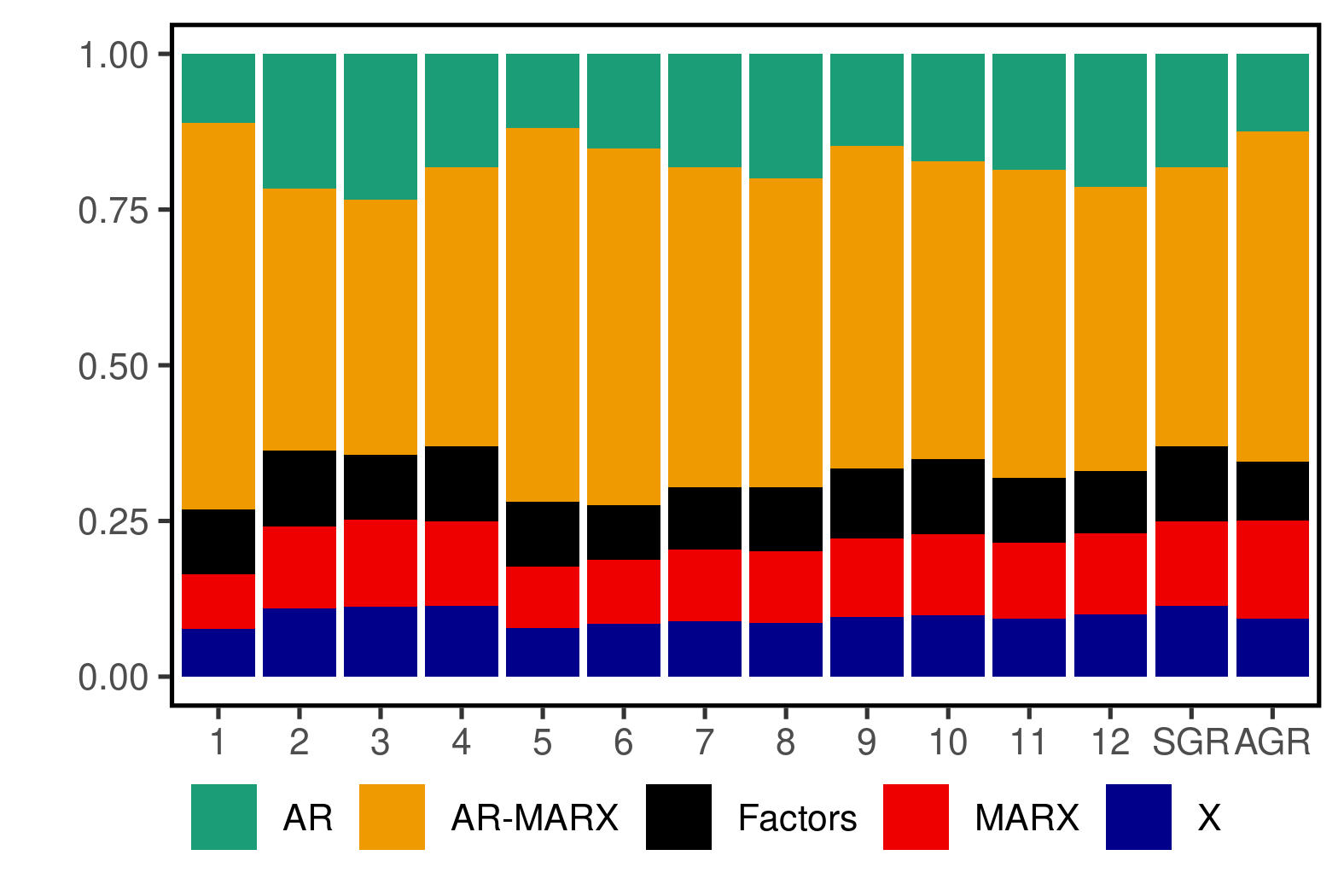}
			\caption*{\hspace{1em} M2 money Stock}
		\end{subfigure}
	\end{center}
	\begin{footnotesize}
		\flushleft
		\vspace{-1em}
		Notes: This figure displays the relative variable importance (VI) measures for the Random Forest F-X-MARX model for horizon $H=12$. Group values are additions of VI for individual series weighted by the share of each groups with the total VI normalized to 1. The first 12 bars reflect horizon-wise differences for the $\hat{y}_{t+h}^{\text{path-avg}}$ models whose forecasts are accumulated and the subsequent bar shows the average importance across those horizons. The last bar displays the equivalent for the $\hat{y}_{t+h}^{\text{direct}}$ model. 
	\end{footnotesize}
\end{figure}

These potential explanations are illustrated using variable importance (VI)  in Figure \ref{VI}. As shown earlier, the path average approach has outperformed the direct one when predicting real activity variables. VI measures in top panels show how models for $\hat{y}_{t+h}^{\text{path-avg}}$ use a much more polarized set of variables whereas those aiming for $\hat{y}_{t+h}^{\text{direct}}$ using a very diverse set of predictors in case of Income and Employment. This shed light on our bet-on-sparsity conjecture, i.e. that $\hat{y}_{t+h}^{\text{path-avg}}$ will have the upper hand if ${\Delta \hat{Y}_{t+h'}}$ predictive problems are quite heterogenous. In both cases, horizon 1 is quite different from 2-3-4, which also differ from the 5-12 block. It is noted in Figures \ref{case:emp_sgrtoagr} and \ref{case:income_sgrtoagr} that $\hat{y}_{t+h}^{\text{path-avg}}$ visibly demonstrate a better capacity for autoregressive behavior (even at $h=12$) which provides it with a clear edge over $\hat{y}_{t+h}^{\text{direct}}$ during recessions. Interestingly, the foundation for this finding is also visible in Figure \ref{VI} for real activity variables: $\hat{y}_{t+h}^{\text{path-avg}}$ reliance on plain AR terms is more than twice that of $\hat{y}_{t+h}^{\text{direct}}$.

The bottom panels show VI measures for CPI inflation and M2 growth. Recall that 
$\hat{y}_{t+h}^{\text{path-avg}} \prec \hat{y}_{t+h}^{\text{direct}}$ was unambiguous for 
those variables. Here again, results are in line with the above arguments. The retained predictors' sets are much more \textit{similar} across the two approaches, which results from the presence of a strong common component over horizons (i.e., persistence which constitutes about 75\% of normalized VI), which favors $\hat{y}_{t+h}^{\text{direct}}$. 

\vspace{0.2em}

\begin{figure}[t!]
\centering
\caption{Distribution of \textit{MAF} Marginal Effects}
\label{fig:MAFwXtornado}
\includegraphics[width=\textwidth,height=0.5\textheight]{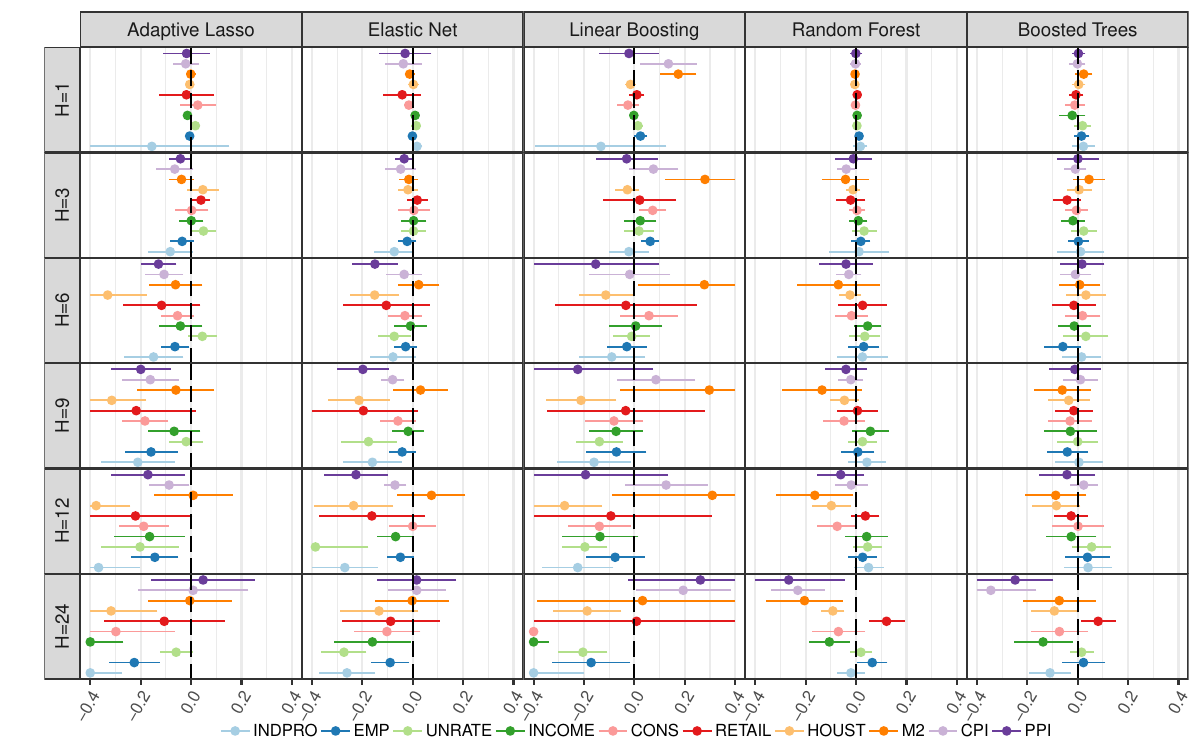}
\flushleft
\begin{scriptsize}
	\vspace{-2em}
	{\singlespacing
	Notes: This figure plots the distribution of ${\alpha}_f^{(h,v)}$ from equation (\ref{e_eq2}) done by $(h,v)$ subsets. That is, it shows the average partial effect on the pseudo-$R^2$ from augmenting the model with \textit{MAF} featuring, keeping everything else fixed. SEs are HAC. These are the 95\% confidence bands. \par}  
\end{scriptsize}
\end{figure}

\noindent {\sc \textbf{MAF.}} Figure \ref{fig:MAFwXtornado} plots the distribution of $\alpha_{\mathit{MAF}}^{(h,v)}$, conditional on including $X$ in the model. The motivation for that is that \textit{MAF}, by construction, summarizes the entirety of $[X_{t-p}]_{p=1}^{p=P_{MAF}}$ with no special emphasis on the most recent information.\footnote{Of course, one could alter the PCA weights in \textit{MAF} to introduce priority on recent lags \`{a} la Minesota-prior, but we leave that possibility for future research.} Thus, it is better-advised to always include the raw $X$ with \textit{MAF}, so recent information may interact with the lag polynomial summary if ever needed. \textit{MAF} contributions are overall more muted than that of \textit{MARX}, except when used with Linear Boosting method. Nevertheless, it is noticed that it shares common gains with the latter as short horizons ($h=3,6$) of real activity variables also benefit from it. More convincing improvements are observed for retail sales at the 2-year horizons for nonlinear methods. 

\begin{figure}[t!]
\centering
\caption{Distribution of \textit{F} Marginal Effects}
\label{fig:FvsXtornado}
\includegraphics[width=\textwidth,height=0.5\textheight]{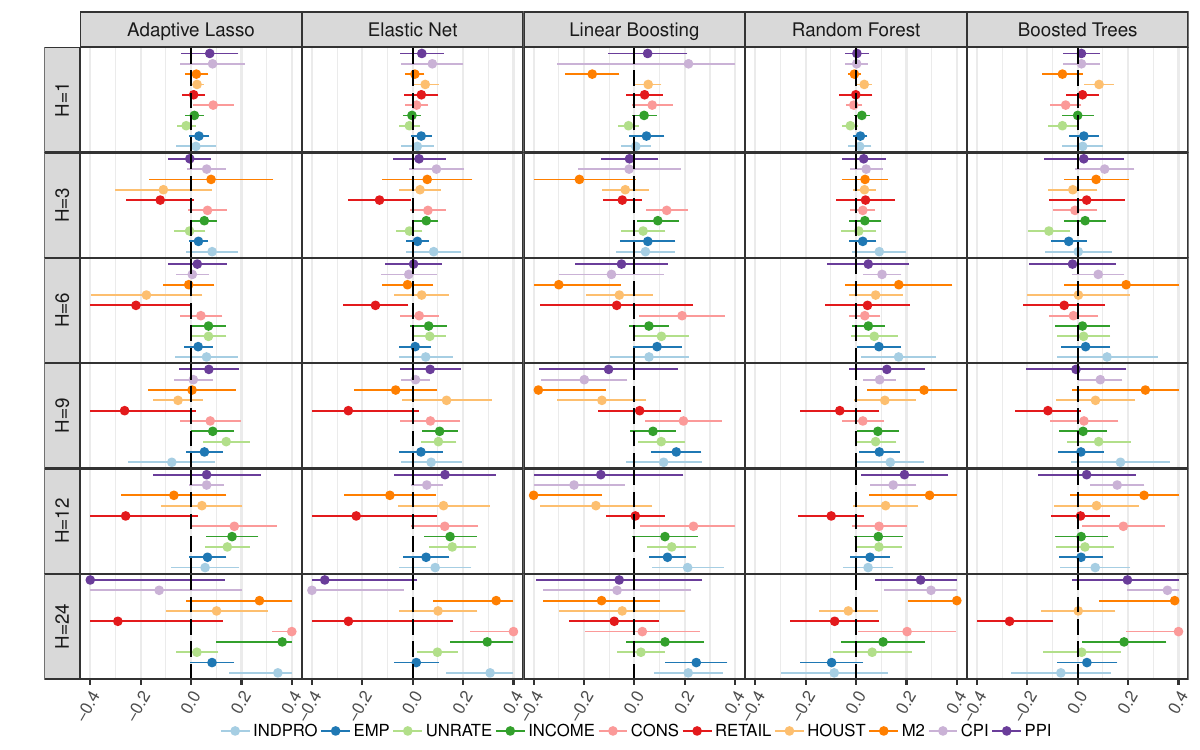}
\flushleft
\begin{scriptsize}
	\vspace{-2em}
	{\singlespacing
	Notes: This figure plots the distribution of ${\alpha}_f^{(h,v)}$ from equation (\ref{e_eq2}) done by $(h,v)$ subsets. That is, it shows the partial effect on the pseudo-$R^2$ from considering only $F$ featuring versus including only observables $X$. SEs are HAC. These are the 95\% confidence bands. \par}  
\end{scriptsize}
\end{figure}

\vspace{0.2em}
\noindent {\sc \textbf{Traditional Factors.}} It has already been documented that factors matter -- and a lot \citep{stock2002forecasting, stock2002macroeconomic}. Figure \ref{fig:FvsXtornado} allows us to evaluate their quantitative effects.  
Including a handful of factors rather than all of (stationary) $X$ improves substantially and significantly forecast accuracy. The case for this is even stronger when those are used in conjunction with nonlinear methods, especially for prediction at longer horizons. This finding supports the view that a factor model is an accurate depiction of the macroeconomy, as originally suggested in the works of \cite{Sargent-Sims(1977)} and \cite{Geweke1977} and later expanded in various forecasting and structural analysis applications \citep{stock2002forecasting,Bernanke-Boivin-Eliasz(2005)}. In this line of thought, transforming $X$ into $F$ is not merely a mechanical dimension reduction step. Rather, it is meaningful feature engineering uncovering true latent factors which contains most, if not all, the relevant information about the current state of the economy. Once $F$'s are extracted, the standard diffusion indexes model of \cite{stock2002macroeconomic} can either be upgraded by using linear methods performing variable selection, or nonlinear functional form approximators such as Random Forests and Boosted Trees.  

\subsection{Case Study}\label{sec:casestudy}

\begin{figure}[h!]
	\begin{center}
		\caption{Cumulative Squared Error (Direct)}\label{cumulerrors}
		\hspace*{-2.1em}
		\includegraphics[width=7.35in, height=3in]{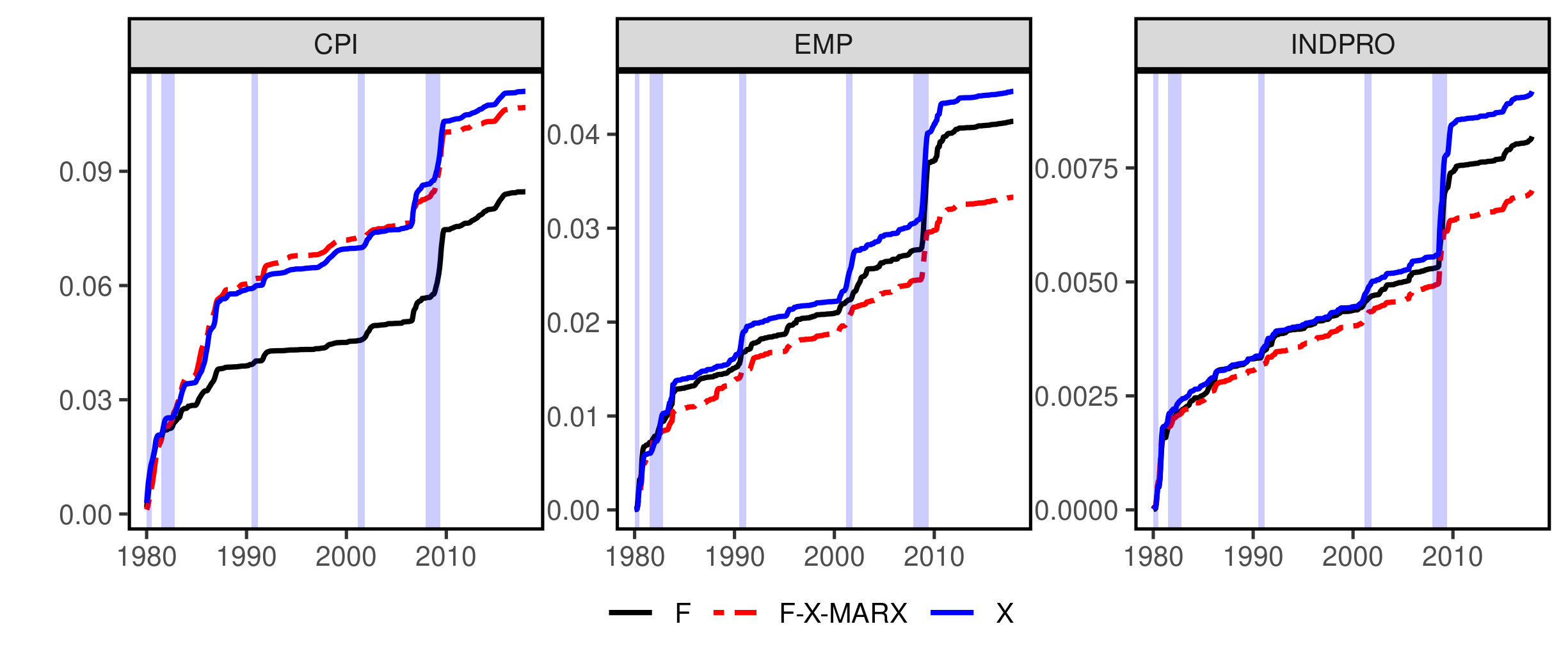}
	\end{center}
\end{figure}

In this section we conduct "event studies" to highlight more explicitly the importance of data pre-processing when predicting real activity and inflation indicators. Figure \ref{cumulerrors} plots cumulative squared errors for three cases where specific transformations stand out. On the left, we compare the performance of RF when predicting industrial production growth three months ahead, using either \textit{F}, \textit{X} or \textit{F-X-MARX} as feature matrix. The middle panel shows the same exercise for employment growth. On the right, we report one-year ahead CPI inflation forecasts. Industrial production and employment examples clearly document the merits of including \textit{MARX}: its cumulatively summed squared errors (when using RF) are always below the ones produced by using \textit{F} and \textit{X}. The gap widens slowly until the Great Recession, after which it increases substantially. As discussed in section \ref{sec:results}, using common factors with RF constitutes the optimal specification for CPI inflation. Figure \ref{cumulerrors} illustrates this finding and shows that the gap between using \textit{F} or \textit{X} widens during the mid-80s, the mid-90s, and just before the Great Recession. To provide a statistical assessment of  the stability of forecast accuracy, we consider the fluctuation test of \cite{Giacomini-Rossi(2010)} in Appendix \ref{sec:stability}.  

\begin{footnotesize}
	\flushleft
	\vspace*{-5em}
	{ \singlespacing
		Notes: Cumulative squared forecast errors for INDPRO and EMP (3 months) and CPI (12 months). All use the Random Forest model and the direct approach. CPI and EMP have been scaled by 100. \par}
\end{footnotesize}

\begin{figure}[p!]
	\caption{Case of Industrial Production (Direct)} \label{case:indpro}
	\begin{subfigure}{\textwidth}
		\centering
		\includegraphics[width=3in, height=2.5in]{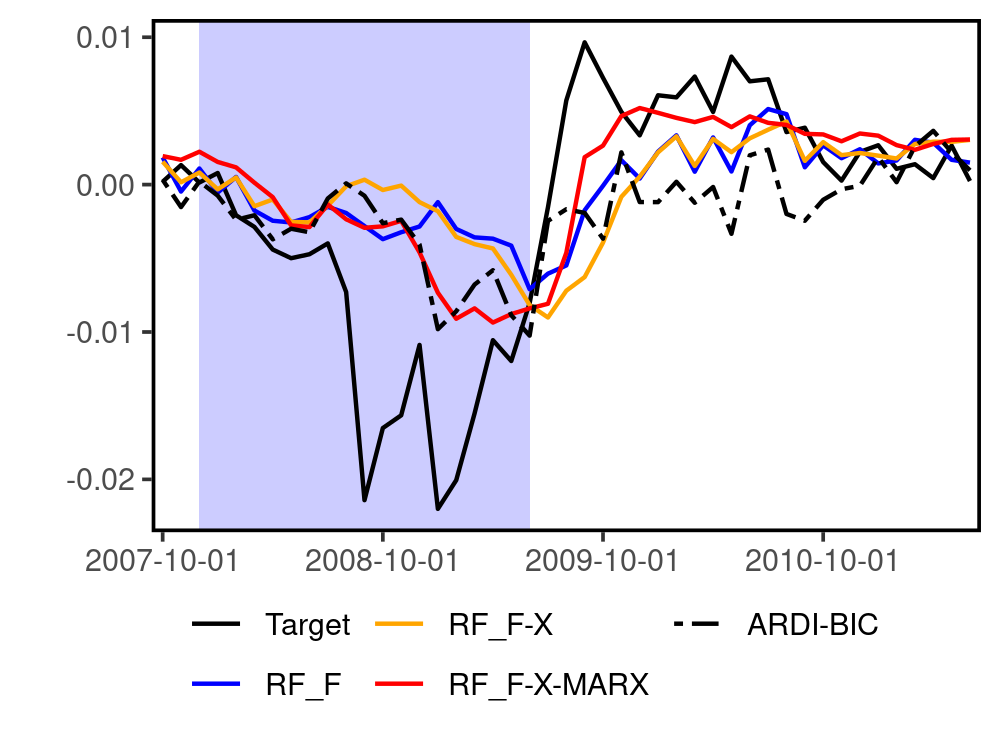}%
		\hspace{3em}
		\includegraphics[width=3in, height=2.5in]{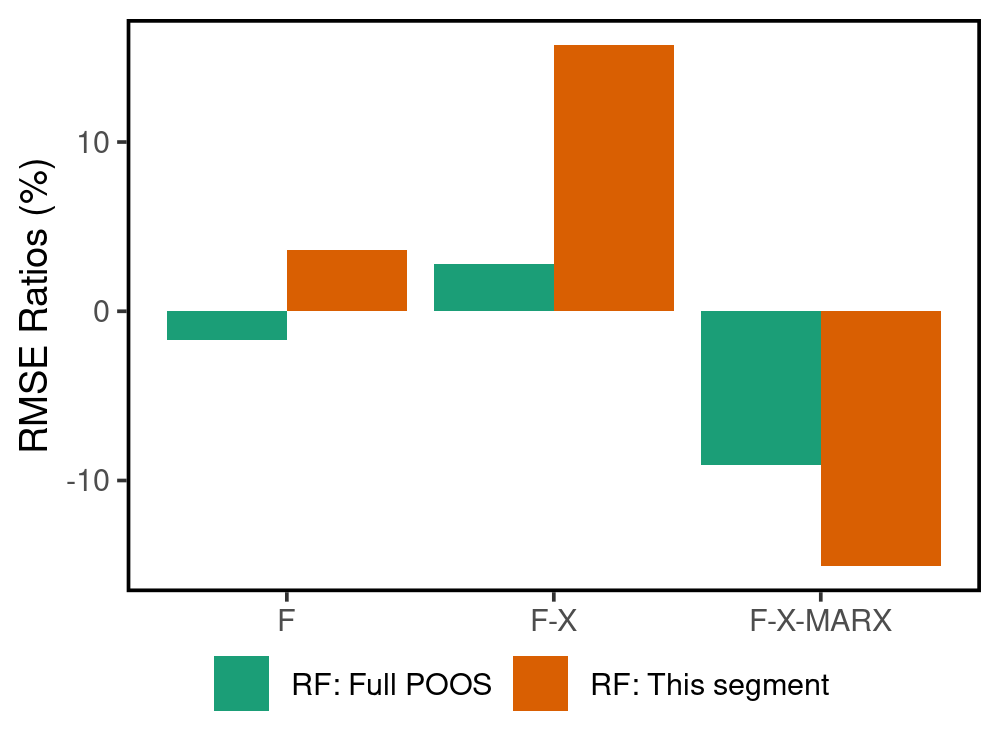}
		\caption{Recession Episode of 2007-12-01}
	\end{subfigure}
	\begin{subfigure}{\textwidth}
		\centering
		\includegraphics[width=3in, height=2.5in]{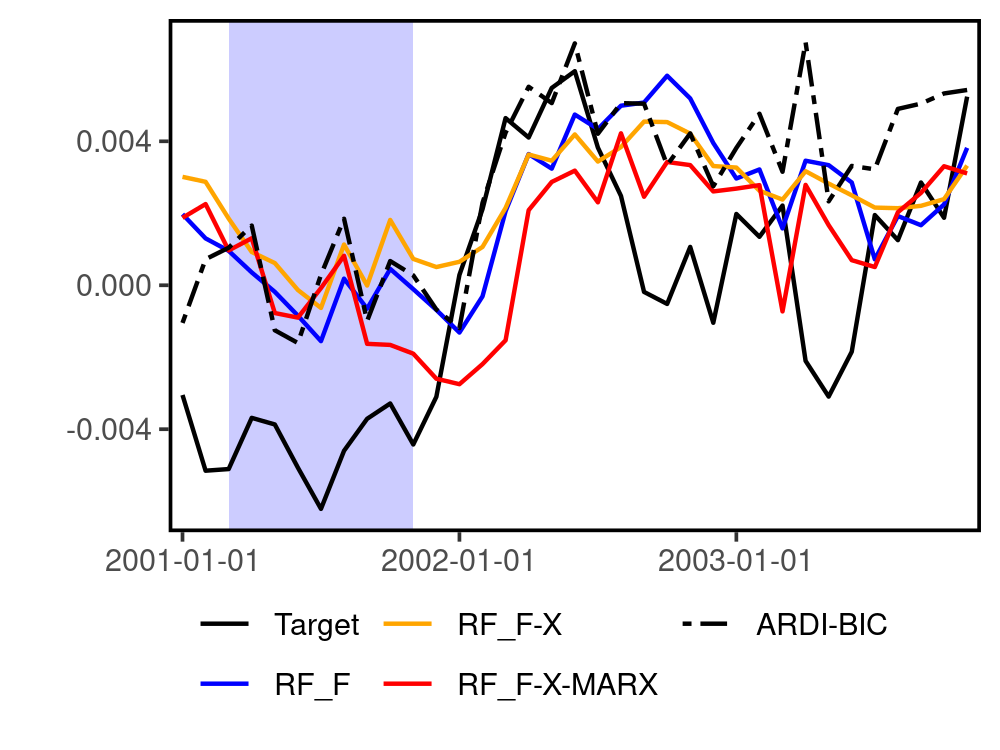}%
		\hspace{3em}
		\includegraphics[width=3in, height=2.5in]{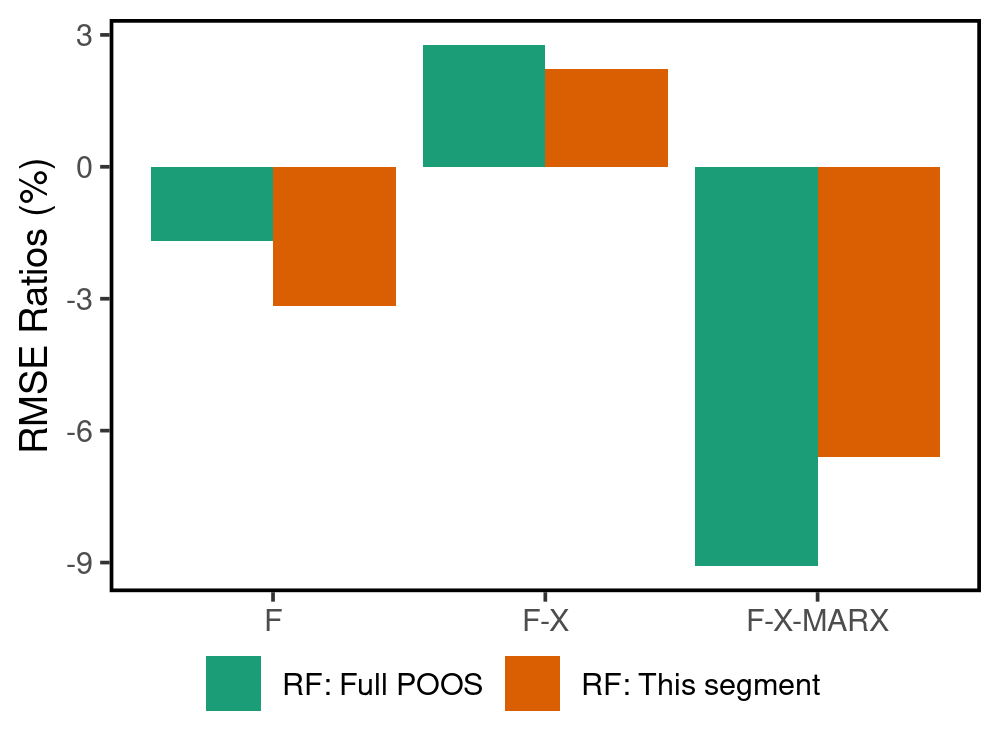}
		\caption{Recession Episode of 2001-03-01}
	\end{subfigure}
	\begin{subfigure}{\textwidth}
		\centering
		\includegraphics[width=3in, height=2.5in]{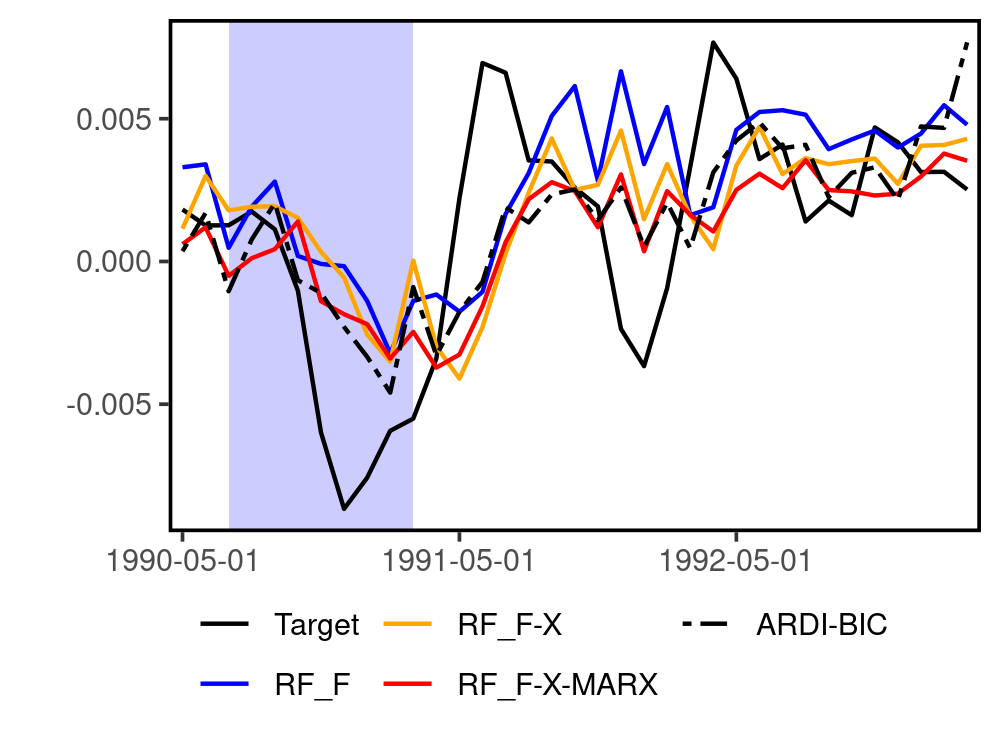}%
		\hspace{3em}
		\includegraphics[width=3in, height=2.5in]{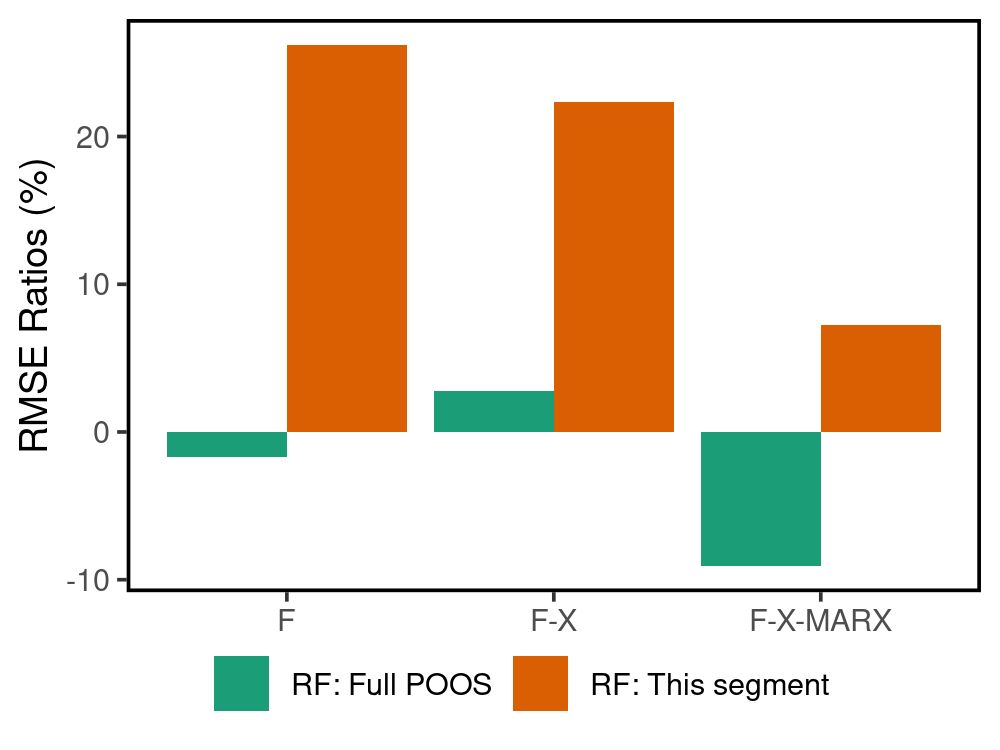}
		\caption{Recession Episode of 1990-07-01}
	\end{subfigure}
	\begin{footnotesize}
		\vspace{-3em}
		\flushleft \singlespacing
		Notes: The figure covers 3 months before and 24 months after the recession. RMSE ratios are relative to FM model and the episode RMSE refers to the visible time period.
	\end{footnotesize}
\end{figure}

In Figure \ref{case:indpro}, we look more closely at each model's forecasts during last three recessions and subsequent recoveries.  Specifically, we plot the 3-month ahead forecasts for the period covering 3 months before, and 24 months after a recession, for industrial production and employment. The forecasting models are all RF-based, and differ by their use of either \textit{F}, \textit{X} or \textit{F-X-MARX}. On the right side, we show the RMSE ratio of each RF specification against the benchmark FM model for the whole POOS and for the episode under analysis. In the case of industrial production, the \textit{F-X-MARX} specification outperforms the others during the Great Recession and its aftermath, and improves even more upon the benchmark model compared to the full POOS period. We observe on the left panel that forecasts made with \textit{F-X-MARX} are much closer to realized values at the end of recession and during the recovery. The situation is qualitatively similar during the 2001 recession but effects are smaller. Including \textit{MARX} also emerges as the best alternative around the 1990-1991 recession, but the benchmark model is more competitive for this particular episode. 

In the case of employment showcased in Figure \ref{case:emp} in Appendix \ref{sec:add_results}, \textit{MARX} again supplants \textit{F} or \textit{X} in all three recessions. For instance, around the Dotcom bubble burst, it displays an outstanding performance,  surpassing the benchmark by 40\%. However, during the Great Recession, it is outperformed by the traditional factor model.  Finally, the \textit{F-X-MARX} combination provides the most accurate forecast during and after the credit crunch recession of the early 1990s.

\begin{figure}[p!]
	\caption{Case of Employment (Path Average)}\label{case:emp_sgrtoagr}
	\begin{subfigure}{\textwidth}
		\centering
		\includegraphics[width=3in, height=2.5in]{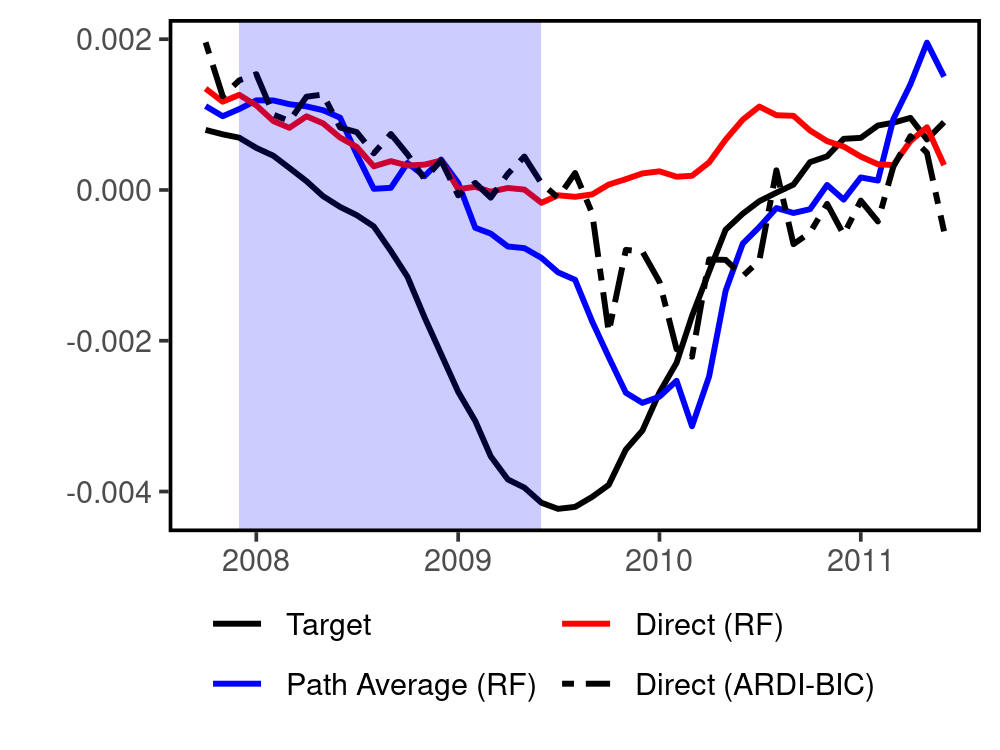}%
		\hspace{3em}
		\includegraphics[width=3in, height=2.5in]{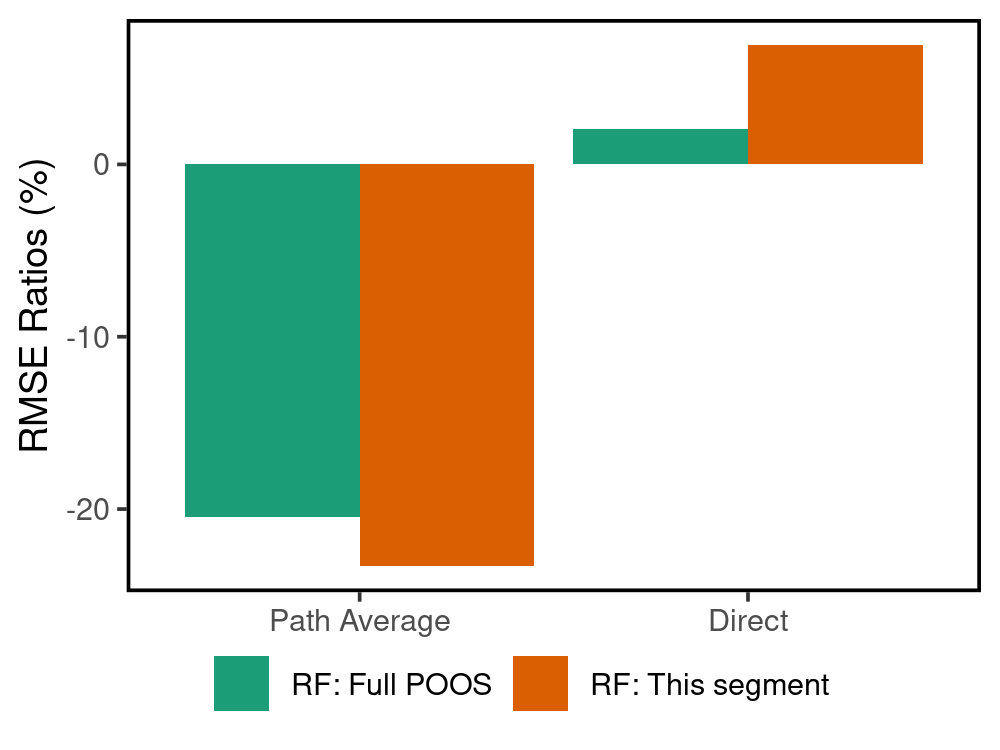}
		\caption{Recession Episode of 2007-12-01}
	\end{subfigure}
	\begin{subfigure}{\textwidth}
		\centering
		\includegraphics[width=3in, height=2.5in]{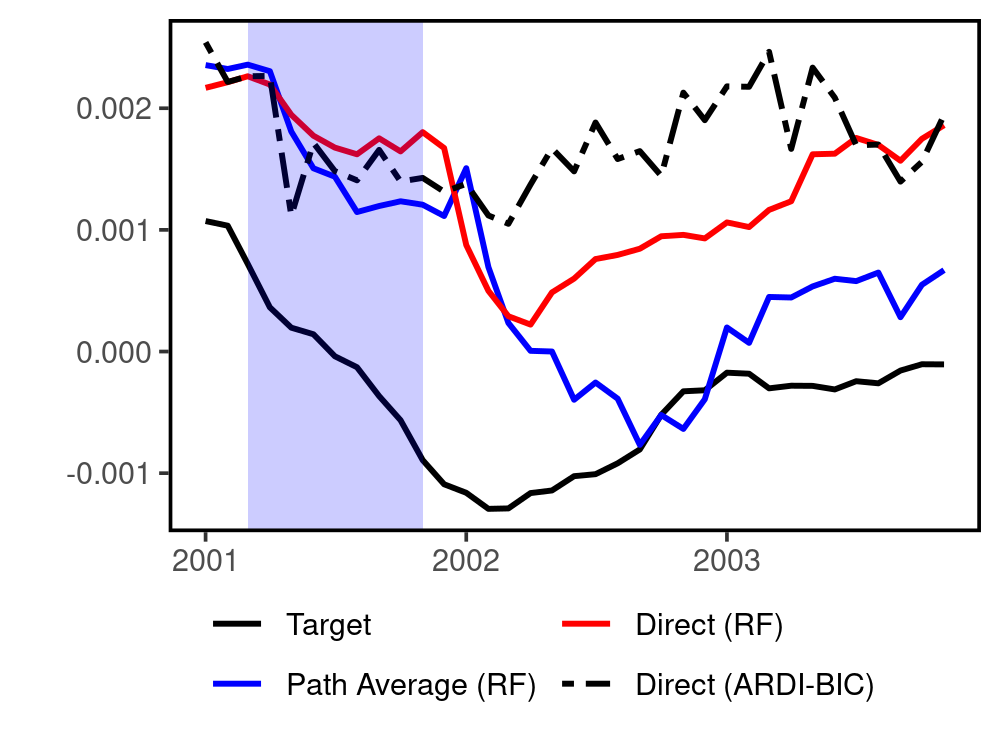}%
		\hspace{3em}
		\includegraphics[width=3in, height=2.5in]{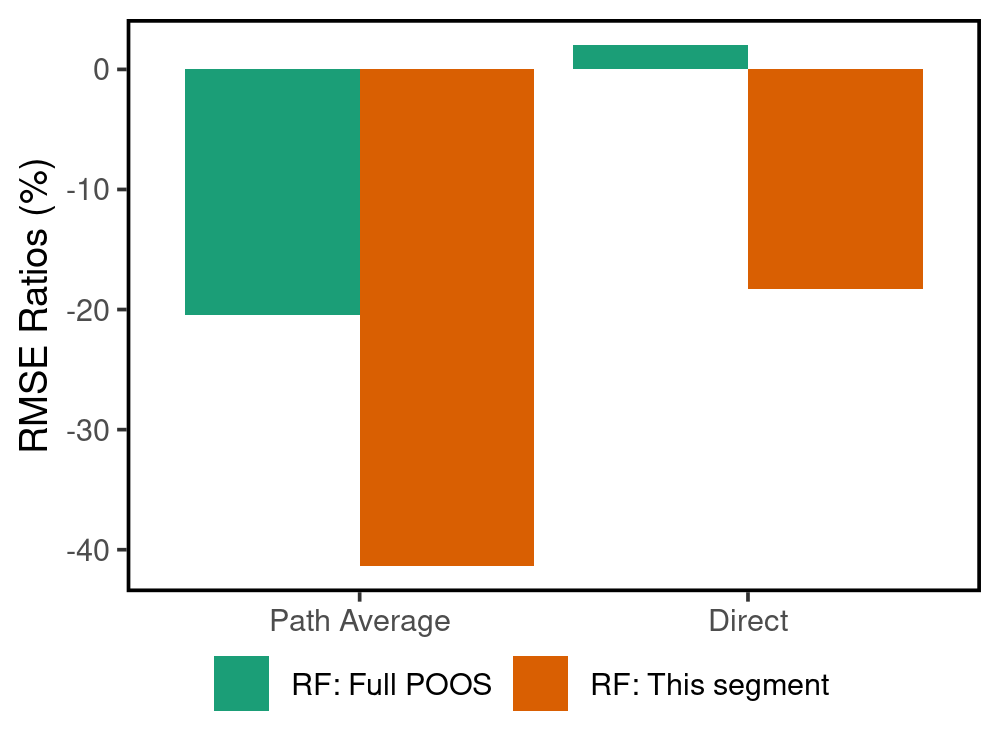}
		\caption{Recession Episode of 2001-03-01}
	\end{subfigure}
	\begin{subfigure}{\textwidth}
		\centering
		\includegraphics[width=3in, height=2.5in]{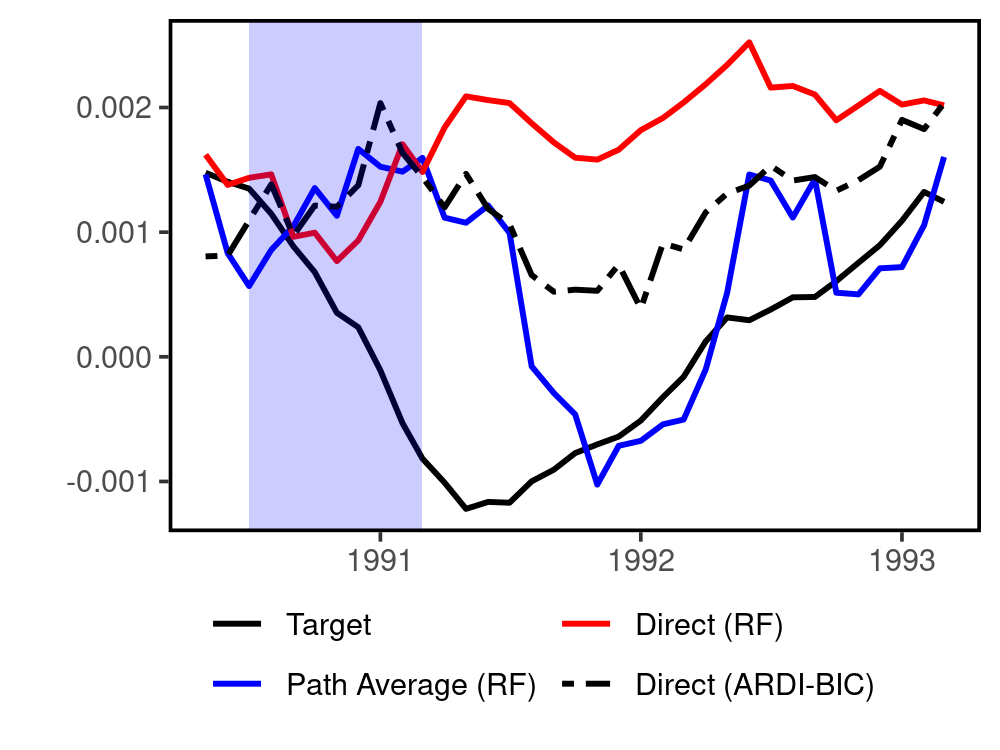}%
		\hspace{3em}
		\includegraphics[width=3in, height=2.5in]{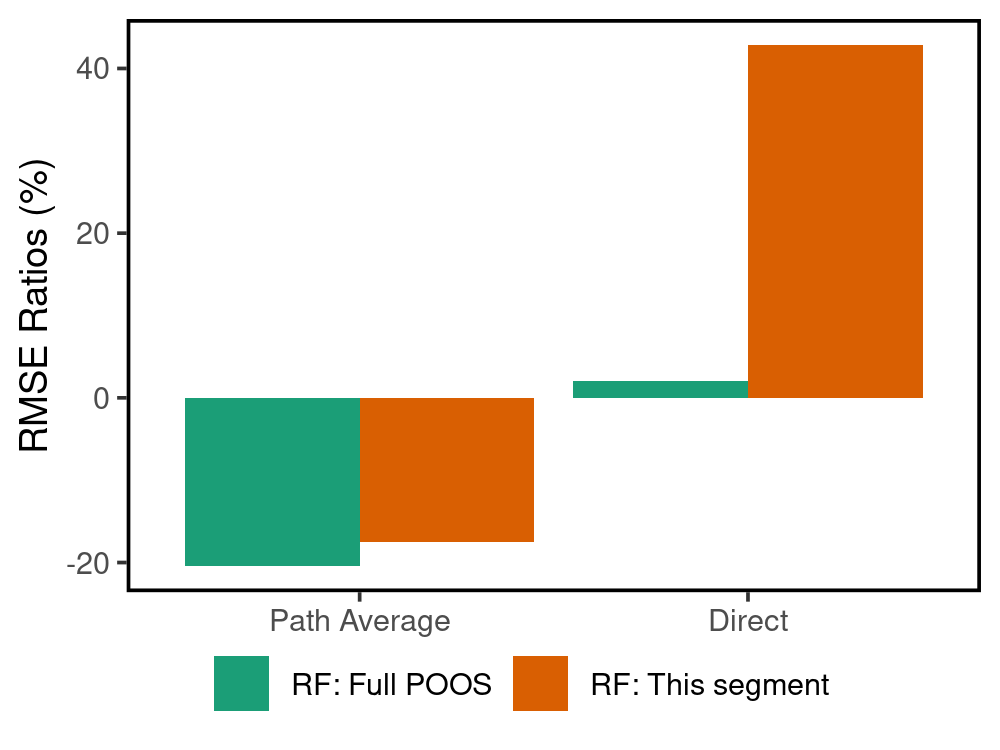}
		\caption{Recession Episode of 1990-07-01}
	\end{subfigure}
	\begin{scriptsize}
		\vspace{-3em}
		{\flushleft \singlespacing
			Notes: The figure plots 12-month ahead forecasts for the period covering 3 months before and 24 months after the recession. RMSE ratios are relative to FM model for average growth rates and the episode RMSE refers to the visible time period and Random Forest models use F-X-MARX. \par}
	\end{scriptsize}	
\end{figure}

Figure \ref{case:emp_sgrtoagr} illustrate the relative performance of the two target transformations for employment and income 12 months ahead. Again, we focus on the three most recent recession episodes. $\hat{y}_{t+h}^{\text{path-avg}}$ dramatically improves performance over $\hat{y}_{t+h}^{\text{direct}}$ and much of that  edge visibly comes from adjusting itself more or less rapidly to new economic conditions. In contrast, $\hat{y}_{t+h}^{\text{direct}}$ is extremely smooth and report something close to the long-run average. Since the last three recessions were characterized by a slow recovery, $\hat{y}_{t+h}^{\text{path-avg}}$ procures much more credible forecasts of employment and income simply by catching up sooner with realized values. This behavior is understandable through the lenses of Figure \ref{VI} where early horizons of $\hat{y}_{t+h}^{\text{path-avg}}$ make a pronounced use of autoregressive terms for both employment (and income, see Figure \ref{case:income_sgrtoagr} in Appendix \ref{sec:add_results}).  

\subsection{Extraneous Transformations}\label{sec:extransfo}

We evaluate four additional data transformation strategies in combination with direct and path average targets. 
First, we accommodate for the presence of error correction terms (ECM) by considering the Factor-augmented ECM approach of \cite{BANERJEE2014589} and include \textit{level} factors estimated from $I(1)$ predictors. Second, we consider volatility factors and data inspired by \cite{nggorodnichenkoVol}, where both factors from $X^2$ and $X^2$ itself are included as predictors. Third, we evaluate the potential predictive gains from including \cite{fhlrDFM}'s dynamic factors in $Z$. 

Figure \ref{fig:LevelFactor_AvgTreatmentEffect}, in Appendix \ref{sec:tornados_add}, reports the distribution of average marginal effects of adding level factors in the predictors' set $Z$. Their impact is generally small and not significant at short horizons, while it depends on methods and forecasting approach at longer horizons. In the case of the direct average approach, as depicted in panel \ref{fig:LevelFactor_AvgTreatmentEffect_direct}, adding level factors generally deteriorates the predictive performance except for M2 with nonlinear methods. The effects are qualitatively similar when the target is achieved by the path average approach, as shown in \ref{fig:LevelFactor_AvgTreatmentEffect_path}.

Adding volatility data and factors is generally harmful with linear methods and has almost no significant impact when random forest and boosted trees are used, see Figure \ref{fig:Volatility_AvgTreatmentEffect}.\footnote{The very weak contribution of volatility terms to BT or RF is expected given that those transformations are locally monotone (i.e, for all points where $X_{k,t}>0$ or $X_{k,t}<0$) and trees are invariant to monotone transformations.} Hence, letting ML methods generate nonlinearities proves to be more resilient than to include simple power terms. This also suggests that volatility or other uncertainty proxies may not be the major sources of nonlinearities for macroeconomic dynamics since they would otherwise be an indispensable form of feature engineering which variable selection algorithms build their predictions from. 

Finally, Figures \ref{fig:DFMAF_TreatmentEffect} and \ref{fig:DFSF_TreatmentEffect} evaluate the marginal predictive content of dynamic factors as opposed to MAF and static factors (PCs) respectively. Considering dynamic factors as opposed to MAF improves the predictability at longer horizons when used to construct $\hat{y}_{t+h}^{\text{direct}}$, while their effects are rather small with $\hat{y}_{t+h}^{\text{path-avg}}$. When it comes to the choice between dynamic and static factors, the results are in general quantitatively small but suggest that standard principal components are preferred, especially in combination with nonlinear methods, which is analogous to the findings of \cite{boivinng2005} in linear environments.

\section{Conclusion}\label{sec:conclusion}

This paper studies the virtues of standard and newly proposed data transformations for macroeconomic forecasting with machine learning. The classic transformations comprise the dimension reduction of stationarized data by means of principal components and the inclusion of level variables in order to take into account low frequency movements. Newly proposed avenues include moving average factors (MAF) and moving average rotation of $X$ (MARX). The last two were motivated by the need to compress the information within a lag polynomial, especially if one desires to keep $X$ close to its original -- interpretable -- space. {In addition to the aforementioned transformations focusing on $X$, we considered two pre-processing alternatives for the target variable, namely the direct and path average approaches.} 


To evaluate the contribution of data transformations for macroeconomic prediction, we have considered three linear and two nonlinear ML methods (Elastic Net, Adaptive Lasso, Linear Boosting, Random Forests and Boosted Trees) in a substantive pseudo-out-of-sample forecasting exercise was done over 38 years for 10 key macroeconomic indicators and 6 horizons. With the different permutations of $f_Z$'s available from the above, we have analyzed a total of 15 different information sets. The combination of standard and non-standard data transformations  (\textit{MARX}, \textit{MAF},  \textit{Level}) is shown to minimize the RMSE, particularly at shorter horizons. Those consistent gains are usually obtained when a nonlinear nonparametric ML algorithm is being used. This is precisely the algorithmic environment we conjectured could benefit most from our proposed $f_Z$'s. Additionally, traditional factors are featured in the overwhelming majority of best information sets for each target. Therefore, while ML methods can handle the high-dimensional $X$ (both computationally and statistically), extracting common factors remains straightforward feature engineering that works. 

{The way the prediction is constructed can make a great difference. The path average approach is more accurate than the direct one for almost all real activity variables (and at various horizons). The gains can be as large as 30\% and are mostly observed when the path average approach is used in conjunction with regularization and/or nonparametric nonlinearity.} 

As the number of researchers and practitioners in the field is ever-growing, we believe those insights constitute a strong foundation on which stronger ML-based systems can be developed to further improve macroeconomic forecasting. 

%
%

\clearpage
\onehalfspace

\setlength\bibsep{5pt}
\bibliographystyle{apalike}
\bibliography{references.bib}

\clearpage
\appendix

\doublespace

\section{ONLINE APPENDIX - Forecasting Models in Details}\label{sec:models}

In this section, we briefly review the basic of the econometric ML methods being used in this paper. For a more complete discussion, see, among other, \cite{hastie2009elements}.

\vskip 0.2cm

{\sc \noindent \textbf{Hyperparameter Optimization}.} We use different means of optimizing hyperparameters depending the problem considered. In each case, the goal is to minimize expected square loss out-of-sample which is approximated using a 5-fold cross-validation. The simplest means of achieving this goal is a \textbf{grid search} which selects the minimizing hyperparameter vector out of a predetermined grid of admissibble candidates. The algorithm is showcased below.

\begin{center}
	\begin{tabular}{l}
		\toprule \toprule
		Grid Search Algorithm \\ \midrule
		1: Randomly assign observations to 5 folds $(Z^{(j)}, y^{(j)})$. Save sizes $\#j$. \\
		2: Define a grid $\mathcal{G}$ for the hyperparameter vector $\tau$ \\
		3: \textbf{For} each $g$ in $\mathcal{G}$ \textbf{do}: \\
		4: \hspace{2em} \textbf{For} $j$ \textbf{in} $1:5$ \textbf{do}: \\
		5: \hspace{4em} Train model on $\{ (Z^{(i)}, y^{(i)}) : \forall i \neq j \}$ using $\tau(g)$ \\
		6: \hspace{4em} Compute prediction: $\hat{y}^{i,j}(\tau(g))$ \\
		7: \hspace{4em} Compute loss: $l_j(\tau(g)) : = (y^{(j)} - \hat{y}^{i,j}(\tau(g)))'(y^{(j)} - \hat{y}^{i,j}(\tau(g)))/ \#j$ \\
		8: \hspace{2em} \textbf{end} \\
		9: \hspace{2em} Average across folds: $l(\tau(g)) = \frac{1}{5} \sum_{j=1}^5 l_j(\tau(g)) $ \\
		10: \textbf{end} \\
		11: \textbf{Return} $\tau(g^*)$ where $g^* = \arg \underset{g \in \mathcal{G}}{\min} \left\{ l(\tau(g)) \right\}$ \\ \bottomrule \bottomrule
	\end{tabular}
\end{center}
Another possible route involves using global optimization heuristics. One example are \textbf{genetic algorithms}. Here, again, we seek to minimize 5-fold squared cross-validation loss as a \textit{proxy} for expected out-of-sample square loss. The algorithm is initialized with a random set (\textit{population}) of admissible candidate hyperparameter vectors (\textit{individuals}). A small fraction of the top performing individuals are kept for the next step (\textit{generation}), while the rest are randomly perturbated. Usually, this is performed by drawing random combinations of many individuals (\textit{parents} having \textit{children}) and by simply randomly perturbating others ({mutation}). A stylized version of such an algorithm is shown here.

\begin{center}
	\begin{tabular}{l}
		\toprule \toprule
		Genetic Algorithm \\ \midrule
		1: Randomly assign observations to 5 folds $(Z^{(j)}, y^{(j)})$. Save sizes $\#j$. \\
		2: Define ranges for hyperparameters: $\tau^r := \{ [\tau_{i,\min}, \tau_{i,\max}] : \tau = \left( \tau_n \right)_{i=1}^n \}$ \\
		4: Randomly draw $P$ vectors $\tau^{(p,0)} \in \tau^r$ \\
		5: \textbf{For} each generation $g$ \textbf{in} $0:G$ \textbf{do}: \\
		6: \hspace{2em} \textbf{For} $p$ \textbf{in} $1:P$ \textbf{do}: \\
		7: \hspace{4em} \textbf{For} $j$ \textbf{in} 1:5 \textbf{do}: \\
		8: \hspace{6em} Train model on $\{ (Z^{(i)}, y^{(i)}) : \forall i \neq j \}$ using $\tau^{(p,g)}$ \\
		9: \hspace{6em} Compute prediction: $\hat{y}^{i,j}(\tau^{(p,g)})$ \\
		10: \hspace{6em} Compute loss: $l_j(\tau^{(p,g)}) : = (y^{(j)} - \hat{y}^{i,j}(\tau^{(p,g)}))'(y^{(j)} - \hat{y}^{i,j}(\tau^{(p,g)}))/ \#j$ \\
		11: \hspace{4em} \textbf{end} \\
		12: \hspace{2em} Average across folds: $l(\tau^{(p,g)}) = \frac{1}{5} \sum_{j=1}^5 l_j(\tau^{(p,g)}) $ \\
		13: \hspace{2em} \textbf{end} \\
		14: Define best $a$\% as $\tau^{g,a}$, rest as $\tilde{\tau}^{g,a}$ \\
		15: Randomnly perturbate the rest $f(\tilde{\tau}^{g,a})$ such that ranges are respected \\
		16: Define $\tau^{g+1} = \{ \tau^{g,a}, f(\tilde{\tau}^{g,a}) \}$ \\
		17: \textbf{end} \\
		18: \textbf{Return} $\tau^* = \arg \min \left\{ l(\tau^{p,G}) \right\}_{p=1}^P$
		\\ \bottomrule \bottomrule
	\end{tabular}
\end{center}
\begin{footnotesize}
	\flushleft
	\singlespacing
	\vspace{-1.8em}
	{
	Note: Details on the nature of random perturbations are available at \url{https://www.mathworks.com/help/gads/how-the-genetic-algorithm-works.html}. We use $G=25$ generations of $P=25$ individuals and keep $5\%$ of elite individuals each generation. The rest are default MATLAB values. \par}
\end{footnotesize}

\vspace{1em}
The last hyperparameter optimization consider is \textbf{Bayesian optimization}. The issue with optimizing hyperparameters is that the function we optimize (5-fold cross-validation average square loss is a function of an hyperparameter vector) is very costly to evaluate. The idea behind bayesian optimization is to work on and update a \textit{surrogate} function which is less costly to evaluate.

The algorithm is initialized by randomnly sampling pairs of cross-validation loss and admissible hyperparameters. The default choice in MATLAB then trains a Gaussian Process using these pairs as the training sample. An \textit{acquisition} function is defined on (1) the current sample, (2) a new set of random draws and (3) the fitted surrogate function. It returns \textit{scores} which evaluate how "promising" each new draw in the new set as a candidate solution. Optimizing the acquisition function returns the most promising point given current information.

Then, we recursively pull new random pairs, optimize the acquisition function, append the selected pair to the current sample of pairs, re-train the surrogate and continue until some criteria are met. We have an example of such an algorithm below.

\begin{center}
	\begin{tabular}{l}
		\toprule \toprule 
		Bayesian Optimization Algorithm \\ \midrule
		1: Randomly assign observations to 5 folds $(Z^{(j)}, y^{(j)})$. Save sizes $\#j$. \\
		2: Define ranges for hyperparameters: $\tau^r := \{ [\tau_{i,\min}, \tau_{i,\max}] : \tau = \left( \tau_n \right)_{i=1}^n \}$ \\
		3: Define model $f(Z_t, \tau)$ \\
		4: Define loss $L(y_{t+h}, Z_t; f, \tau) = \sum_{t=1}^T(y_{t+h} - f(Z_t, \tau))^2/T$ \\
		5: Define $j$ fold-trained model $f^{j}(Z_t, \tau)$ \\
		6: Define validation loss $\tilde{L} = \sum_{k=1}^5 L(y_{t+h}^{(k\neq j)}, Z_t^{(k\neq j)}; f^{(j)}, \tau)/5$ where $\tau \in \tau^r$ \\
		7: Define drawing function $D^{(l)} = \{ ( \tilde{L}^{(i)}, \tau^{(i)}) \}_{i=1}^N$ where $\tau^{(i)} \in \tau^r$ \\
		8: Define surrogate model $g(D^{(l)})$ \\
		9: Define acquisition function $s^{} = a(D^{(l)}, \tilde{D}^{(l)}, g)$ (returns scores for draw $\tilde{D}$) \\
		10: Random draws $D^{(1)}$ \\
		11: \textbf{For} $l$ \textbf{in} $1:l_{\max}$ \textbf{do}: \\
		12: \hspace{2em} Random draws $\tilde{D}^{(l)}$ \\
		13: \hspace{2em} Train surrogate $g(D^{(l)})$ \\
		14: \hspace{2em} New point $D^* = (L^{s^*}, \tau^{s^*})$ such that $s^* = \arg \max a(D^{(l)}, \tilde{D}^{(l)}, g)$ \\
		15: \hspace{2em} Append data $D^{(l+1)} = \{ D^{(l)}, D^* \}$ \\
		16: \textbf{end} \\
		17: \textbf{Return} approximate optimal parameter vector $\tau^{s^*}$
		\\ \bottomrule \bottomrule
	\end{tabular}
\end{center}
\begin{footnotesize}
	\flushleft 
	\singlespacing
	\vspace{-1em}
	{
	Note: In MATLAB, the default surrogate model is a Gaussian Process. The acquisition sampling method and other options are all set to default values. \par}
\end{footnotesize}

\vskip 0.2cm

{\sc \noindent \textbf{Linear Models}.} We consider the autoregressive model (AR), as well as the factor model of \cite{stock2002forecasting, stock2002macroeconomic}. Let $Z_t := \left[ y_t, ..., L^{P_y} y_t, F_t, ..., L^{P_f} F_t  \right]$ be our feature matrix, then the factor model is given by \vspace{-1em}
\begin{align}
y_{t+h} = \beta Z_t + \epsilon_{t+h}
\end{align}
where aforementioned factors are extracted by principal components from $X_t$ and parameters are estimated by OLS. The AR model is obtained by imposing $\beta_{k} = 0$ for all $k's$ tied to latent factors and their lagged values.

\vskip 0.2cm

{\sc \noindent \textbf{Elastic Net and Adaptive Lasso}.} The Elastic Net algorithm forecast the target variable $y_{t+h}$ using a linear combination of the $K$ features contained in $Z_t$ whose weights $\beta := (\beta_k)_{k=1}^K$ solve the following penalized regression problem
\begin{equation}
\hat{\beta} := \text{arg} \underset{\beta}{\min} \sum_{t=1}^T \left( y_{t+h} - Z_t \beta \right)^2 + \lambda \sum_{k=1}^K \left( \alpha \hat{w}_k |\beta_k| + (1-\alpha) \beta_k^2 \right)
\end{equation}
and where $(\alpha, \lambda)$ are hyperparameters and $\hat{w}$ is a weight vector. The Ridge estimator obtains with $\alpha = 0$, while LASSO is the case where $\alpha = 1$ and $\hat{w}_k = 1$ for all $k \in \{1,...,K\}$. The Adaptive Lasso of \cite{Zou2006} uses $\hat{w} = 1/|\hat{\beta}^\gamma|$ where $\hat{\beta}$ is a $\sqrt{T}$-consistent estimator for the above regression such as the OLS estimator (or the Ridge estimator as suggested by \cite{Zou2006} when collinearity is an issue).Theoretical restrictions on $\gamma$ for which consistent variable selection is justified can be found in \cite{Zou2006}. We make the common choice of $\gamma = 1$ and use a first step ridge estimator with hyperparameter selection performed by a genetic algorithm. The algorithms we used for Adaptative LASSO and Elastic Net are provided below.

\begin{center}
	\begin{tabular}{l}
		\toprule \toprule
		Adaptative LASSO Algorithm \\ \midrule
		1: Set $\alpha =0,  \hat{w} = 1$ \\
		2: Using $(Z,y)$ identify $\lambda_{max} = \min \{ \lambda: \hat{\beta}_{k,LASSO} = 0, \forall k > 1 \}$ \\
		3: \textbf{Ridge regression step:} \\
		4: \hspace{2em} Define range $[0, \tilde{\lambda}]$ \\
		5: \hspace{2em} Apply GA search to get $\lambda^R \in [0,\tilde{\lambda}]$ \\
		6: \hspace{2em} Using $(Z,y)$, estimate $\hat{\beta}_R(\lambda_R)$ by ridge regression \\
		7: Set penalty weights $\hat{w}_k = 1/|\hat{\beta}_{R,k}(\lambda_R)|$, $\alpha = 1$ \\
		8: \textbf{LASSO step:} \\
		9: \hspace{2em} Define 100 equally-log-spaced points in $[0, \lambda_{max}]$ \\
		10: \hspace{2em} Apply Grid Search to get $\lambda_L \in [0,\lambda_{max}]$ \\
		11: Using $(Z,y)$, estimate $\hat{\beta}_{LASSO}(\lambda_L)$ \\
		12: \textbf{Return} prediction $\hat{\beta}_{LASSO}(\lambda_L)' Z_T$ \\ \midrule \midrule
		Elastic Net \\ \midrule
		1: Using $(Z,y), \alpha=1$ identify $\lambda_{max} = \max \{ \lambda: \exists k > 1 s.t. \hat{\beta}_{1,LASSO}, \hat{\beta}_{k,LASSO} \neq 0 \}$ \\
		2: Define 100 equally-log-spaced points in $[0, \lambda_{max}]$ \\
		3: Define 100 equally spaced points in $[0.01, 1]$ \\
		4: Apply Grid Search to get $(\alpha^*, \lambda^*)$ \\
		5: Using $(Z,y)$, estimate $\hat{\beta}(\alpha^*, \lambda^2)$ \\
		6: \textbf{Return} prediction $\hat{\beta}(\alpha^*, \lambda^2)'Z_T$ \\ \bottomrule \bottomrule
	\end{tabular}
\end{center}
\begin{footnotesize}
	\flushleft
	\singlespacing
	\vspace{-1.8em}
	{
	Note: We use $\tilde{\lambda} = \infty$ for the Ridge regression step. The first coefficient ($k=1$) is the constant. $\lambda_{max}$ is the largest penalty which leaves at least one variable and the constant in the model. \par}
\end{footnotesize}

\vspace{1em}
Note that in both cases, we decided to take advantage of the variable selection ability of the LASSO penalty and of the shrinkage ability of the Ridge penalty and opted not to cross-validate the autoregressive lag order $P_y$, the factor lag order $P_f$ and the number of factors $k$. We imposed $(P_y,P_f,k) = (12,12,8)$ where relevant.

\vskip 0.2cm

\vskip 0.2cm

{\sc \noindent \textbf{Random Forests}.} This algorithm provides a means of approximating nonlinear functions by combining regression trees. Each regression tree partitions the feature space defined by $Z_t$ into distinct regions and, in its simplest form, uses the region-specific mean of the target variable $y_{t+h}$ as the forecast, i.e. for $M$ leaf nodes
\begin{align}
\hat{y}_{t+h} = \sum_{m=1}^M c_m I_{(Z_t \in R_m)}
\end{align}
where $R_1,...,R_M$ is a partition of the feature space. To circumvent some of the limitations of regression trees, \cite{Breiman2001} introduced Random Forests. Random Forests consist in growing many trees on subsamples (or nonparametric bootstrap samples) of observations. A random subset of features is eligible for the splitting variable, further decorrelating them. The final forecast is obtained by averaging over the forecasts of all trees. The algorithm is showcased below.
\begin{center}
	\begin{tabular}{l}
		\toprule \toprule
		Random Forest Algorithm \\ \midrule
		1: \textbf{For} b \textbf{in} 1:200 \textbf{do} \\
		2: \hspace{2em} Randomly draw a subsample $(Z^{(b)}, y^{(b)})$ of size $N$ from training sample $(Z,y)$ \\
		3: \hspace{2em} Grow regression tree $T_b$ on  $(Z^{(b)}, y^{(b)})$: \\
		4: \hspace{4em} \textbf{While} terminal node size $n > 5$ \textbf{do}: \\
		5: \hspace{6em} Randomly draw $\#Z/3$ regressors from $Z^{(b)}$ \\
		6: \hspace{6em} Find the variable-threshold pair minimizing MSE in daughter regions \\
		7: \hspace{6em} (Prediction in daughter regions is mean of $y^{(b)}$ in said region) \\
		8: \hspace{6em} Repeat for each terminal node and split accordingly \\
		9: \hspace{4em} \textbf{end} \\
		10: Compute $c_m^{(b)} = avg\{ y^{(b)} | Z_t^{(b)} \in R_m^{(b)} \}$ where $R_m^{(b)}$ is region $m$ of tree $T_b$ \\
		10: \textbf{end} \\
		11: Define prediction of Tree $T_b$ with $M$ nodes: $f(Z_t, T_b) = \sum_{m=1}^M c_m^{(b)} I(Z_t \in R_m)$ \\
		12: \textbf{Return} prediction $\frac{1}{200}\sum_{b=1}^{200} f(Z_t, T_b)$ \\ \bottomrule \bottomrule
	\end{tabular}
\end{center}
\begin{footnotesize}
	\flushleft
	\singlespacing
	\vspace{-1.8em}
	{
	Note: We do not cross-validate any hyperparameters for Random Forests. Bootstrap sample size $N$ is the default MATLAB value. \par}
\end{footnotesize}

\vspace{1em}
Given that we imposed $(P_y,P_f,k) = (12,12,8)$ where relevant here as well, the attentive reader will note that no hyperparameters optimization has been performed with Random Forests. By averaging over "randomized trees," we cannot induce overfit by using "too many" of them.

\vskip 0.2 cm

{\sc \noindent \textbf{Boosted Trees}.} This algorithm provides an alternative means of approximating nonlinear functions by additively combining regression trees in a sequential fashion. Let $\eta \in [0,1]$ be the learning rate and $\hat{y}_{t+h}^{(n)}$ and $e_{t+h}^{(n)} := y_{t+h} - \eta \hat{y}_{t+h}^{(n)}$ be the step $n$ predicted value and pseudo-residuals, respectively. Then, for square loss, the step $n+1$ prediction is obtained as \vspace{-1em}
\begin{align}
\hat{y}_{t+h}^{(n+1)} = \hat{y}_{t+h}^{(n)} + f(Z_t, c_{n+1} )
\end{align}
where $c_{n+1} := \text{arg}\underset{c}{\min} \sum_{t=1}^T \left( e_{t+h}^{(n)} - f(Z_t, c_{n+1}) \right)^2$ and $c_{n+1} := \left( c_{n+1,m} \right)_{m=1}^M$ are the parameters of a regression tree. In other words, it recursively fits trees on pseudo-residuals. We select the number of steps and $\eta \in [0,1]$ with Bayesian optimization. We imposed $(P_y,P_f,k) = (12,12,8)$ where relevant here as well. The algorithm is provided here.

\begin{center}
	\begin{tabular}{l}
		\toprule \toprule
		Boosted Trees Algorithm \\ \midrule
		1: Set learning rate $\eta \in (0,1)$ and maximal step $N$ \\
		2: Define $ f^{(1)}(Z_t) = \bar{y}_{t+h}$\\
		3: \textbf{For} $n$ \textbf{in} 1:N \textbf{do}: \\
		4: \hspace{2em} Define pseudo-residuals $e_{t+h}^{(n)} = y_{t+h} - \eta f^{(n)}$ \\
		5: \hspace{2em} Optimize $\theta_{n+1} := \arg \underset{\theta}{\min} \sum_{t=1}^T \left( e_{t+h}^{(n)} - T(Z_t, \theta_{n+1}) \right)^2$ \\
		6: \hspace{2em} where $\theta_{n+1} := \left(R_{m,n+1}, c_{m,n+1} \right)_{m=1}^M$ are the parameters of a regression trees \\
		7: \hspace{2em} Set $f^{(n+1)}(Z_t) = f^{(n)}(Z_t) + T(Z_t, \theta_{n+1})$ \\
		8: \textbf{end} \\
		9: \textbf{Return} prediction $f^{(N+1)}(Z_T)$
		\\ \bottomrule \bottomrule
	\end{tabular}
\end{center}
\begin{footnotesize}
	\flushleft
	\singlespacing
	\vspace{-1.8em}
	{Note: We consider individual trees with (1) a maximal depth of 5 splits and (2) where $\#Z_t/3$ features are available at each split in individual trees. The learning rate $\eta \in (0,1)$ and the number of boosting steps $N \in \{1, \dots, 500\}$ are obtained by Bayesian optimization with 5 fold cross-validation using MATLAB default values. \par}
\end{footnotesize}

\vspace{1em}

{\sc \noindent \textbf{Component-wise $L_2$ boosting}.} 
Linear boosting algorithms are convenient methods to fit models when the number of potential predictors is large. Many linear models are estimated and combined iteratively using a single regressor at a time chosen so that it reduces the most the loss considered. We specifically follow \cite{BaiNg2009} and consider all features in $Z_t$ as separate predictors. The algorithm is provided here.

\begin{center}
\begin{tabular}{l}
		\toprule \toprule
		Component-wise $L_2$ Boosting Algorithm \\ \midrule 
		1: Define $f^{(1)}(Z_t) = \bar{y}_{t+h}$ \\
		2: \textbf{For} $n$ \textbf{in} $1:N$ \textbf{do}: \\
		3: \hspace{2em} Define $e_{t+h}^{(n)} = y_{t+h} - f^{(n)}(Z_t) $\\
		4: \hspace{2em} Randomnly draw $m$ features from $Z_t$ \\
		5: \hspace{2em} OLS regressions $e_{t+h}^{(n)} = \beta^{(i)} Z_{i,t} + \nu_{i,t+h}$ and residuals $\hat{\nu}_{i,t+h} = e_{t+h} - \hat{\beta}^{(i)} Z_{i,t}$ \\
		6: \hspace{2em} Compute $SSR_i : \hat{\nu}_i'\hat{\nu}_i$ \\
		7: \hspace{2em} Select $i^* := \arg \min SSR_i$ \\
		8: \hspace{1.5em} Define $g(Z_t) = Z_{i^*} \hat{\beta}^{i^*}$ \\
		9: \hspace{1.5em} Update $f^{(n+1)}(Z_t) = f^{(n)}(Z_t) + \eta g(Z_t)$\\
		10: \textbf{end} \\
		11: \textbf{Return} prediction $f^{(N+1)}(Z_t)$
		\\ \bottomrule \bottomrule
	\end{tabular}
\end{center}
\begin{footnotesize}
	\flushleft
	\singlespacing
	\vspace{-1.8em}
	{
	Note: $m=\min(200,\frac{\# Z_t}{3})$. $N \in \{1,2,...,500\}$ and $\eta \in [0,1]$ are selected with a genetic algorithm of 25 generations of 25 individuals. All other values are default values for MATLAB. \par}
\end{footnotesize}

\clearpage
\onehalfspace
\section{ONLINE APPENDIX - Detailed Relative RMSE Results}\label{sec:rmse_results}

\subsection{Average growth targets $\hat{y}_{t+h}^{\text{direct}}$}

\onehalfspace
\begin{tiny}
\begin{ThreePartTable}
\begin{longtable}{l|cccccccccc}
\caption{Relative RMSE - H=1} \\ 
\hline
 & INDPRO & EMP & UNRATE & INCOME & CONS & RETAIL & HOUST & M2 & CPI & PPI \\ 
\hline
\endhead
\hline
\endfoot
\hline
\multicolumn{ 1 }{l|}{\textbf{ Benchmarks }} \\
FM (RMSE) & 0.006 & 0.001 & 0.148 & \textbf{0.007} & \textbf{\underline{0.004}} & \textbf{\underline{0.011}} & \textbf{0.072} & \textbf{0.003} & \textbf{0.002} & 0.006 \\ 
  AR & 1.06* & 1.03 & 1.04* & 1.06** & \textbf{1.03}* & \textbf{1.02} & \textbf{1.01} & \textbf{1.01} & \textbf{1.03} & 1.01 \\ 
\hline
\multicolumn{ 1 }{l|}{\textbf{ Adaptive Lasso }} \\
  F & \textbf{0.96}*** & \textbf{0.97}** & 0.97** & \textbf{1.00} & \textbf{1.03}* & \textbf{1.04}* & \textbf{1.02} & \textbf{0.98}* & \textbf{0.98} & 0.98 \\ 
  F-X & \textbf{0.95}*** & 1.03 & \textbf{0.96}** & 1.01 & \textbf{1.08}* & \textbf{1.09}** & \textbf{1.02} & \textbf{0.99} & 1.06 & 1.00 \\ 
  F-MARX & \textbf{0.95}** & 0.99 & \textbf{0.95}** & \textbf{1.00} & \textbf{1.04}* & \textbf{1.02} & \textbf{1.01} & \textbf{0.99} & \textbf{0.96} & \textbf{0.93}*** \\ 
  F-MAF & \textbf{0.94}*** & 0.99 & \textbf{0.95}*** & 1.01 & \textbf{1.04}* & \textbf{1.05}* & \textbf{1.02}* & \textbf{1.00} & 1.05 & 1.02 \\
\end{longtable}
\end{ThreePartTable}
\end{tiny}

\clearpage

\begin{tiny}
\begin{ThreePartTable}
\begin{longtable}{l|cccccccccc}
\caption{Relative RMSE - H=3} \\ 
\hline
 & INDPRO & EMP & UNRATE & INCOME & CONS & RETAIL & HOUST & M2 & CPI & PPI \\ 
\hline
\endhead
\hline
\endfoot
\hline
\multicolumn{ 1 }{l|}{\textbf{ Benchmarks }} \\
FM (RMSE) & 0.004 & 0.001 & 0.088 & \textbf{0.003} & \textbf{0.002} & \textbf{0.005} & 0.033 & \textbf{0.003} & \textbf{0.002} & \textbf{0.004} \\ 
  AR & 1.08 & 1.03 & 1.04 & 1.09** & \textbf{1.06}* & \textbf{1.01} & \textbf{0.96} & \textbf{1.00} & \textbf{1.01} & \textbf{1.00} \\ 
\hline
\multicolumn{ 1 }{l|}{\textbf{ Adaptive Lasso }} \\
  F & \textbf{0.95}** & \textbf{0.91}*** & 0.94* & \textbf{0.98} & \textbf{0.99} & 1.07 & 1.05 & \textbf{0.98} & \textbf{1.01} & \textbf{0.99} \\ 
  F-X & \textbf{0.99} & 0.98 & 0.95* & 1.01 & \textbf{1.03} & \textbf{1.00} & \textbf{0.96} & \textbf{1.04} & 1.07** & \textbf{0.99} \\ 
  F-MARX & 1.06 & 1.02 & 0.89*** & 1.09 & \textbf{1.06} & \textbf{1.05} & \textbf{0.97} & \textbf{0.99} & 1.07* & \textbf{0.98} \\ 
  F-MAF & 1.10* & 1.03 & 0.90** & 1.01 & \textbf{1.04} & \textbf{1.03} & \textbf{0.97} & \textbf{1.03} & 1.12** & 1.03 \\
\end{longtable}
\end{ThreePartTable}
\end{tiny}

\clearpage

\begin{tiny}
\begin{ThreePartTable}
\begin{longtable}{l|cccccccccc}
\caption{Relative RMSE - H=6} \\ 
\hline
 & INDPRO & EMP & UNRATE & INCOME & CONS & RETAIL & HOUST & M2 & CPI & PPI \\ 
\hline
\endhead
\hline
\endfoot
\hline
\multicolumn{ 1 }{l|}{\textbf{ Benchmarks }} \\
FM (RMSE) & 0.004 & \textbf{0.001} & \textbf{0.077} & \textbf{0.002} & \textbf{0.002} & \textbf{0.004} & \textbf{0.024} & \textbf{0.002} & 0.002 & \textbf{0.004} \\ 
  AR & 1.03 & 1.07 & 1.09* & 1.04 & \textbf{0.92}* & \textbf{0.98} & \textbf{0.94} & \textbf{0.94} & \textbf{0.96} & \textbf{0.95} \\ 
\hline
\multicolumn{ 1 }{l|}{\textbf{ Adaptive Lasso }} \\
  F & \textbf{0.94}** & \textbf{0.93}** & \textbf{0.95} & \textbf{0.96}** & \textbf{0.97}* & \textbf{1.05} & \textbf{1.03} & \textbf{0.96} & 0.99 & \textbf{1.00} \\ 
  F-X & 0.96* & \textbf{0.97} & \textbf{0.99} & \textbf{1.01} & \textbf{0.99} & \textbf{0.95} & \textbf{0.91} & \textbf{0.96} & 1.00 & \textbf{0.99} \\ 
  F-MARX & 1.01 & \textbf{1.04} & \textbf{0.94} & 1.05 & \textbf{1.00} & \textbf{\underline{0.92}}* & \textbf{1.05} & \textbf{1.02} & 1.09* & 1.11* \\ 
  F-MAF & 1.17** & 1.16** & \textbf{0.94} & 1.09* & \textbf{1.04} & \textbf{1.09} & 1.48*** & 1.06 & 1.12** & 1.16*** \\
\end{longtable}
\end{ThreePartTable}
\end{tiny}

\clearpage

\begin{tiny}
\begin{ThreePartTable}
\begin{longtable}{l|cccccccccc}
\caption{Relative RMSE - H=9} \\ 
\hline
 & INDPRO & EMP & UNRATE & INCOME & CONS & RETAIL & HOUST & M2 & CPI & PPI \\ 
\hline
\endhead
\hline
\endfoot
\hline
\multicolumn{ 1 }{l|}{\textbf{ Benchmarks }} \\
FM (RMSE) & \textbf{0.004} & \textbf{0.001} & \textbf{0.076} & \textbf{0.002} & 0.002 & 0.004 & \textbf{0.021} & 0.002 & \textbf{0.002} & 0.003 \\ 
  AR & \textbf{1.01} & \textbf{1.07} & \textbf{1.11}** & \textbf{1.02} & \textbf{0.90}* & 0.96 & \textbf{0.92} & 0.92* & \textbf{1.02} & \textbf{0.94} \\ 
\hline
\multicolumn{ 1 }{l|}{\textbf{ Adaptive Lasso }} \\
  F & \textbf{0.95} & \textbf{\underline{0.95}}* & \textbf{0.96} & \textbf{0.97}* & \textbf{0.96}** & 1.03 & 1.04 & 0.96 & \textbf{1.00} & 0.99 \\ 
  F-X & \textbf{0.97} & \textbf{1.00} & \textbf{1.04} & \textbf{1.02} & 1.01 & 0.92* & \textbf{1.01} & 0.95 & \textbf{1.02} & 1.02 \\ 
  F-MARX & \textbf{1.07} & 1.14* & \textbf{1.00} & 1.10* & 1.07 & 0.97 & 1.23*** & 1.03 & 1.05 & 1.07 \\ 
  F-MAF & 1.28*** & 1.27*** & 1.13** & 1.15** & 1.21*** & 1.16* & 1.44*** & 1.01 & 1.31** & 1.19** \\
\end{longtable}
\end{ThreePartTable}
\end{tiny}

\clearpage

\begin{tiny}
\begin{ThreePartTable}
\begin{longtable}{l|cccccccccc}
\caption{Relative RMSE - H=12} \\ 
\hline
 & INDPRO & EMP & UNRATE & INCOME & CONS & RETAIL & HOUST & M2 & CPI & PPI \\ 
\hline
\endhead
\hline
\endfoot
\hline
\multicolumn{ 1 }{l|}{\textbf{ Benchmarks }} \\
FM (RMSE) & \textbf{0.003} & \textbf{0.001} & \textbf{0.077} & \textbf{0.002} & 0.002 & 0.003 & \textbf{0.019} & 0.002 & \textbf{0.001} & \textbf{0.003} \\ 
  AR & \textbf{1.01} & \textbf{1.06} & 1.10** & \textbf{1.02} & \textbf{0.95} & 0.95* & \textbf{0.92} & 0.92* & \textbf{1.06} & \textbf{0.97} \\ 
\hline
\multicolumn{ 1 }{l|}{\textbf{ Adaptive Lasso }} \\
  F & \textbf{0.96}* & \textbf{0.96}* & \textbf{0.95} & \textbf{0.97} & \textbf{0.96}*** & 1.02 & \textbf{0.97} & 0.99 & \textbf{1.01} & 1.05* \\ 
  F-X & \textbf{0.99} & \textbf{1.01} & \textbf{1.03} & \textbf{1.05}* & 1.05 & \textbf{0.92} & \textbf{1.00} & 0.96 & \textbf{1.05}* & 1.06 \\ 
  F-MARX & 1.20*** & 1.07 & \textbf{1.02} & \textbf{1.05} & 1.14** & 1.14 & 1.15* & 1.01 & 1.12** & 1.13** \\ 
  F-MAF & 1.32*** & 1.19** & 1.41** & 1.21*** & 1.22*** & 1.16 & 1.40*** & 0.99 & 1.19** & 1.21** \\
\end{longtable}
\end{ThreePartTable}
\end{tiny}

\clearpage

\begin{tiny}
\begin{ThreePartTable}
\begin{longtable}{l|cccccccccc}
\caption{Relative RMSE - H=24} \\ 
\hline
 & INDPRO & EMP & UNRATE & INCOME & CONS & RETAIL & HOUST & M2 & CPI & PPI \\ 
\hline
\endhead
\hline
\endfoot
\hline
\multicolumn{ 1 }{l|}{\textbf{ Benchmarks }} \\
FM (RMSE) & 0.003 & 0.001 & 0.068 & \textbf{0.002} & 0.002 & 0.003 & 0.014 & 0.002 & 0.002 & 0.003 \\ 
  AR & \textbf{0.98} & 1.03 & 1.08* & \textbf{0.98} & \textbf{0.85}** & 0.93*** & 0.92 & \textbf{0.90}*** & 0.95 & \textbf{0.87} \\ 
\hline
\multicolumn{ 1 }{l|}{\textbf{ Adaptive Lasso }} \\
  F & \textbf{0.93}*** & \textbf{0.95}*** & \textbf{0.93}** & \textbf{0.98} & 0.91** & 1.00 & 0.94* & 0.99 & 1.21* & 1.06* \\ 
  F-X & 1.08** & \textbf{0.98} & 0.95*** & 1.11** & 1.11*** & \textbf{0.89} & 0.95 & 1.08 & 1.07 & \textbf{0.88} \\ 
  F-MARX & 1.22** & 1.13** & 1.17*** & 1.12* & 1.08 & 1.06 & 1.17** & 1.00 & 1.05 & 1.01 \\ 
  F-MAF & 1.33*** & 1.19*** & 1.03 & 1.31*** & 1.23*** & 0.98 & 1.30*** & 1.03 & 1.20** & 0.98 \\
\end{longtable}
\end{ThreePartTable}
\end{tiny}

\clearpage
\onehalfspace

\subsection{Path Averages ($\hat{y}_{t+h}^{\text{path-avg}}$)}\label{sec:rmse_results_svgtoavg}



\begin{tiny}
\begin{ThreePartTable}
\begin{longtable}{l|cccccccccc}
\caption{Relative RMSE - H=1, SGR} \\ 
\hline
 & INDPRO & EMP & UNRATE & INCOME & CONS & RETAIL & HOUST & M2 & CPI & PPI \\ 
\hline
\endhead
\hline
\endfoot
\hline
\multicolumn{ 1 }{l|}{\textbf{ Benchmarks }} \\
FM-AGR (RMSE) & 0.006 & 0.001 & 0.148 & 0.007 & \textbf{\underline{0.004}} & \textbf{\underline{0.011}} & \textbf{0.072} & \textbf{0.003} & \textbf{0.002} & 0.006 \\ 
  FM-SGR & 1.06* & 1.03 & 1.04* & 1.06** & \textbf{1.03}* & \textbf{1.02} & \textbf{1.01} & 1.01 & \textbf{1.03} & 1.01 \\ 
  AR & 1.00 & 1.00 & 1.00 & 1.00 & \textbf{1.00} & \textbf{1.00}* & \textbf{1.00} & \textbf{1.00} & \textbf{1.00} & 1.00 \\ 
\hline
\multicolumn{ 1 }{l|}{\textbf{ Adaptive Lasso }} \\
  F & \textbf{0.96}*** & \textbf{0.97}** & 0.97** & 1.00 & \textbf{1.03}* & \textbf{1.04}* & \textbf{1.02} & \textbf{0.98}* & \textbf{0.98} & \textbf{0.98} \\ 
  F-X & \textbf{0.95}*** & 1.03 & \textbf{0.96}** & 1.01 & 1.08* & \textbf{1.09}** & \textbf{1.02} & \textbf{0.99} & 1.06 & 1.00 \\ 
  F-MARX & \textbf{0.95}** & 0.99 & \textbf{0.95}** & 1.00 & \textbf{1.04}* & \textbf{1.02} & \textbf{1.01} & \textbf{0.99} & \textbf{0.96} & \textbf{0.93}*** \\
\end{longtable}
\end{ThreePartTable}
\end{tiny}

\clearpage

\begin{tiny}
\begin{ThreePartTable}
\begin{longtable}{l|cccccccccc}
\caption{Relative RMSE - H=3, SGR} \\ 
\hline
 & INDPRO & EMP & UNRATE & INCOME & CONS & RETAIL & HOUST & M2 & CPI & PPI \\ 
\hline
\endhead
\hline
\endfoot
\hline
\multicolumn{ 1 }{l|}{\textbf{ Benchmarks }} \\
FM-AGR (RMSE) & 0.004 & 0.001 & 0.088 & \textbf{0.003} & \textbf{0.002} & \textbf{0.005} & 0.033 & 0.003 & \textbf{0.002} & 0.004 \\ 
  FM-SGR & 1.08 & 1.04 & 1.06 & 1.13*** & \textbf{1.07}* & \textbf{1.02} & 0.96 & 1.02 & \textbf{1.02} & 1.00 \\ 
  AR & 0.97** & 0.96** & 0.96* & \textbf{1.00} & \textbf{1.03}** & \textbf{0.98}* & 0.98 & 1.01 & \textbf{1.02} & 1.00 \\ 
\hline
\multicolumn{ 1 }{l|}{\textbf{ Adaptive Lasso }} \\
  F & \textbf{0.92}*** & 0.90*** & 0.92** & \textbf{1.00} & \textbf{1.04} & \textbf{1.03} & \textbf{0.92}* & \textbf{0.97}* & \textbf{0.98} & \textbf{0.98} \\ 
  F-X & 1.00 & 0.99 & 0.94* & 1.03 & 1.16* & 1.07 & \textbf{0.92}* & \textbf{0.95}*** & \textbf{1.00} & 1.00 \\ 
  F-MARX & \textbf{0.91}*** & 0.92** & \textbf{0.87}*** & \textbf{1.02} & \textbf{1.06}* & \textbf{1.01} & \textbf{0.91}** & \textbf{0.96}** & \textbf{0.99} & \textbf{0.94} \\
\end{longtable}
\end{ThreePartTable}
\end{tiny}

\clearpage

\begin{tiny}
\begin{ThreePartTable}
\begin{longtable}{l|cccccccccc}
\caption{Relative RMSE - H=6, SGR} \\ 
\hline
 & INDPRO & EMP & UNRATE & INCOME & CONS & RETAIL & HOUST & M2 & CPI & PPI \\ 
\hline
\endhead
\hline
\endfoot
\hline
\multicolumn{ 1 }{l|}{\textbf{ Benchmarks }} \\
FM-AGR (RMSE) & 0.004 & 0.001 & 0.077 & 0.002 & 0.002 & 0.004 & 0.024 & 0.002 & \textbf{0.002} & \textbf{0.004} \\ 
  FM-SGR & 1.00 & 1.00 & 1.05 & 1.14** & 1.04 & 1.01 & 0.99 & 0.98 & \textbf{0.96} & \textbf{0.96} \\ 
  AR & 0.93* & 0.93*** & 0.95 & 0.97* & 1.01 & \textbf{0.95}** & 1.02 & 0.99 & \textbf{0.97} & \textbf{0.97} \\ 
\hline
\multicolumn{ 1 }{l|}{\textbf{ Adaptive Lasso }} \\
  F & \textbf{0.86}*** & 0.87*** & 0.90*** & 0.95** & 1.00 & 1.03 & \textbf{0.90} & 0.95 & \textbf{0.93} & \textbf{0.94} \\ 
  F-X & 0.96 & 0.94* & 0.93* & 1.01 & 1.14 & 1.10 & \textbf{0.89} & \textbf{0.91}** & \textbf{0.93} & \textbf{0.97} \\ 
  F-MARX & \textbf{0.87}*** & 0.87*** & \textbf{0.84}*** & 0.96* & 1.03 & 1.00 & \textbf{0.90} & 0.95 & \textbf{0.96} & \textbf{0.92} \\
\end{longtable}
\end{ThreePartTable}
\end{tiny}

\clearpage

\begin{tiny}
\begin{ThreePartTable}
\begin{longtable}{l|cccccccccc}
\caption{Relative RMSE - H=9, SGR} \\ 
\hline
 & INDPRO & EMP & UNRATE & INCOME & CONS & RETAIL & HOUST & M2 & CPI & PPI \\ 
\hline
\endhead
\hline
\endfoot
\hline
\multicolumn{ 1 }{l|}{\textbf{ Benchmarks }} \\
FM-AGR (RMSE) & 0.004 & 0.001 & 0.076 & 0.002 & 0.002 & \textbf{0.004} & 0.021 & 0.002 & \textbf{0.002} & 0.003 \\ 
  FM-SGR & 0.99 & 0.96 & 1.05 & 1.12** & 1.03 & \textbf{1.02} & 1.01 & 0.98 & \textbf{1.01} & \textbf{0.95} \\ 
  AR & 0.95 & 0.89*** & 0.95 & 0.95* & 1.02 & \textbf{0.97} & 1.06 & 1.01 & \textbf{1.04} & \textbf{0.97} \\ 
\hline
\multicolumn{ 1 }{l|}{\textbf{ Adaptive Lasso }} \\
  F & \textbf{0.86}*** & 0.85*** & 0.89*** & 0.92** & \textbf{0.96}** & \textbf{1.04} & \textbf{0.90} & 0.94 & \textbf{0.99} & \textbf{0.95} \\ 
  F-X & 0.96 & 0.92* & 0.93** & 0.98 & 1.13 & \textbf{1.16} & \textbf{0.90} & \textbf{0.89}** & \textbf{0.96} & \textbf{0.95} \\ 
  F-MARX & \textbf{0.88}*** & 0.86** & \textbf{0.85}*** & 0.94** & 0.99 & \textbf{1.00} & 0.91 & 0.94 & \textbf{1.04} & \textbf{0.94} \\
\end{longtable}
\end{ThreePartTable}
\end{tiny}

\clearpage

\begin{tiny}
\begin{ThreePartTable}
\begin{longtable}{l|cccccccccc}
\caption{Relative RMSE - H=12, SGR} \\ 
\hline
 & INDPRO & EMP & UNRATE & INCOME & CONS & RETAIL & HOUST & M2 & CPI & PPI \\ 
\hline
\endhead
\hline
\endfoot
\hline
\multicolumn{ 1 }{l|}{\textbf{ Benchmarks }} \\
FM-AGR (RMSE) & 0.003 & 0.001 & 0.077 & 0.002 & 0.002 & \textbf{0.003} & 0.019 & 0.002 & \textbf{0.001} & \textbf{0.003} \\ 
  FM-SGR & 1.00 & 0.94 & 1.02 & 1.08** & 1.01 & \textbf{1.01} & 1.07 & 0.98 & \textbf{1.05} & \textbf{0.99} \\ 
  AR & 0.97 & 0.88** & 0.94 & 0.90*** & 1.00 & \textbf{0.98} & 1.15*** & 1.02 & \textbf{1.09} & \textbf{1.03} \\ 
\hline
\multicolumn{ 1 }{l|}{\textbf{ Adaptive Lasso }} \\
  F & \textbf{0.87}** & 0.84*** & \textbf{0.86}*** & 0.89*** & \textbf{0.91}** & \textbf{1.03} & 0.90 & 0.95 & \textbf{1.03} & \textbf{0.98} \\ 
  F-X & \textbf{0.96} & 0.89** & 0.90** & 0.94** & 1.04 & \textbf{1.17} & 0.92 & \textbf{0.89}** & \textbf{0.98} & \textbf{0.98} \\ 
  F-MARX & \textbf{0.90}** & 0.86** & \textbf{0.84}** & 0.90** & 0.96 & \textbf{0.99} & 0.93 & 0.94 & \textbf{1.08} & \textbf{1.02} \\
\end{longtable}
\end{ThreePartTable}
\end{tiny}

\clearpage

\begin{tiny}
\begin{ThreePartTable}
\begin{longtable}{l|cccccccccc}
\caption{Relative RMSE - H=24, SGR} \\ 
\hline
 & INDPRO & EMP & UNRATE & INCOME & CONS & RETAIL & HOUST & M2 & CPI & PPI \\ 
\hline
\endhead
\hline
\endfoot
\hline
\multicolumn{ 1 }{l|}{\textbf{ Benchmarks }} \\
FM-AGR (RMSE) & \textbf{0.003} & \textbf{0.001} & \textbf{0.068} & 0.002 & 0.002 & 0.003 & 0.014 & 0.002 & \textbf{0.002} & 0.003 \\ 
  FM-SGR & \textbf{1.06} & \textbf{0.95} & 1.09 & 1.05 & 0.89** & 1.02 & 1.18** & 0.93 & \textbf{1.00} & 0.94 \\ 
  AR & 1.15** & \textbf{0.94} & 1.13* & \textbf{0.96} & 0.97 & 1.05 & 1.48*** & 1.02 & 1.04 & 0.99 \\ 
\hline
\multicolumn{ 1 }{l|}{\textbf{ Adaptive Lasso }} \\
  F & \textbf{1.02} & \textbf{0.93} & \textbf{0.99} & \textbf{0.91}* & \textbf{0.82}*** & 1.06** & 1.01 & 0.95 & \textbf{0.99} & 0.96 \\ 
  F-X & \textbf{1.02} & \textbf{0.94} & 1.01 & \textbf{0.93} & 0.96 & 1.18 & 1.05 & 0.89* & \textbf{0.91} & 0.94 \\ 
  F-MARX & \textbf{1.07} & \textbf{0.95} & 1.03 & \textbf{0.92} & \textbf{0.84}*** & 1.02 & 1.09 & 0.96 & \textbf{0.98} & 1.01 \\
\end{longtable}
\end{ThreePartTable}
\end{tiny}

\section{ONLINE APPENDIX - Stability of Predictive Performance}\label{sec:stability}

\begin{figure}[H]
\caption{Giacomini-Rossi Fluctuation Test}
\label{fig:GRfluctuationtest}\centering
\caption*{Single to Average Growth Rate $\hat{y}_{t+h}^{\text{path-avg}}$}
\vspace{-.5em}
\begin{subfigure}[b]{0.49\textwidth}
	\includegraphics[width=\textwidth,height=0.45\textheight]{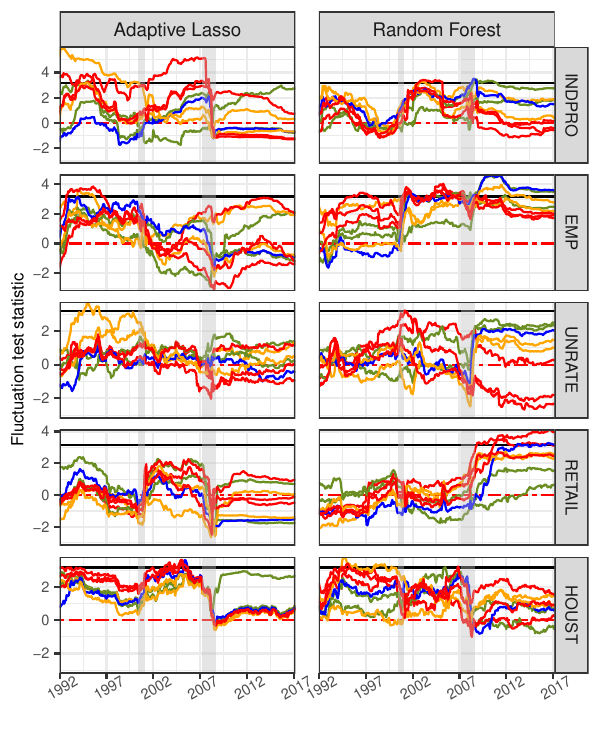}
\end{subfigure}
\begin{subfigure}[b]{0.49\textwidth}
	\includegraphics[width=\textwidth,height=0.45\textheight]{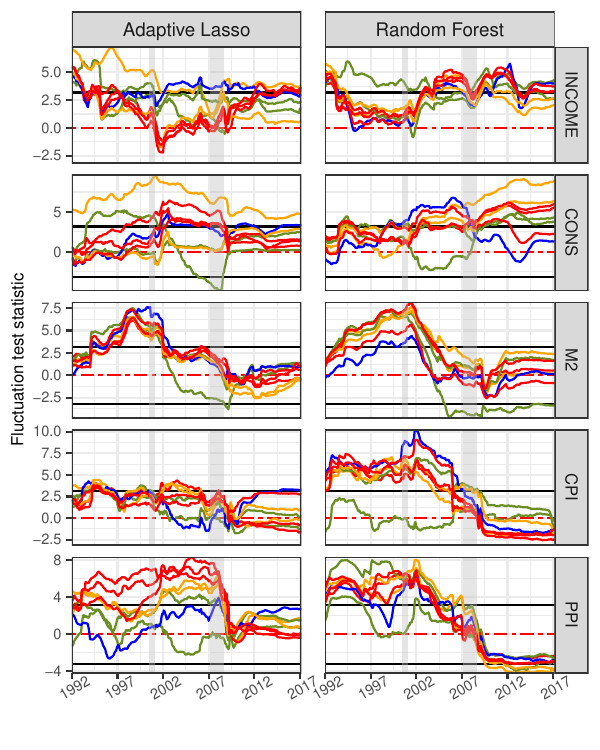}
\end{subfigure}
\vspace{-1.5em}
\caption*{Average Growth Rate $\hat{y}_{t+h}^{\text{direct}}$}
\vspace{-.5em}
\begin{subfigure}[b]{0.49\textwidth}
    \includegraphics[width=\textwidth,height=0.45\textheight]{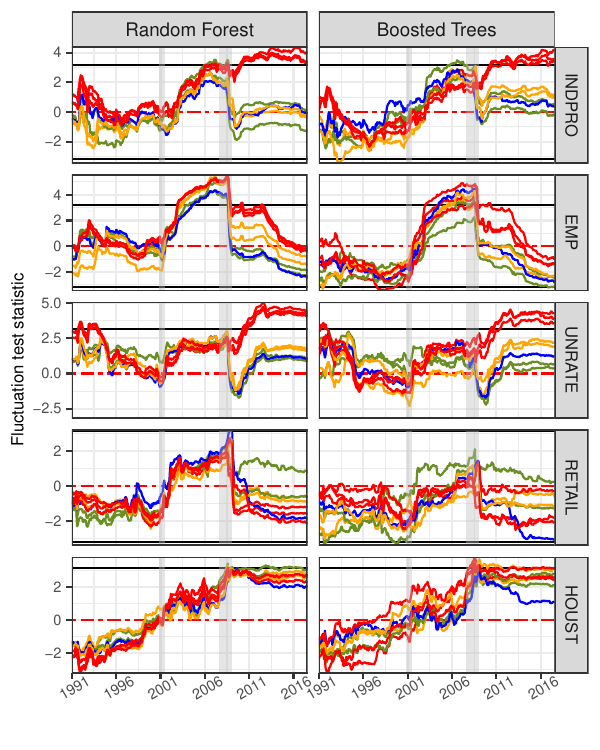}
\end{subfigure}
\begin{subfigure}[b]{0.49\textwidth}
    \includegraphics[width=\textwidth,height=0.45\textheight]{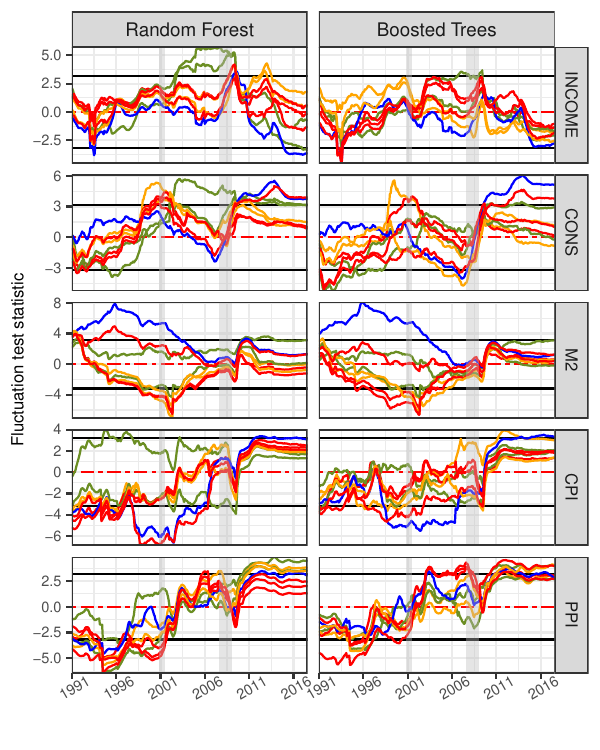}
\end{subfigure}
\flushleft
\begin{scriptsize}
\vspace{-3.5em}
	 {\singlespacing
	Note: The figure shows the Giacomini-Rossi fluctuation tests. The top panel uses the $\hat{y}_{t+h}^{\text{direct}}$ version of each model as benchmark while the bottom panel uses the factor model as a benchmark. The horizontal lines depict the 10\% critical values. A model is significantly better than the benchmark if the test statistic is above the upper critical value line. Colors represent selected data transformations included with each nonlinear forecasting model: {\color{ForestGreen}\textit{F,F-X}}, {\color{Red}\textit{F-MARX,F-X-MARX,F-X-MARX-Level}}, {\color{blue}\textit{F-X-Level}}, {\color{orange}\textit{F-MAF,F-X-MAF}}. \par}
\end{scriptsize}
\end{figure}

In order to examine the stability of forecast accuracy, we consider the fluctuation test of \cite{Giacomini-Rossi(2010)}. Figure \ref{fig:GRfluctuationtest} shows the results for a few selected cases. Following the simulation results in \cite{Giacomini-Rossi(2010)}, the moving average of the standardized difference of MSEs is produced with a 136-month window, which corresponds to 30\% of the out-of-sample size. 

The top panels compares the predictive performance of the path average versus direct approach, in combination with Adaptive Lasso and Random Forests models using different data transformation combinations. The bottom panels compare the performance of nonlinear methods using data transformations against the standard factor model. 

There is a fair amount of instability. The path average approach becomes preferable to the direct approach after 2007 when combined with Random Forest and for real activity variables. In the case of M2 growth and CPI and PPI inflation rates, combining $h$ simple growth rate problems does better during the first half of the pseudo-out-of-sample, but the situation completely inverses in the second part.  

When looking at the bottom panel, it is worth noting that in the case of INDPRO with RF, the data combinations including the \textit{MARX} transformation dominates the benchmark and the alternatives most of the time, but takes off even more significantly and substantially since the Great Recession. A similar pattern is observed with unemployment rate, while in the case of employment the improvements are not significant since 2010. 

\clearpage

\section{ONLINE APPENDIX - Additional Results on Marginal Contribution of Data Pre-processing}\label{sec:tornados_add}

\clearpage 

\begin{figure}[H]
\caption{Distribution of Average Marginal Treatment Effects of Factors in Levels}
	\label{fig:LevelFactor_AvgTreatmentEffect}
\begin{subfigure}[b]{0.99\textwidth}
	\centering
	\includegraphics[width=\textwidth,height=0.42\textheight]{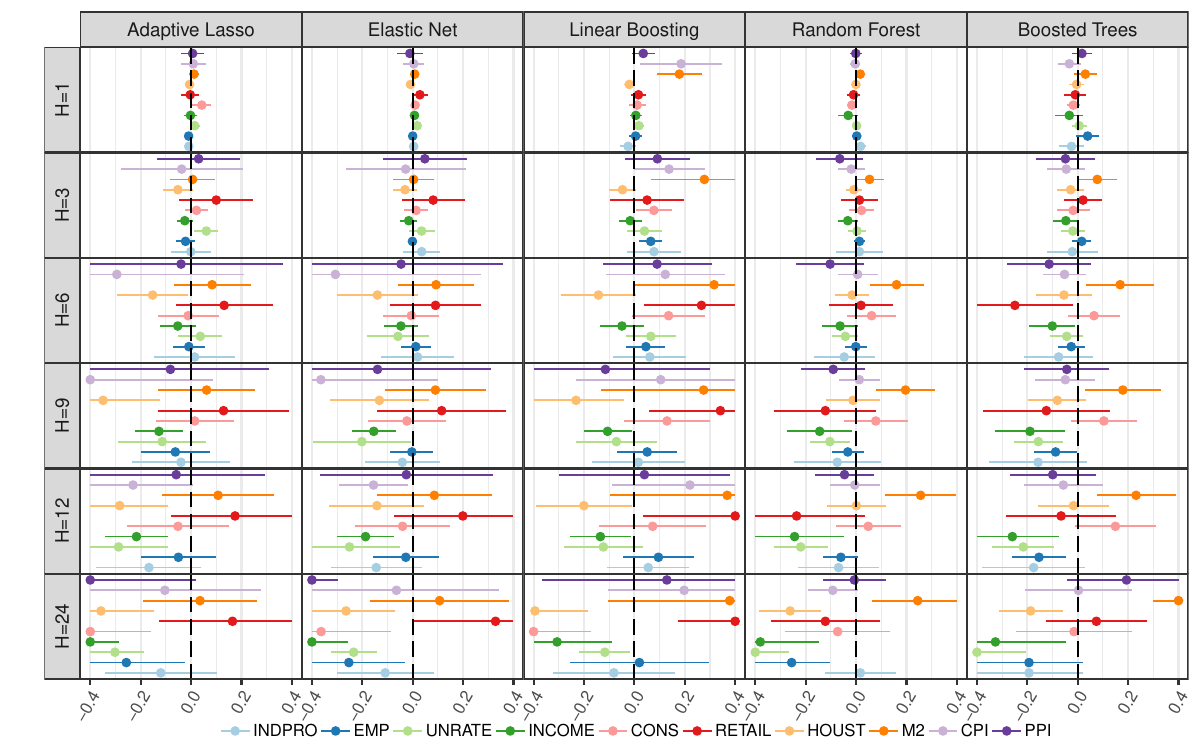}
	\caption{Direct Approach ($\hat{y}_{t+h}^{\text{direct}}$)}
	\label{fig:LevelFactor_AvgTreatmentEffect_direct}
\end{subfigure}
\\
\begin{subfigure}[b]{0.99\textwidth}
	\centering
	\includegraphics[width=\textwidth,height=0.42\textheight]{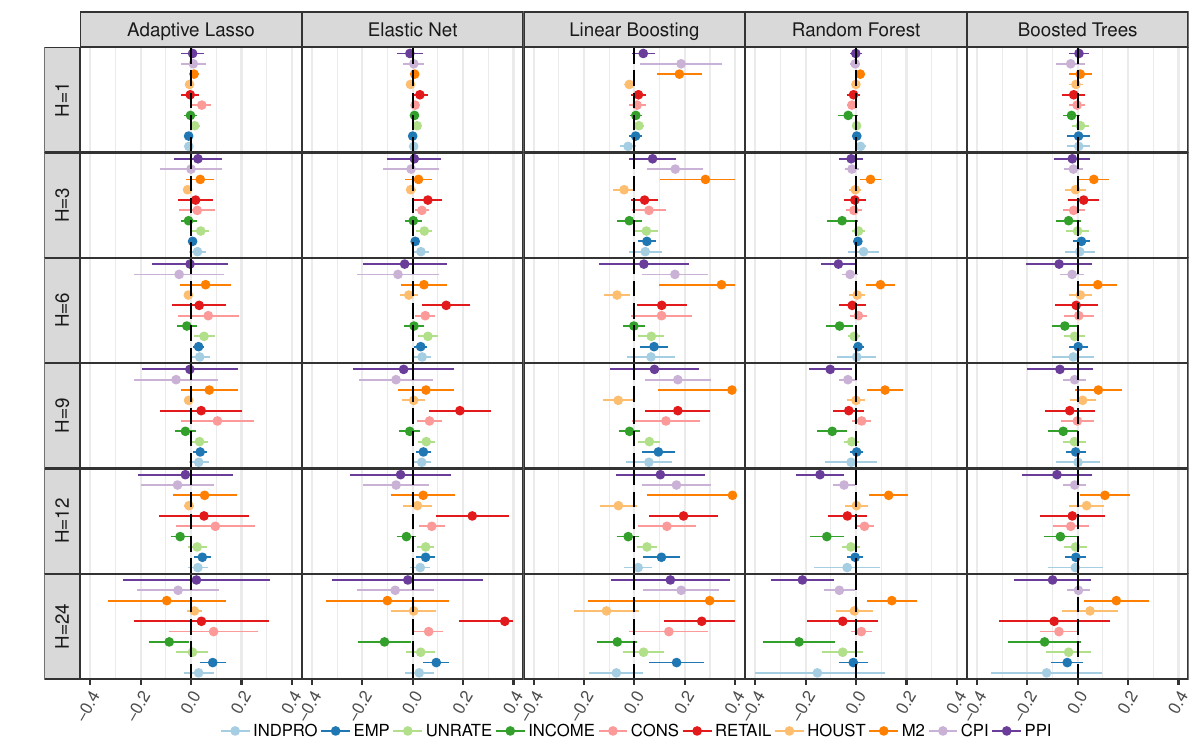}
	\caption{Path Average Approach ($\hat{y}_{t+h}^{\text{path-avg}}$)}
	\label{fig:LevelFactor_AvgTreatmentEffect_path}
\end{subfigure}
	\flushleft
	\begin{scriptsize}
		\vspace{-2.8em}
		{\singlespacing
			Note: This figure plots the distribution of ${\alpha}_f^{(h,v)}$ from equation (\ref{e_eq2}) done by $(h,v)$ subsets. It shows the average partial effect on the pseudo-$R^2$ from augmenting the model with factors in levels featuring, keeping everything else fixed. SEs are HAC. These are the 95\% confidence bands. \par}  
	\end{scriptsize}
\end{figure}

\clearpage 

\begin{figure}[H]
\caption{Distribution of Average Marginal Treatment Effects of Volatility}
	\label{fig:Volatility_AvgTreatmentEffect}
\begin{subfigure}[b]{0.99\textwidth}
	\centering
	\includegraphics[width=\textwidth,height=0.42\textheight]{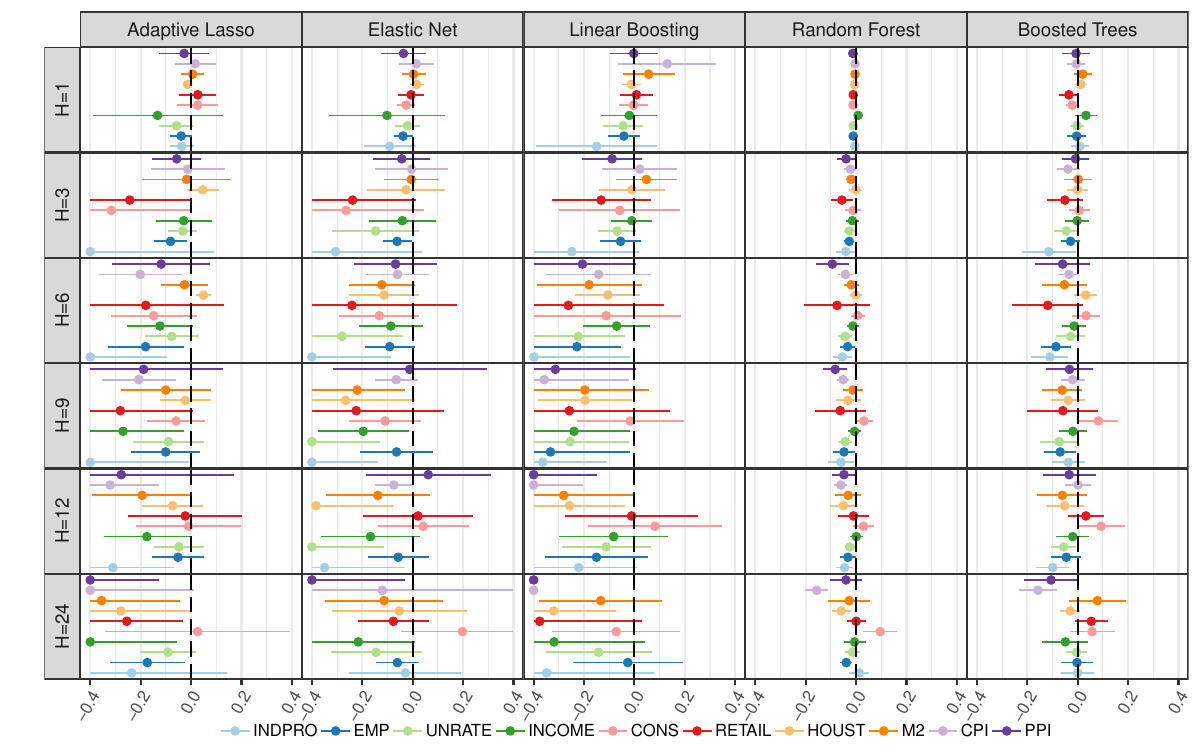}
	\caption{Direct Approach ($\hat{y}_{t+h}^{\text{direct}}$)}
	\label{fig:Volatility_AvgTreatmentEffect_direct}
\end{subfigure}
\\
\begin{subfigure}[b]{0.99\textwidth}
	\centering
	\includegraphics[width=\textwidth,height=0.42\textheight]{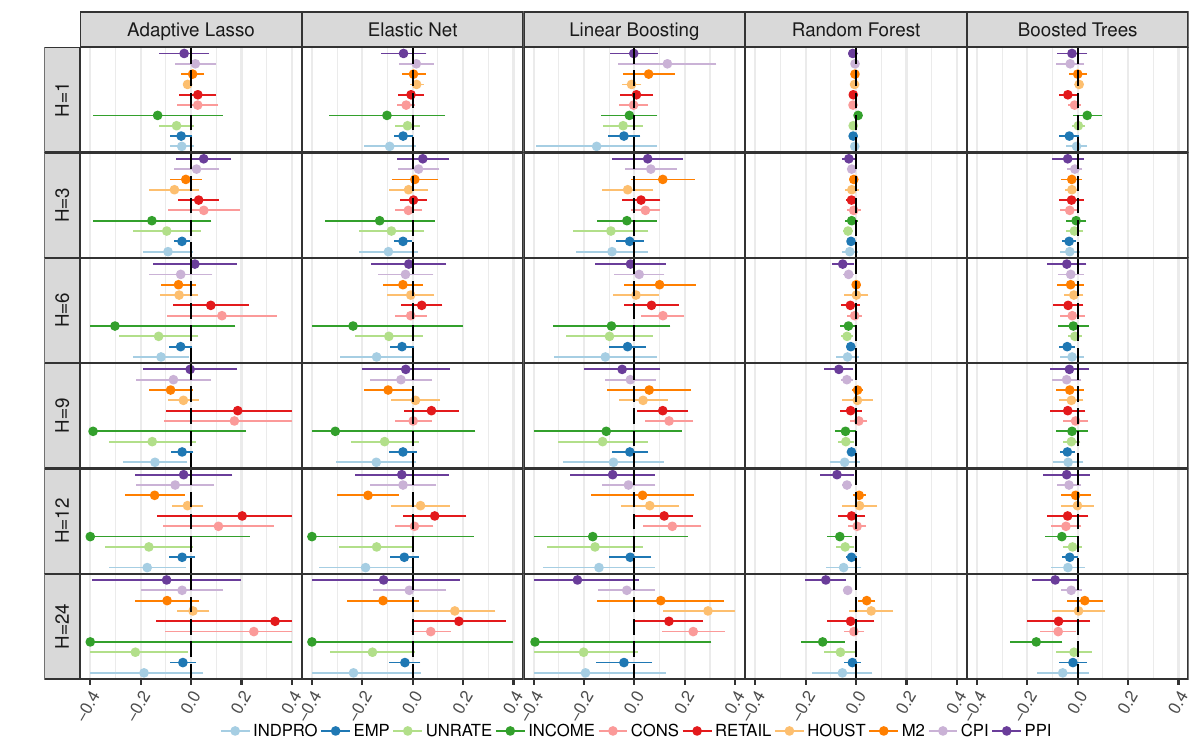}
	\caption{Path Average Approach ($\hat{y}_{t+h}^{\text{path-avg}}$)}
	\label{fig:Volatility_AvgTreatmentEffect_path}
\end{subfigure}
	\flushleft
	\begin{scriptsize}
		\vspace{-2.8em}
		{\singlespacing
			Note: This figure plots the distribution of ${\alpha}_f^{(h,v)}$ from equation (\ref{e_eq2}) done by $(h,v)$ subsets. It shows the average partial effect on the pseudo-$R^2$ from augmenting the model with $X^2$ and corresponding factors featuring, keeping everything else fixed. SEs are HAC. These are the 95\% confidence bands. \par}  
	\end{scriptsize}
\end{figure}

\clearpage 

\begin{figure}[H]
\caption{Distribution of Marginal Treatment Effects of Dynamic Factors vs MAF}
	\label{fig:DFMAF_TreatmentEffect}
\begin{subfigure}[b]{0.99\textwidth}
	\centering
	\includegraphics[width=\textwidth,height=0.42\textheight]{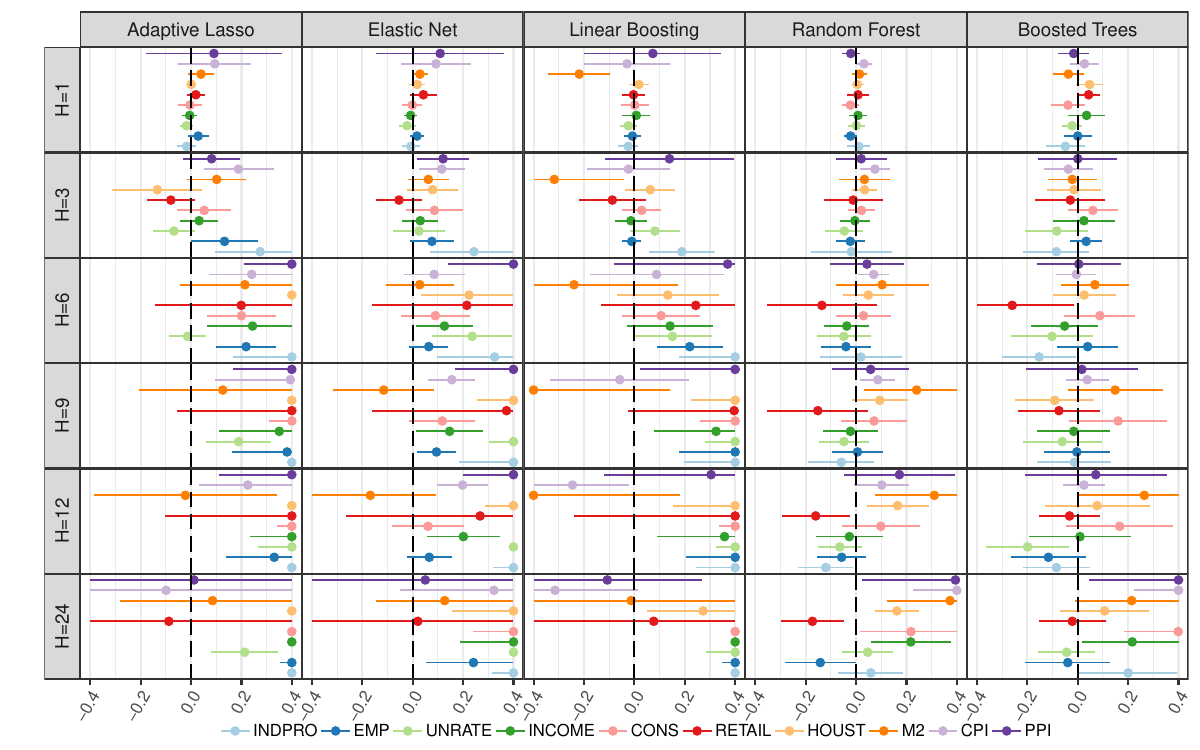}
	\caption{Direct Approach ($\hat{y}_{t+h}^{\text{direct}}$)}
	\label{fig:DFMAF_TreatmentEffect_direct}
\end{subfigure}
\\
\begin{subfigure}[b]{0.99\textwidth}
	\centering
	\includegraphics[width=\textwidth,height=0.42\textheight]{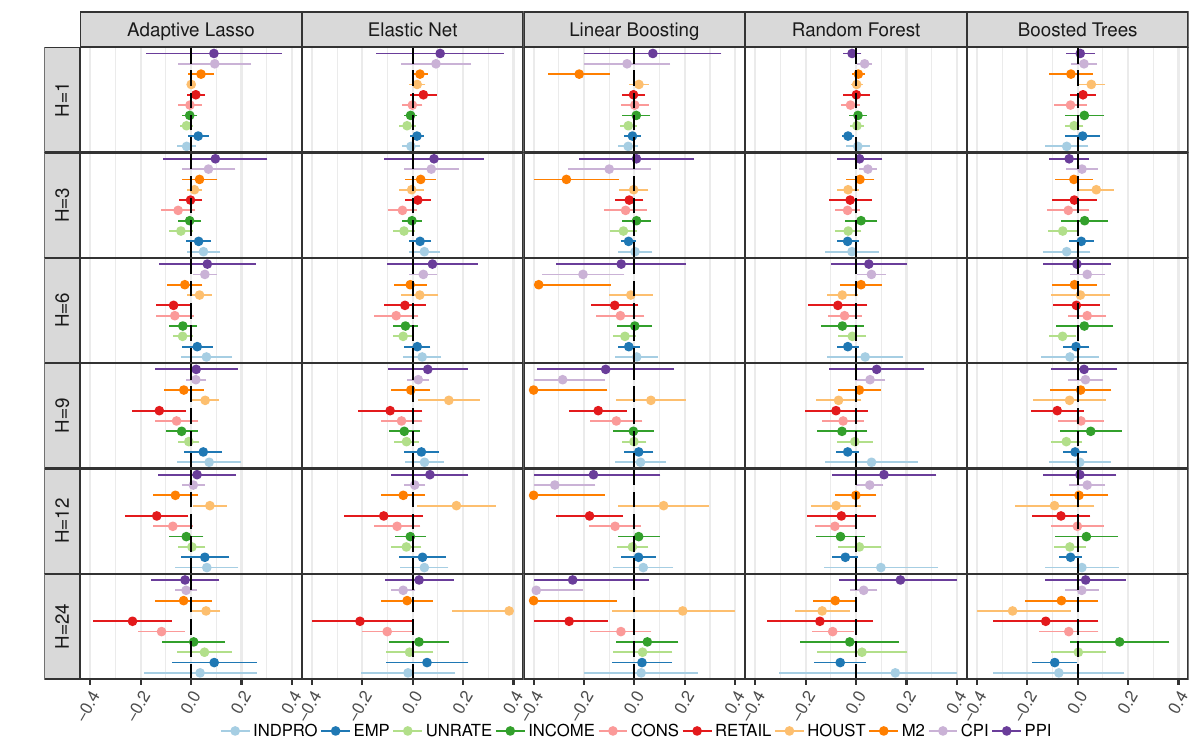}
	\caption{Path Average Approach ($\hat{y}_{t+h}^{\text{path-avg}}$)}
	\label{fig:DFMAF_TreatmentEffect_path}
\end{subfigure}
	\flushleft
	\begin{scriptsize}
		\vspace{-2.8em}
		{\singlespacing
			Note: This figure plots the distribution of ${\alpha}_f^{(h,v)}$ from equation (\ref{e_eq2}) done by $(h,v)$ subsets. It shows the average partial effect on the pseudo-$R^2$ from considering dynamic factors versus \textit{MAF}, keeping everything else fixed. SEs are HAC. These are the 95\% confidence bands. \par}  
	\end{scriptsize}
\end{figure}

\clearpage 

\begin{figure}[H]
\caption{Distribution of Marginal Treatment Effects of Dynamic Factors vs Static Factors}
	\label{fig:DFSF_TreatmentEffect}
\begin{subfigure}[b]{0.99\textwidth}
	\centering
	\includegraphics[width=\textwidth,height=0.42\textheight]{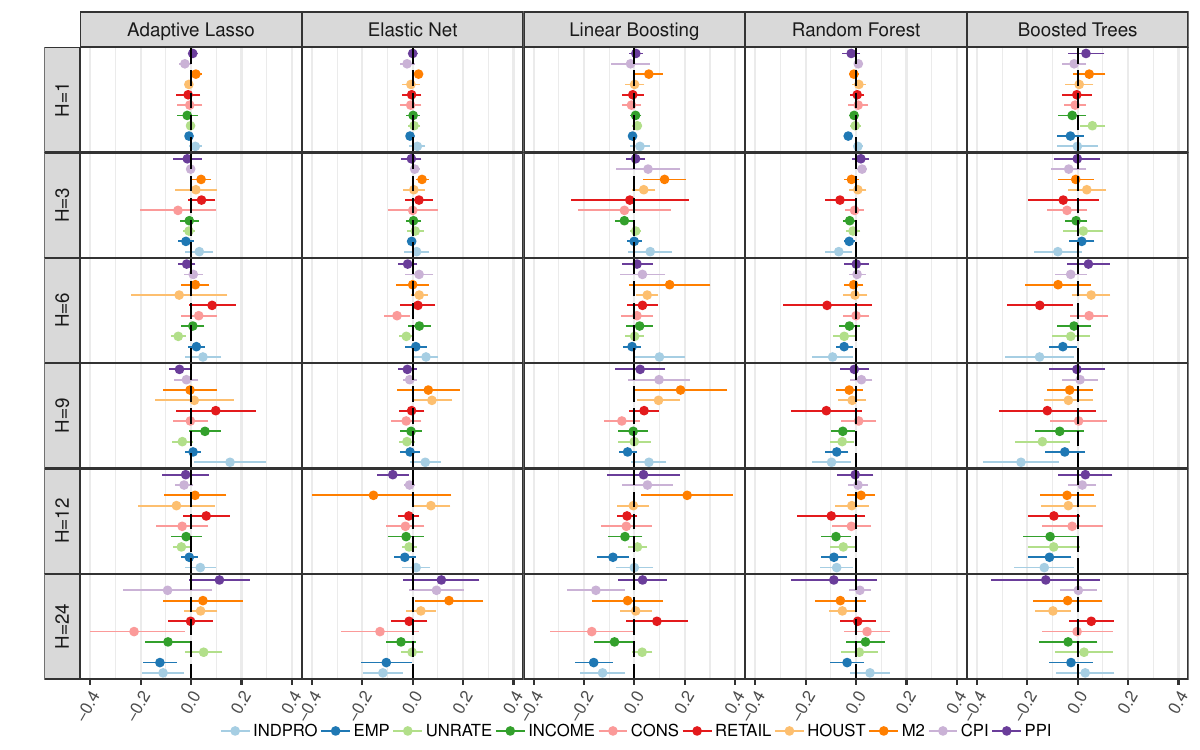}
	\caption{Direct Approach ($\hat{y}_{t+h}^{\text{direct}}$)}
	\label{fig:DFSF_TreatmentEffect_direct}
\end{subfigure}
\\
\begin{subfigure}[b]{0.99\textwidth}
	\centering
	\includegraphics[width=\textwidth,height=0.42\textheight]{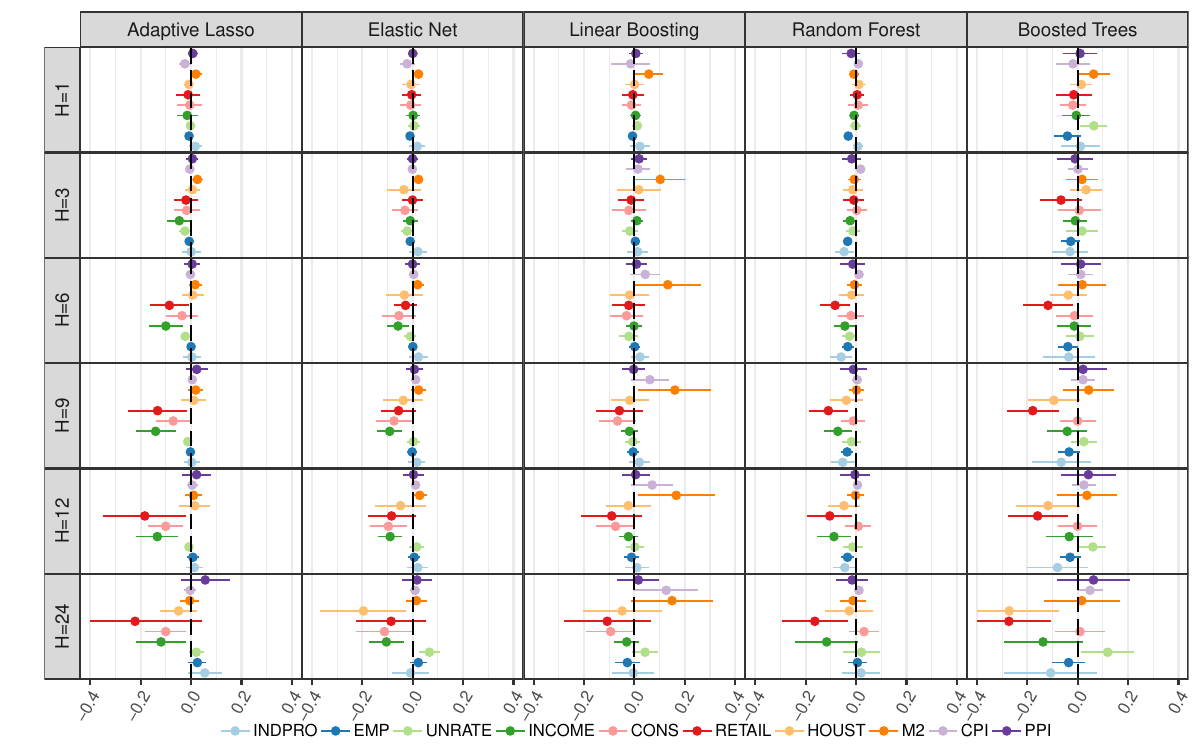}
	\caption{Path Average Approach ($\hat{y}_{t+h}^{\text{path-avg}}$)}
	\label{fig:DFSF_TreatmentEffect_path}
\end{subfigure}
	\flushleft
	\begin{scriptsize}
		\vspace{-2.8em}
		{\singlespacing
			Note: This figure plots the distribution of ${\alpha}_f^{(h,v)}$ from equation (\ref{e_eq2}) done by $(h,v)$ subsets. It shows the average partial effect on the pseudo-$R^2$ from considering dynamic factors versus static factors, keeping everything else fixed. SEs are HAC. These are the 95\% confidence bands. \par}  
	\end{scriptsize}
\end{figure}

\section{ONLINE APPENDIX - Additional Case Studies}\label{sec:add_results}

\begin{figure}[t!]
	\caption{Case of Employment (Direct)}\label{case:emp}
	\begin{subfigure}{\textwidth}
		\centering
		\includegraphics[width=3in, height=2.5in]{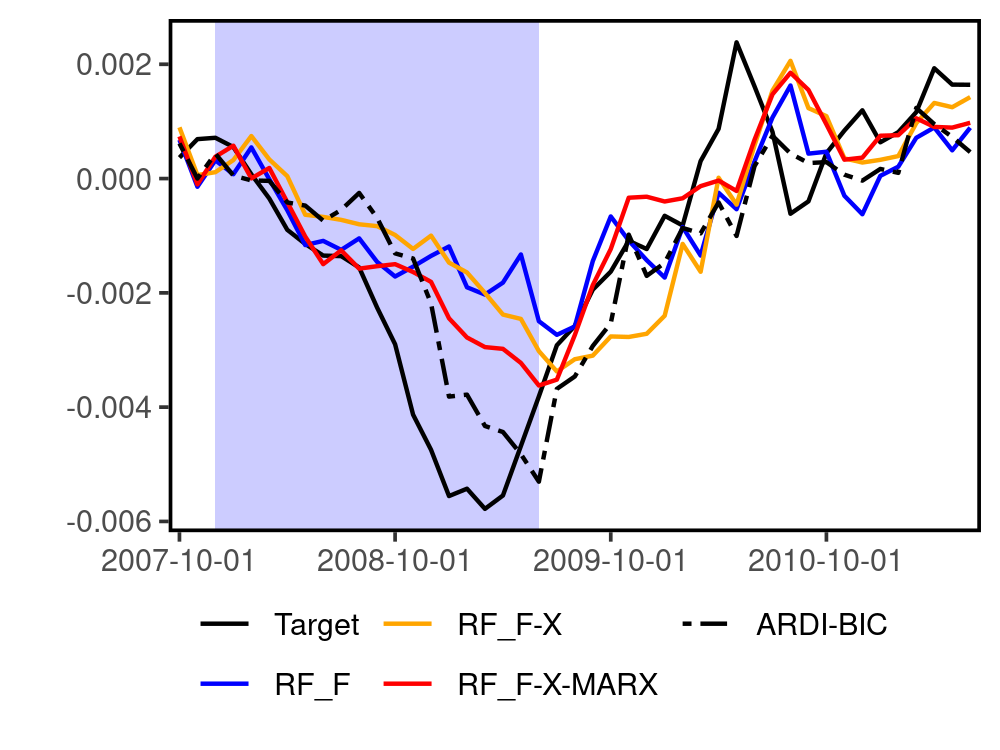}%
		\hspace{3em}
		\includegraphics[width=3in, height=2.5in]{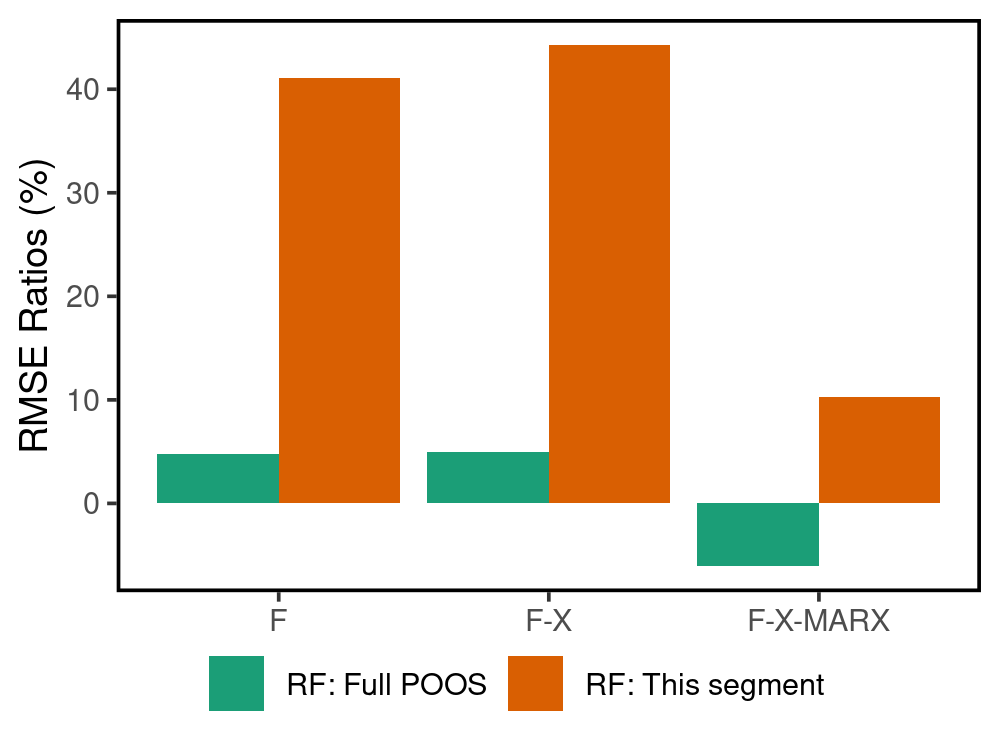}
		\caption{Recession Episode of 2007-12-01}
	\end{subfigure}
	\begin{subfigure}{\textwidth}
		\centering
		\includegraphics[width=3in, height=2.5in]{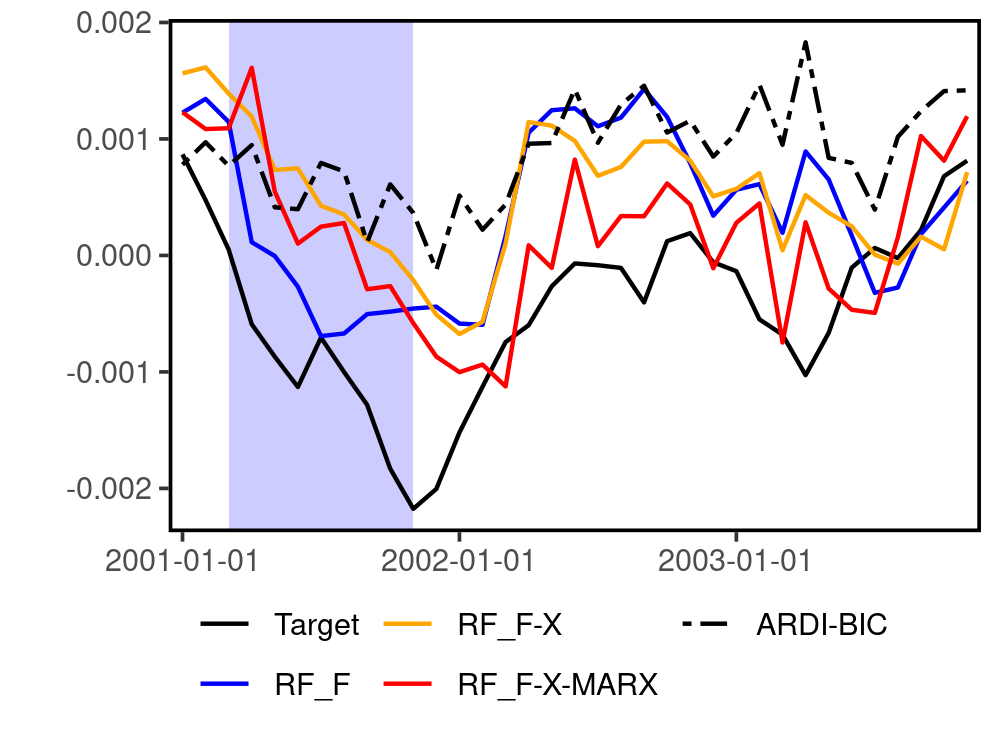}%
		\hspace{3em}
		\includegraphics[width=3in, height=2.5in]{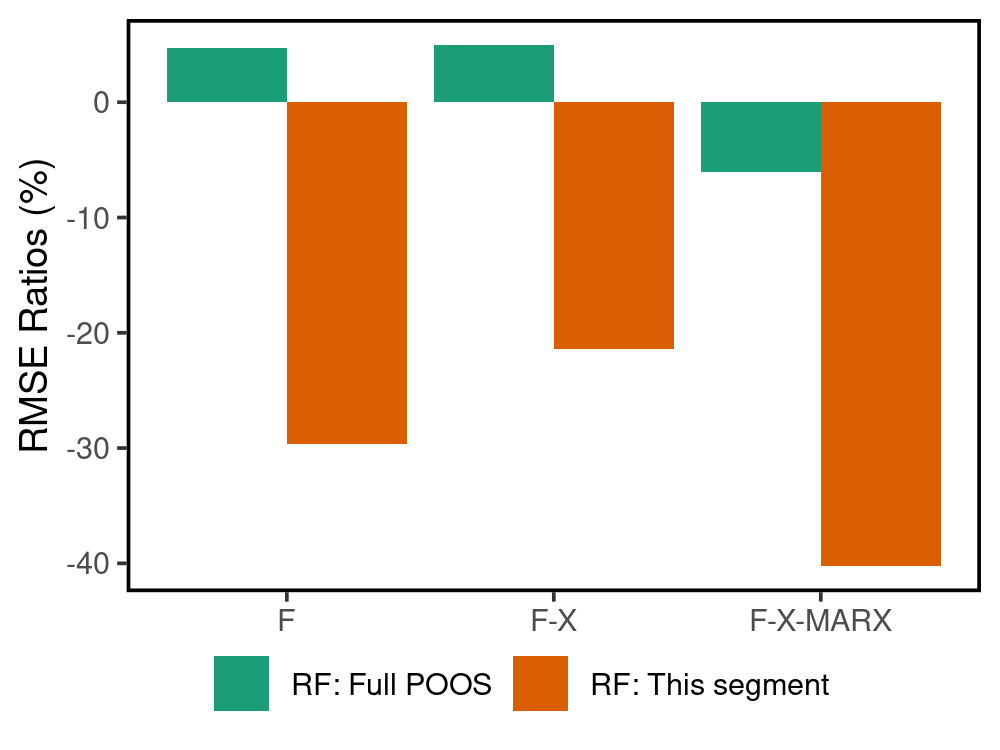}
		\caption{Recession Episode of 2001-03-01}
	\end{subfigure}
	\begin{subfigure}{\textwidth}
		\centering
		\includegraphics[width=3in, height=2.5in]{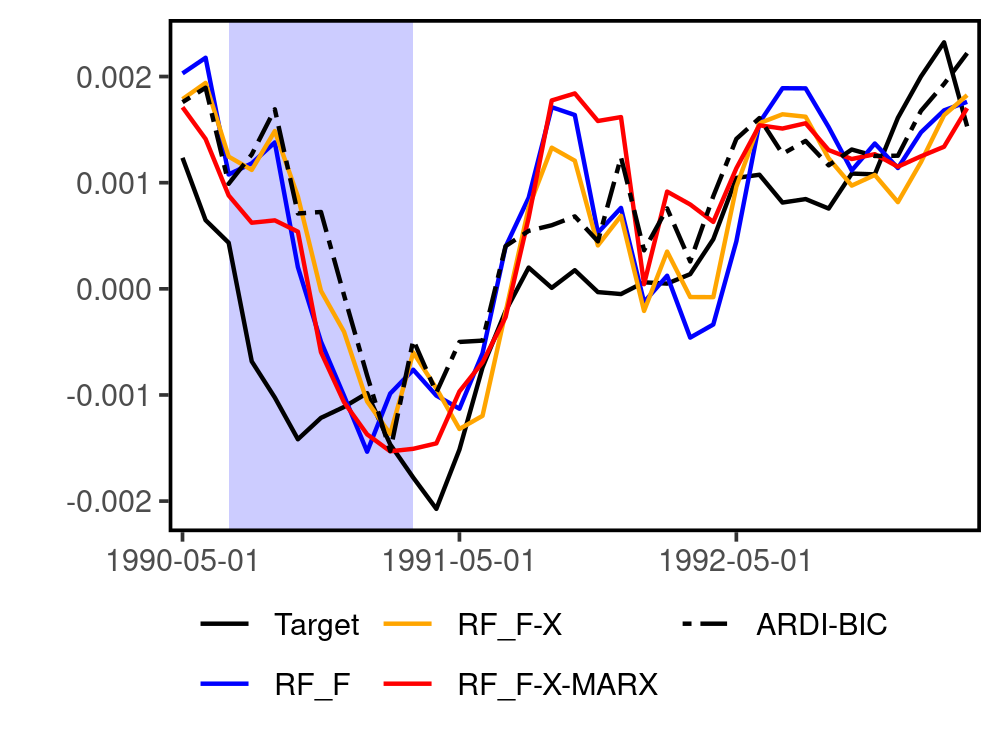}%
		\hspace{3em}
		\includegraphics[width=3in, height=2.5in]{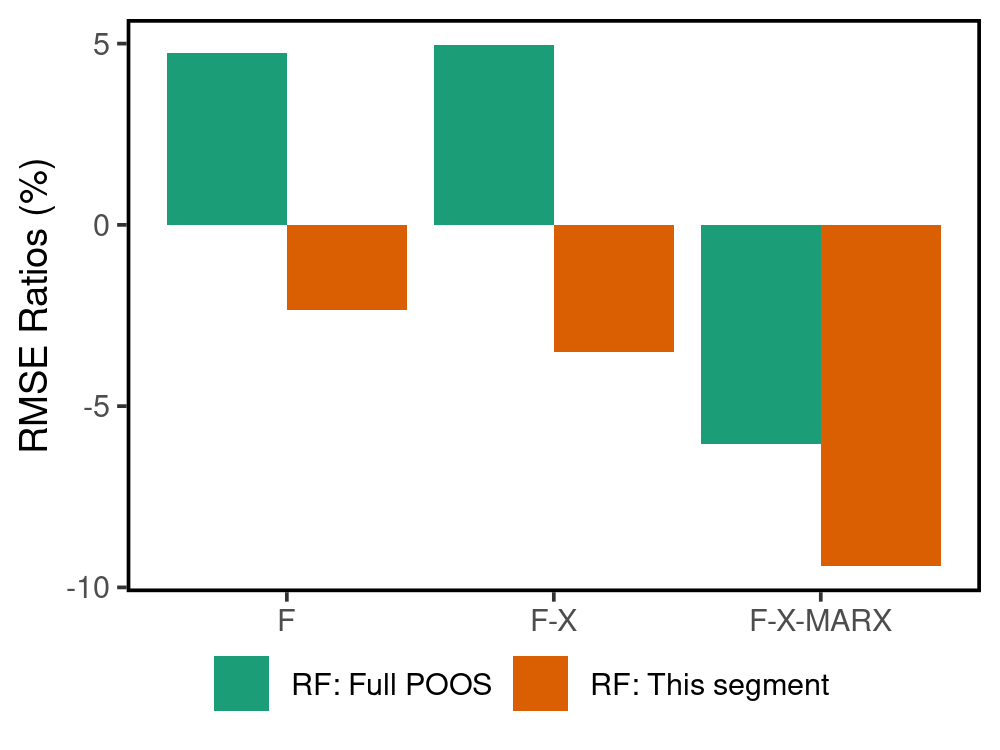}
		\caption{Recession Episode of 1990-07-01}
	\end{subfigure}
	\begin{scriptsize}
		\vspace{-3em}
		{\flushleft \singlespacing
			Note: The figure plots 3-month ahead forecasts for the period covering 3 months before and 24 months after the recession. RMSE ratios are relative to FM model and the episode RMSE refers to the visible time period. \par}
	\end{scriptsize}	
\end{figure}

\begin{figure}[t!]
	\caption{Case of Income (Path Average)}\label{case:income_sgrtoagr}
	\begin{subfigure}{\textwidth}
		\centering
		\includegraphics[width=3in, height=2.5in]{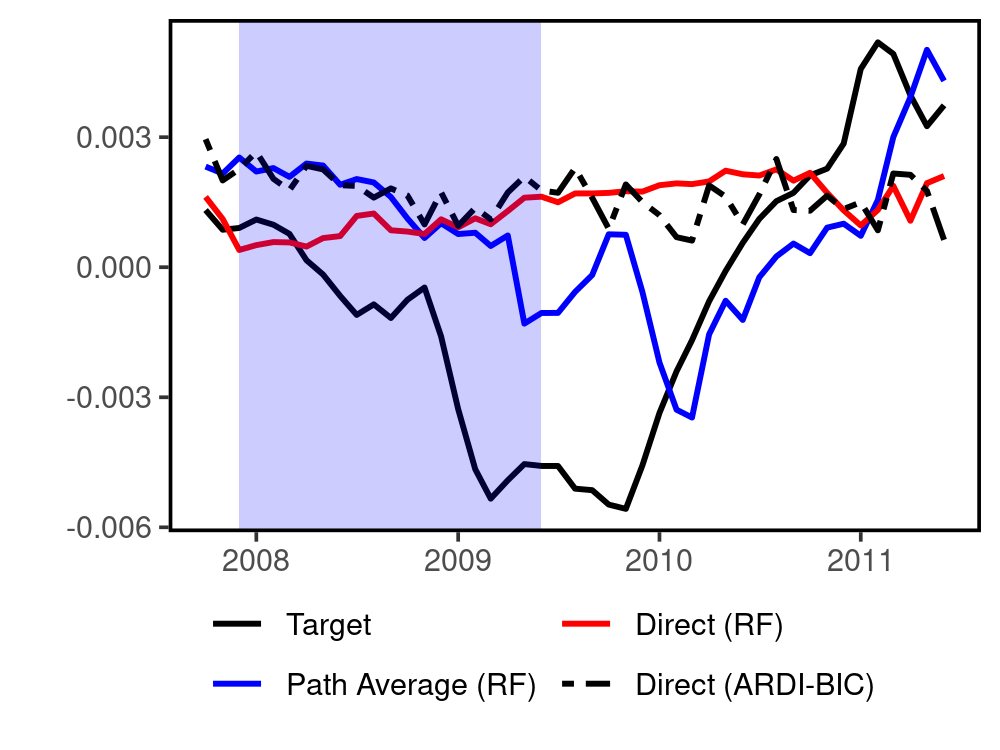}%
		\hspace{3em}
		\includegraphics[width=3in, height=2.5in]{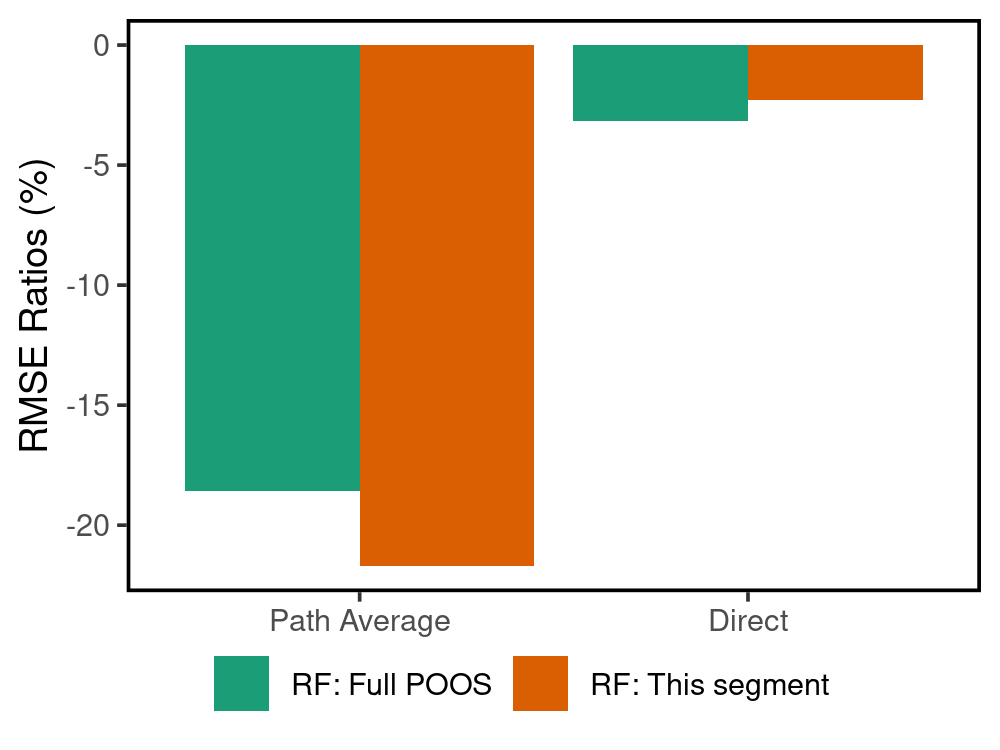}
		\caption{Recession Episode of 2007-12-01}
	\end{subfigure}
	\begin{subfigure}{\textwidth}
		\centering
		\includegraphics[width=3in, height=2.5in]{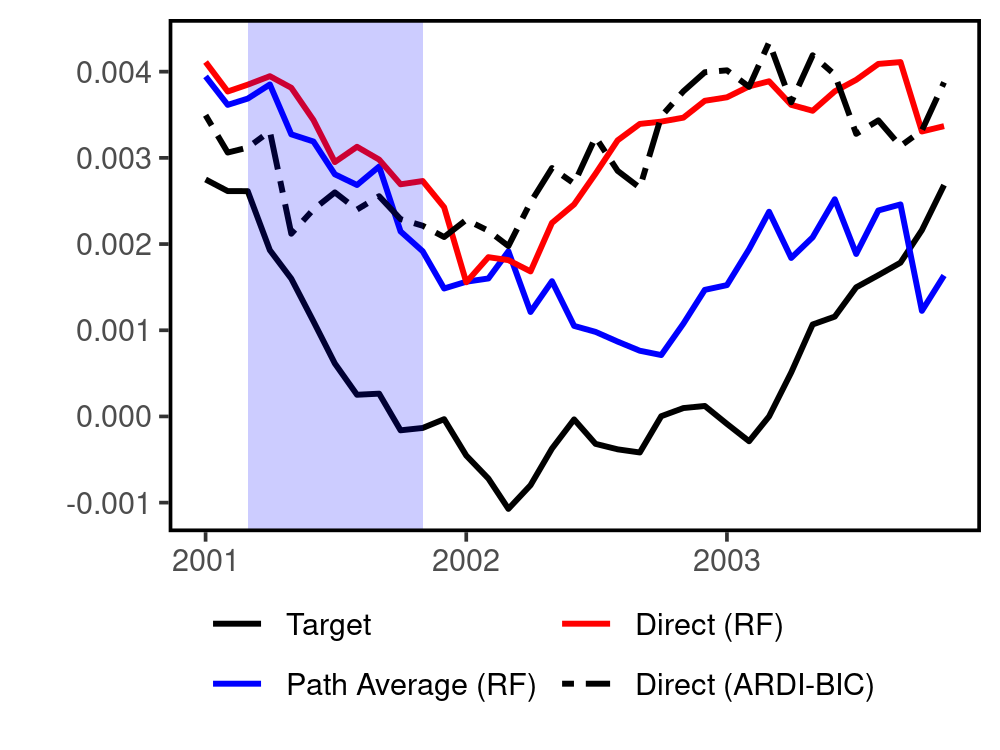}%
		\hspace{3em}
		\includegraphics[width=3in, height=2.5in]{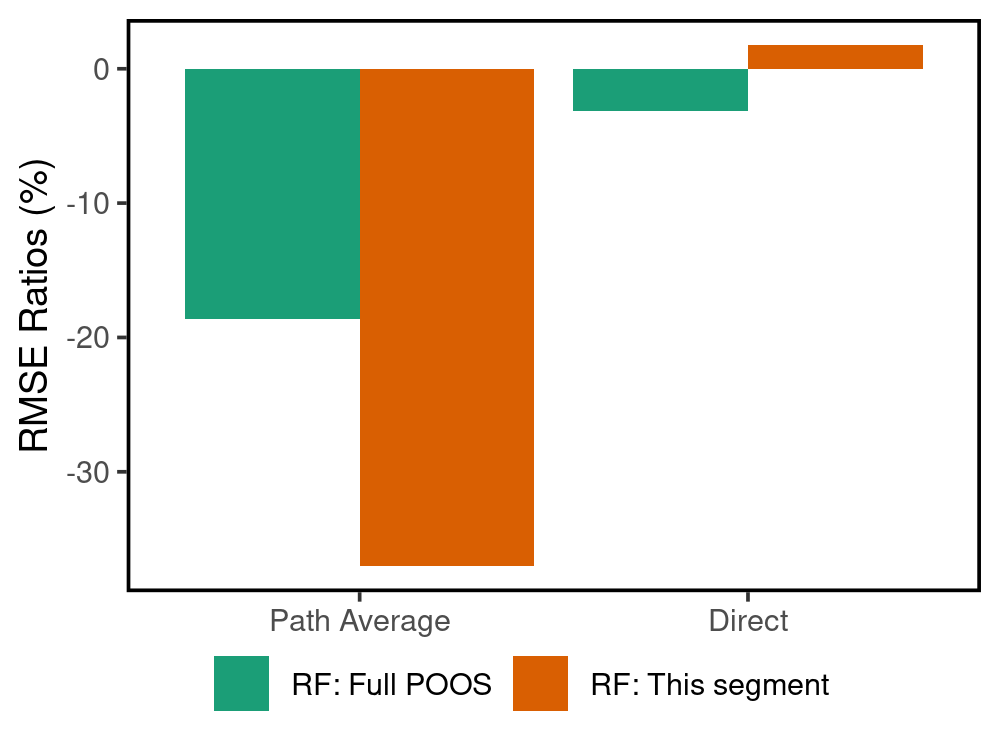}
		\caption{Recession Episode of 2001-03-01}
	\end{subfigure}
	\begin{subfigure}{\textwidth}
		\centering
		\includegraphics[width=3in, height=2.5in]{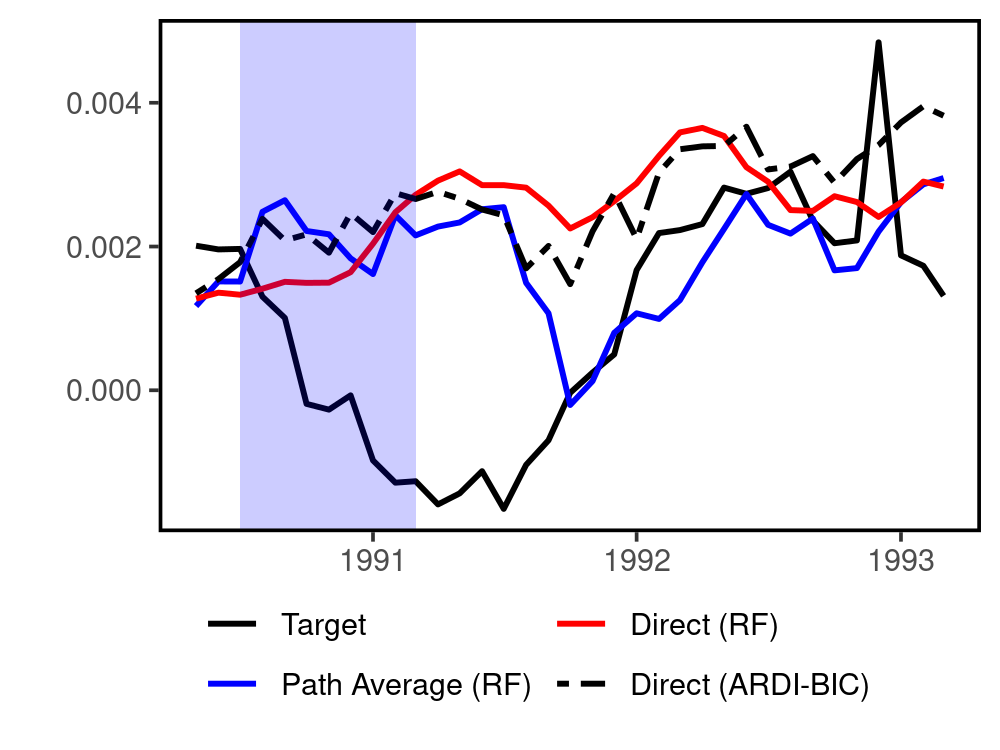}%
		\hspace{3em}
		\includegraphics[width=3in, height=2.5in]{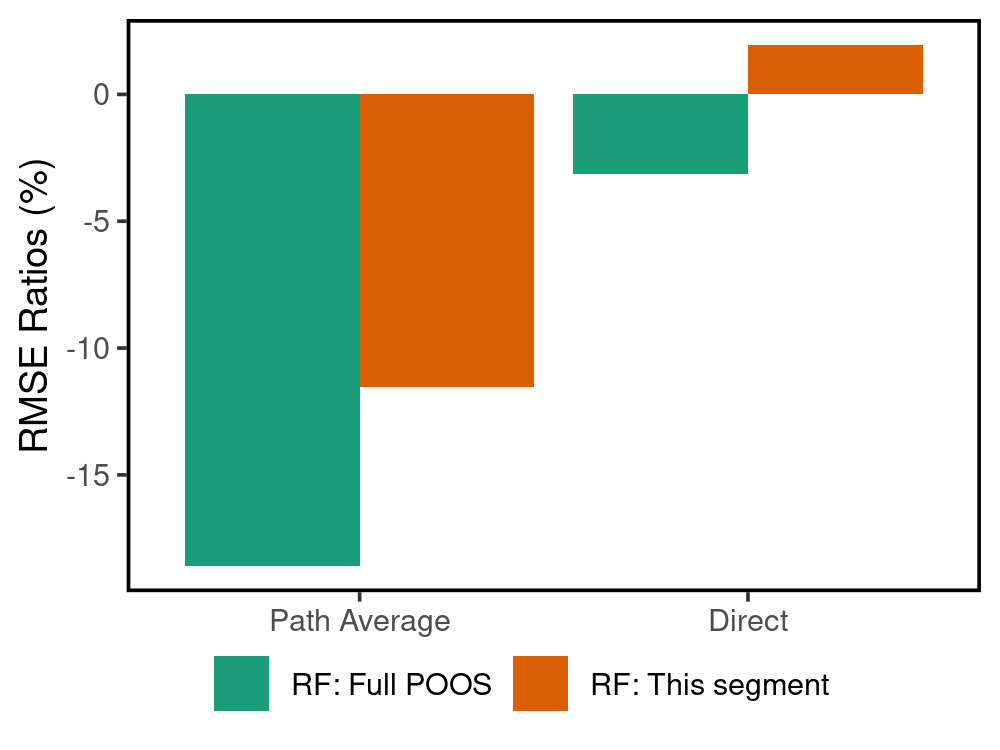}
		\caption{Recession Episode of 1990-07-01}
	\end{subfigure}
	\begin{scriptsize}
		\vspace{-3em}
		{\flushleft \singlespacing
			Note: The figure plots 12-month ahead forecasts for the period covering 3 months before and 24 months after the recession. RMSE ratios are relative to FM model for average growth rates and the episode RMSE refers to the visible time period and Random Forest models use F-X-MARX. \par}
	\end{scriptsize}	
\end{figure}

\end{document}